\documentclass[9pt, a4paper, twocolumn]{article}
\usepackage[a4paper, total={6.5in, 9.5in}]{geometry}
\usepackage{amssymb}
\usepackage{amsmath}
\usepackage{mathtools}
\usepackage{xurl}
\usepackage{xr}
\usepackage[numbers]{natbib}
\usepackage{abstract}
\usepackage{authblk}
\usepackage{color}
\usepackage{xcolor}
\usepackage[normalem]{ulem}
\usepackage{gensymb}
\usepackage[labelfont=bf]{caption}

\usepackage{lipsum}
\usepackage{graphicx}
\usepackage{hyperref}
\hypersetup{colorlinks=false, urlcolor=blue, linkcolor=blue, citecolor=blue}
\usepackage[switch]{lineno} 
\usepackage{titletoc}

\usepackage[utf8]{inputenc}

\usepackage{amsthm}
\usepackage{amsfonts}

\usepackage{comment}

\usepackage{upgreek}
\usepackage{siunitx}
\usepackage{ulem}
\usepackage{cancel}

\usepackage{setspace}		
\usepackage[english]{babel}
\usepackage{float}
\usepackage{booktabs}
\usepackage[hang]{footmisc} 	
\usepackage{subcaption}
\usepackage{xr}
\usepackage{authblk}
\usepackage[labelfont=bf]{caption}

\newcommand{\subr}[1]{_{\mathrm{#1}}}

\title{Intrinsically nonlinear nanomechanical resonator toward the single-phonon Kerr regime}

\author[1,3]{S. Klingel}
\author[1,3]{T. Althuon}
\author[1]{C. Wassmer}
\author[1]{C. Sürgers}
\author[2]{T. Cubaynes}
\author[1,2]{W. Wernsdorfer}
\affil[1]{PHI, Karlsruhe Institute of Technology, Kaiserstr. 12, 76131 Karlsruhe, Germany}
\affil[2]{IQMT, Karlsruhe Institute of Technology, Kaiserstr. 12, 76131 Karlsruhe, Germany}
\affil[3]{These authors contributed equally to this work.}

\begin{document}


\twocolumn[
\begin{@twocolumnfalse}
\maketitle

Manipulating a physical system at the single-quantum level typically requires an anharmonic energy spectrum. Nanomechanical resonators, however, are commonly admitted to be linear objects in the single-phonon regime, thus motivating most efforts to employ external sources of nonlinearity at the cost of added relaxation and decoherence channels. Here, we take a radically different approach by directly exploiting strong intrinsic mechanical nonlinearities. We quantify this nonlinearity using a novel method based on the coupling between bending modes. Using this technique, we establish an upper bound of $34.0 \pm 9.7$ phonons for the onset of intrinsically mechanical bistability. In addition, we highlight a new tuning mechanism based on a mechanical polarization angle of an upward-buckled nanotube resonator allowing us to reach an optimal working point. Extrapolating this level of intrinsic nonlinearity to realistically shorter resonators shows that the single-phonon Kerr regime is within reach, allowing for the realization of purely mechanical Schrödinger-cat states and highly coherent mechanical qubits.
\vspace{5mm}
\end{@twocolumnfalse}]
 

Mechanical systems are central to many quantum technologies. Their strong  appeal largely arises from their exceptional isolation from the environment, which makes them ideal candidates for storing quantum information \cite{hann2019hardware}. This directly reflects in the record high quality factors measured across a wide variety of nanomechanical platforms \cite{maccabe2020nano, ghadimi2018elastic, moser2014nanotube, huttel2009carbon, laird2012high, cupertino2024centimeter, dania2024ultrahigh}.

In order to attain an ultimate control of a mechanical degree of freedom some degree of anharmonicity in the mechanical energy spectrum is required. One way to express this requirement is via the so-called single-phonon Kerr regime, in which the addition of a single phonon shifts the mechanical resonance frequency by an amount that exceeds the resonator’s linewidth.

A natural strategy to reach this regime would be to engineer a nanomechanical resonator with sufficient intrinsic anharmonicity to remain nonlinear even at the level of a single phonon. However, realizing such a mechanical system has proven extremely challenging, which is why nanomechanical systems in the quantum regime are commonly regarded as linear resonators \cite{chu2017quantum, yang2024mechanical, pistolesi2021proposal}. As a result, most experimental platforms to date have relied on hybrid architectures, in which the mechanical mode inherits its nonlinearity from an external element rather than from its own dynamics.

In particular, interfacing mechanical resonators with superconducting qubits \cite{chu2017quantum} has opened the route toward quantum control of motion. Such hybrid architectures have enabled high-fidelity readout of mechanical quantum states \cite{cohen2015phonon, viennot2018phonon, arrangoiz2019resolving, von2022parity, sletten2019resolving}, the preparation of non-classical states of motion \cite{o2010quantum, wollack2022quantum, chu2018creation, satzinger2018quantum}, and even the generation of hybrid mechanical Schrödinger cat states \cite{bild2023schrodinger} and mechanical qubits \cite{yang2024mechanical}. In the case of carbon-nanotube mechanical resonators, another hybrid approach relying on the mechanical coupling to single-electron tunneling has recently been proposed \cite{pistolesi2021proposal} and experimentally explored \cite{samanta2023nonlinear}.

However, by inheriting the anharmonicity from an external object, these hybrid platforms inevitably exhibit additional dissipation paths, compromising the very coherence that makes nanomechanical systems attractive in the first place. Therefore, realizing a strongly nonlinear mechanical resonator without degrading its coherence remains an outstanding challenge. From a broader perspective, using intrinsic mechanical nonlinearity is also at the heart of new self-sustaining schemes for micromechanical oscillators \cite{chen2016self}.

\begin{figure*}[h!]
\centering
\includegraphics[width=1\linewidth]{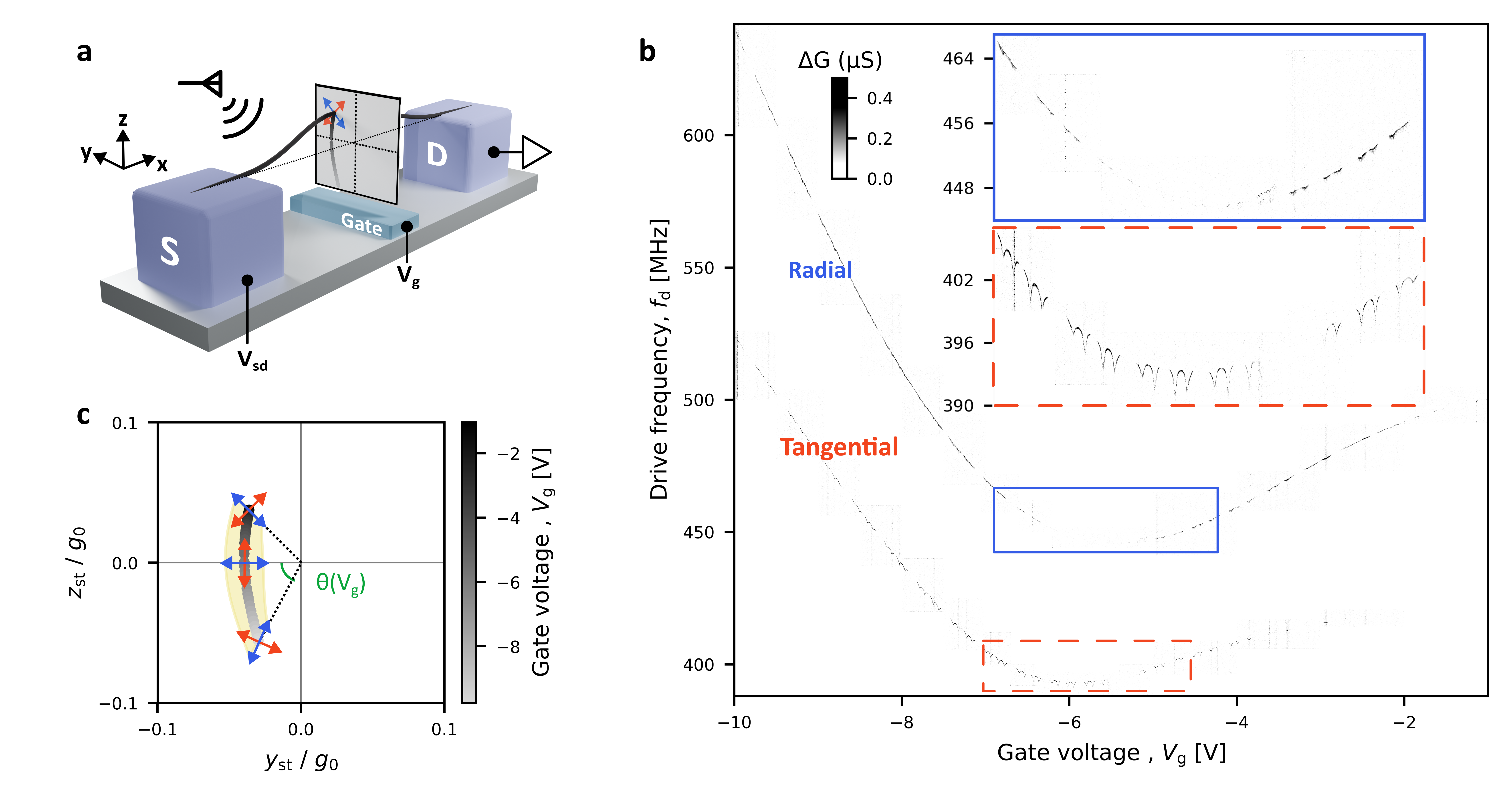}
\caption{\textbf{3D nanomechanics in an upward-buckled carbon nanotube.}
\textbf{a} Schematic of the nanomechanical device. The nanotube is 385$\pm 10\,$nm  long and upward-buckled at zero gate voltage. The buckling direction is tuned by applying a gate voltage $V_\mathrm{g}$ as represented on the $y$-$z$ plane. RF radiation is applied by a nearby antenna.
\textbf{b}
Differential conductance map as a function of the frequency $f_\mathrm{d}$ applied on the RF antenna and the gate voltage $V_\mathrm{g}$. The RF driving power is $P_\mathrm{d}=-40\,$dBm for the  radial  mode and $P_\mathrm{d}=-65\,$dBm for the  tangential  mode. The two insets are zooms in the vicinity of $V_\mathrm{g} = -6\,$V, highlighting the signal suppression and the enhanced spring softening effect for the  radial  and the  tangential  modes, respectively.
\textbf{c}
Static midpoint deflection of the nanotube in the $y$-$z$ plane as a function of the gate voltage $V_\mathrm{g}$. This displacement is scaled to the height difference between contact and gate electrodes $g_0$ and is obtained from the fit of the  radial  resonance frequency (Supplementary Section \ref{SI-th-fitting}). The initial buckling coordinates are: $y_0 = 2.3\pm 1.2\,$nm, and $z_0 = 2.5 \pm 1.2\,$nm. The light yellow area corresponds to the uncertainty obtained from a sensitivity analysis performed on the Euler-Bernoulli model (see Supplementary Section \ref{SI-th-sensitivity-analysis}). The blue and red arrows indicate the oscillation directions at different gate voltages of  radial  and  tangential  modes, respectively.
}
\label{fig1}
\end{figure*}


In this work, we overcome this limitation, and demonstrate a strongly nonlinear nanomechanical resonator solely based on intrinsic mechanical nonlinearities. For this purpose, we use a short and upward-buckled suspended carbon nanotube, a platform known to generate strong tension-induced intrinsic nonlinearity \cite{xu2017global, gibson2007vibrations, meerwaldt2012carbon}. The use of an upward-buckled nanotube also allows us to find an optimal working point where a maximum of nonlinearity coincides with a minimum of the resonance frequency. By measuring the nonlinear coupling between the two fundamental bending modes, we demonstrate the presence of a strong intrinsic Duffing nonlinearity corresponding to a bifurcation onset at only $34.0\pm 9.7$  phonons. This is two orders of magnitude lower than other platforms based solely on mechanical nonlinearity \cite{Eichler12}, and similar to recent hybrid mechanical systems \cite{samanta2023nonlinear}. Finally, we show that with modest optimization of the device dimensions, our mechanical platform could reach the single-phonon Kerr regime.

\subsection*{3D Nanomechanics}

The device consists of a $385\pm10\,$nm long semiconducting carbon nanotube suspended in between a source and a drain contact electrode. A gate electrode is placed below the carbon nanotube allowing us to form a single quantum dot circuit as represented in Fig.\ref{fig1}a. The mechanical motion of the carbon nanotube is actuated by a separate RF antenna. All measurements are performed in a dilution cryostat at the base temperature of $40\,$mK.

By measuring the conductance of the device as a function of the RF drive frequency, one can detect the mechanical bending modes of the carbon nanotube \cite{sazonova2004tunable}. Two distinct mechanical resonances are visible in the range $400-600 \,$MHz (Fig.\ref{fig1}b, and Supplementary Section \ref{SI-EMcharac-Frequencysweep}).
The low-frequency mode corresponds to the dangling motion of the nanotube, and its spring constant mainly comes from bending energy. In the following, we will refer to it as the  tangential  mode \cite{wang2021visualizing}. The higher frequency mode has a bouncing motion, and its larger spring constant is due to its dual origin, namely the bending and the tension energies. We will refer to it as the  radial  mode \cite{wang2021visualizing}.


From the evolution of both resonance frequencies as a function of the gate voltage $ V_\mathrm{g} $ one can infer further information about the nature of the two modes (Fig.\ref{fig1}b and Supplementary Section \ref{SI-EMcharac-Frequencygatemaps}). Aiming for a stronger read-out signal, we operate the device in the Coulomb blockade regime ($V_\mathrm{g} < 0$). The resonance frequencies of both modes display an unusual non-monotonic gate voltage dependence. From $V_\mathrm{g}=0\,$V to $V_\mathrm{g} \simeq-6\,$V, the two resonance frequencies first decrease, and then increase for more negative gate voltages. This behavior can be explained by an upward-buckling of the carbon nanotube \cite{rechnitz2022mode}. In the following more features are presented, supporting the scenario of an upward-buckled carbon nanotube. The capacitive attraction force between the nanotube and the gate electrode first induces compression in the nanotube. When the nanotube lies in a plane parallel to the substrate, the compression reaches a maximum, corresponding to a minimum of both resonance frequencies. A further increase in gate voltage induces axial tension in the nanotube and consequently leads to a rise of the resonance frequencies of the two bending modes \cite{ustunel2005modeling}. This behavior is well captured by an Euler-Bernoulli beam model in three dimensions, including built-in tension (Supplementary Section \ref{SI-th}). 

From this model one can also extract the static displacement of the carbon nanotube [$y_\mathrm{st}(V_\mathrm{g})$, $z_\mathrm{st}(V_\mathrm{g})$],  with $y_\mathrm{st}$ and $z_\mathrm{st}$ the carbon nanotube midpoint deflection coordinates. The nanotube describes a partial rotation in the $y$-$z$ plane as a function of the gate voltage (Fig.\ref{fig1}c).

Importantly, this displacement is accompanied by a change in oscillation direction of the two modes, which is captured by the polarization angle $\theta = \text{arctan}(\frac{z_\mathrm{st}}{y_\mathrm{st}})$ (Fig.\ref{fig1}c). This change in direction strongly affects the mechanical modulation of the dot-gate capacitance $C_\mathrm{g}$ by an amount  $\delta C_\mathrm{g} = \frac{\mathrm{d}C_\mathrm{g}}{\mathrm{d}z} \delta z (\theta)$, which has two direct implications. First, the amplitude of the signal can be strongly suppressed, and second, the strength of the spring softening effect (due to the coupling of the mechanical motion to single-electron tunneling \cite{steele2009strong,lassagne2009coupling}) can be changed. 

Both of these effects manifest clearly around $V_\mathrm{g} = -6\,$V, where the polarization angle $\theta \simeq 0°$. For the  radial  mode, a complete suppression of the signal is visible (upper inset of Fig.\ref{fig1}b). This is due to the oscillation of the mode being solely along the $y$-axis, resulting in $\delta C_\mathrm{g} = 0$.  Conversely, in case of the  tangential  mode, the motion is mostly along the $z$-axis, which leads to a large $\delta C_\mathrm{g}$ and consequently an enhanced spring softening effect. This is visible as more pronounced dips in the resonance frequency as a function of the gate voltage (lower insets of Fig.\ref{fig1}b). The control of the oscillation direction plays a key role in the tunability of the nonlinearities, as we will see in the following.

\subsection*{Nonlinearity in an upward-buckled nanotube}

Because the carbon nanotube resonator studied in this work is doubly-clamped, strong axial strain can build up along the nanotube. The resulting elastic potential energy is the dominant source of intrinsic mechanical nonlinearity \cite{Lif08}. In the following, we focus on the nonlinearity of the  radial  mode, which is expected to exhibit stronger intrinsic nonlinearities (Supplementary Section \ref{SI-th}). This is because the nanotube undergoes greater elongation during the  radial  motion than during the  tangential  motion. By retaining only the lowest-order term in the Taylor expansion of the axial strain, the Euler-Bernoulli beam equation results in the following equation of motion of the  radial  mode (Supplementary Section \ref{SI-th}):

\begin{equation}
    \ddot{u}_\mathrm{r} + \omega^2_\mathrm{r} u_\mathrm{r} + \alpha_{2,\mathrm{r}} u_\mathrm{r}^2 + \alpha_{3,\mathrm{r}} u_\mathrm{r}^3 = F_\mathrm{r} .
\label{eq:EOM1}
\end{equation}
Here, $u_\mathrm{r}$ is the amplitude of motion of the  radial  mode, and $\alpha_{2,\mathrm{r}}$ and $\alpha_{3,\mathrm{r}}$ are the second- and third-order nonlinearity parameters of the  radial  mode, which can be written as:
\begin{align}
    \alpha_{2,\mathrm{r}} & = c_0(\theta) y_\mathrm{st} + c_1(\theta) z_\mathrm{st} \\
    \alpha_{3,\mathrm{r}} & = \cos^2(\theta) \alpha_{3,y} + \sin^2(\theta) \alpha_{3,z}.
\label{eq:alpha23}
\end{align}
$c_i(\theta)$ are some geometric factors dependent on the polarization angle $\theta$ (see Supplementary Section \ref{SI-th-derivationequations}). The form of $\mathrm{\alpha_{3,\mathrm{r}}}$ assumes an anisotropy ($\alpha_{3,z} \neq \alpha_{3,y}$), the origin of which will be discussed in the next section.


Besides this purely mechanical nonlinearity originating from axial tension, another important source of nonlinearity in carbon nanotube devices arises from the strong coupling to single-electron tunneling \cite{samanta2023nonlinear, vigneau2022ultrastrong, steele2009strong, sevitz2025sources}. This so-called electromechanical nonlinearity is a direct consequence of the spring softening effect \cite{steele2009strong}. In order to extract solely the purely mechanical nonlinearity, we propose to use a novel extraction method which allows us to unambiguously distinguish the intrinsic mechanical nonlinearity from its electromechanical counterpart \cite{sevitz2025sources}.
This method relies on the fact that the intrinsic and the electromechanical nonlinearities have very distinct behaviors as a function of the amplitude of oscillations. This effect can be directly observed by measuring the nonlinear coupling between the  tangential  and the  radial  mode as a function of the amplitude of oscillations of one of the two modes. 


\begin{figure*}[h!]
\centering
\includegraphics[width=1\linewidth]{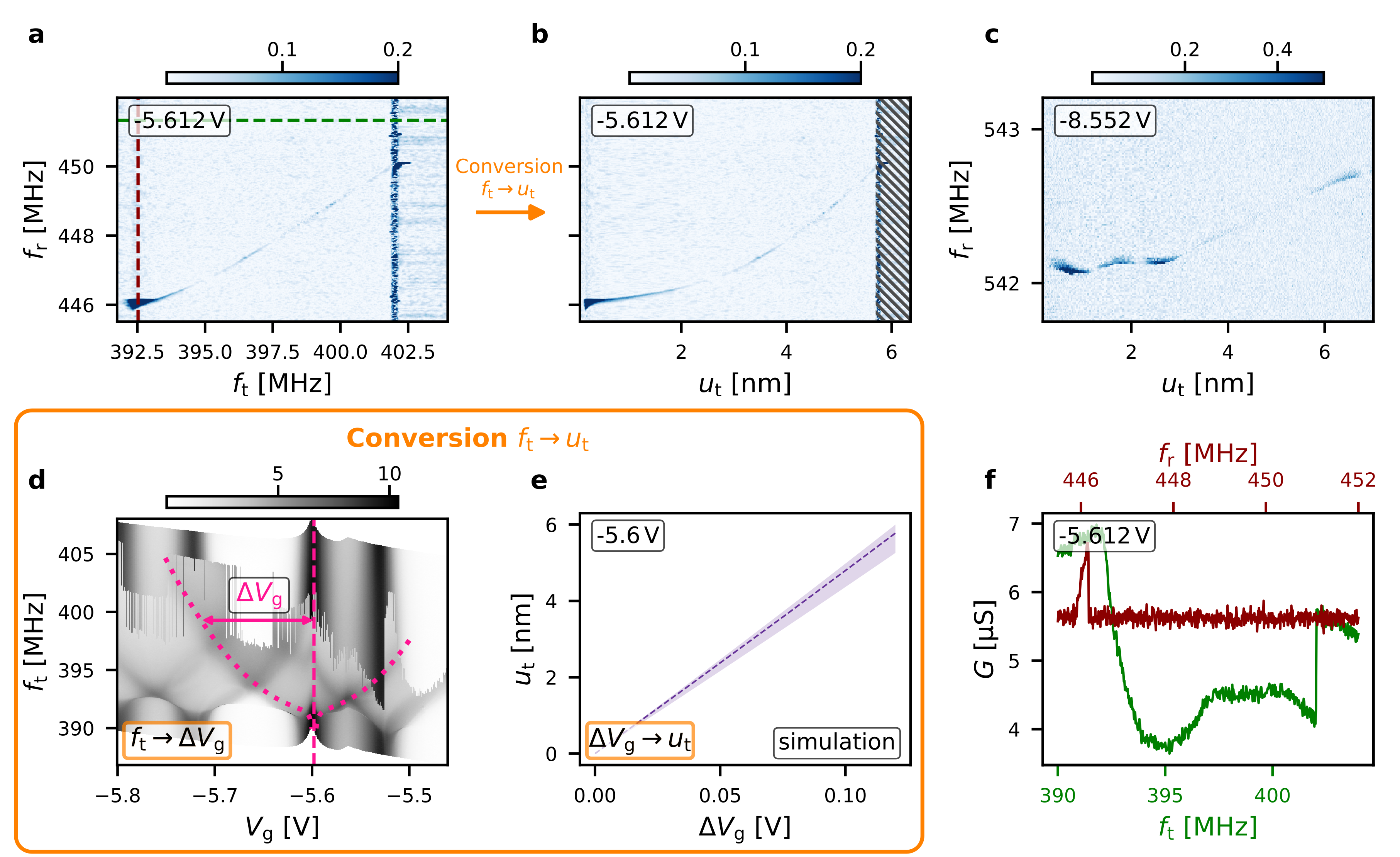}
\caption{
\textbf{Two-tone spectroscopy and amplitude calibration method.}
The method to calibrate the amplitude is illustrated  here  on the dataset around  $V_\mathrm{g}=-5.6\,$V.
\textbf{a}  a two-tone spectroscopy of the  radial  mode taken from the dataset at $V_\mathrm{g}=-5.6\,$V. The antenna RF driving powers of the  radial  and  tangential  modes are $P_\mathrm{r}=-40\,$dBm and $P_\mathrm{t}=-38.5\,$dBm, respectively. To improve visibility of the  radial  resonance, the variation of the conductance $|\Delta G| = | \frac{\partial I}{\partial V_\text{sd}} - \langle \frac{\partial I}{\partial V_\text{sd}} \rangle_\text{sweep}|$ is displayed.
\textbf{b} Same two-tone spectroscopy as in \textbf{a}  after conversion of the frequency axis $f_\mathrm{t}$ into a motion amplitude axis $u_\mathrm{t}$ . In the hatched area, the  tangential  mode ceases to oscillate and the amplitude axis $u_\mathrm{t}$ is not valid anymore.
\textbf{c} Converted two-tone spectroscopy extracted from another dataset at $V_\mathrm{g}=-8.5\,$V.
The stronger electromechanical contribution at this gate voltage results in local dips of $\omega_\mathrm{r, tot}/2\pi$ in direct contrast to \textbf{b}.
\textbf{d} Differential conductance map as a function of $f_\mathrm{t}$ and $V_\mathrm{g}$ at large RF driving power ($P_\mathrm{t}=-38.5\,$dBm). The vertical pink dashed line  is the position of the Coulomb peak in absence of mechanical oscillations . The pink dotted line represents the shift of the Coulomb peak $\Delta V_\mathrm{g}$ as a function of $f_\mathrm{t}$, caused by the large  tangential  mode displacement.  Such measurement allows to determine the relation $\Delta V_\mathrm{g}$ - $f_\mathrm{t}$ for a given Coulomb peak (here at $V_\mathrm{g}=-5.6\,$V) .
\textbf{e} Relation between the  tangential mode displacement $u_\mathrm{t}$ and the shift of the Coulomb peak $\Delta V_\mathrm{g}$ obtained from finite-element simulation of the electric potential in the device.  From the data presented in \textbf{d} and \textbf{e}, one determine the relation $u_\mathrm{t}$ - $f_\mathrm{t}$ which was used for the x-axis conversion in \textbf{a} and \textbf{b}. 
\textbf{f} Resonance curves of the  radial  and the  tangential  mode extracted from  the dashed lines in  \textbf{a}.
}
\label{fig2}
\end{figure*}

\subsection*{Mode-coupling–based nonlinearity extraction method}





When actuating the  tangential  and the  radial  motions simultaneously, additional terms will emerge in the equation of motion of the  radial  mode (Supplementary Section \ref{SI-th-derivationequations}). In particular, here we exploit the fact that the terms $ \alpha_{3,\mathrm{r}} u_\mathrm{r}^3$ and $ \alpha_{3,\mathrm{r}} u_\mathrm{r} u_\mathrm{t}^2$ share the same prefactor. This offers a novel way to measure the Duffing nonlinearity $\alpha_{3,\mathrm{r}}$ relying on the nonlinear coupling between the  tangential  and the  radial  modes.

By properly driving the two modes one can isolate the contribution of the nonlinear coupling term $u_\mathrm{r} u_\mathrm{t}^2$ and therefore perform a direct measurement of the Duffing nonlinearity. In practice, it means to drive strongly the  tangential  mode, to induce large variation of $u_\mathrm{t}$, and to apply a weak probe-tone at the  radial  mode frequency. This probe must remain sufficiently weak so that the term $\alpha_{3,\mathrm{r}} u_\mathrm{r}^3$ stays negligible, while being able to detect variations of $\omega_\mathrm{r}$.

Under such driving conditions, the term $u_\mathrm{r} u_\mathrm{t}^2$ leads to the following shift of $\mathrm{\omega_{r}}$:


\begin{equation}
    \mathrm{\omega_{r, int}} = \mathrm{\omega_{r,0}} \sqrt{1 + \frac{\mathrm{\alpha_{3,r}}}{2} \Big( \frac{u_\mathrm{t}}{\mathrm{\omega_{r,0}}} \Big)^2}.
\label{eq:omegat}
\end{equation}

The remaining coupling terms, $u_\mathrm{r} u_\mathrm{t}$ and $u_\mathrm{r}^2 u_\mathrm{t}$ do not contribute to the measured frequency shift, as they are averaged-out over the measurement integration time \cite{ari2018intermodal} (Supplementary Section \ref{SI-th-derivationequations}). 

An example of such a two-tone spectroscopy measurement is presented in Fig.\ref{fig2}.a. The validity of the assumption $u_\mathrm{t} \gg u_\mathrm{r}$ is confirmed by the large difference of mechanical signal between the  tangential  and the  radial  mode, as shown in Fig.\ref{fig2}.f. In particular, the  tangential  resonance extends over more than $10\,$MHz, while the  radial  resonance is limited to a few $100\,$kHz (more details in Supplementary Section \ref{SI-approx_validity}).

We then developed an analysis method (Supplementary Section \ref{SI-alpha3t}) to reliably extract the intrinsic Duffing parameter $\mathrm{\alpha_{3,r}}$ from the two-tone spectroscopy measurements. In order to fit the dependence of $ \mathrm{\omega_{r}}(u_\mathrm{t})$ using the expression in equation \ref{eq:omegat}, one has to translate the frequency $f_\mathrm{t}$-axis into an amplitude $u_\mathrm{t}$-axis (see Fig.\ref{fig2}.a and \ref{fig2}.b). This conversion is central to our analysis method and is performed in two steps. First, we perform a single-tone measurement of the  tangential  mode with the exact same driving power as the one used in the two-tone spectroscopy from which we want to convert the axis. In Fig.\ref{fig2}.d an example of such a measurement is shown. When the  tangential  mode is driven into resonance, Coulomb peaks are split due to the capacitance modulation produced by the mechanical oscillation of the nanotube.  The splitting of the Coulomb peaks is visible as arch-like structures (see Fig.\ref{fig2}.d).  This allows us to convert the frequency $f_\mathrm{t}$ into an effective gate voltage shift $\Delta V_\mathrm{g}$ which is defined as half of the full Coulomb peak splitting (see Fig.\ref{fig2}.d).

In a second step, this effective gate voltage shift $\Delta V_\mathrm{g}$ is converted into a mechanical displacement $u_\mathrm{t}$. This step relies on the precise knowledge of the electrical potential produced by the gate electrode which is estimated using finite-element simulations (COMSOL) (see Fig.\ref{fig2}.e and Supplementary Section \ref{SI-alpha3t-DeltaVgus}). The accuracy of the $\Delta V_\mathrm{g}$ to $u_\mathrm{t}$ conversion strongly depends on the uncertainty on the dimensions of the circuit. To minimize this uncertainty, we measure the lateral and vertical dimensions of the circuit at the exact location of the nanotube using scanning electron microscopy and atomic force microscopy, respectively. More details on the uncertainty estimations and their propagation are given in Supplementary Section \ref{SI-estimationuncertainty}. Moreover, the predictions of the simulation are validated by gate lever-arm measurements (see Supplementary Section \ref{SI-FEM_benchmarking}).

Importantly, this method is only weakly dependent on the static displacement of the nanotube predicted by the Euler-Bernoulli model. This is due to the large gate-nanotube distance ($g_0=65\,$nm) compared to the total static displacement variation ($\Delta z_\text{st} = 7\,$nm).


In general, the observed variations of $ \mathrm{\omega_{r}}(u_\mathrm{t})$ have contributions from the intrinsic and the electromechanical nonlinear couplings, and can be written as: $ \mathrm{\omega_{r}}(u_\mathrm{t}) = \mathrm{\omega_{r,int}}(u_\mathrm{t}) + \mathrm{\omega_{r,elec}}(u_\mathrm{t})$. Both contributions depend on the gate voltage which can lead to very different $\mathrm{\omega_{r}}(u_\mathrm{t})$. The two-tone spectroscopies shown in Fig.\ref{fig2}.b and \ref{fig2}.c are taken at different gate voltages and highlight the two main scenarios. In Fig.\ref{fig2}.b, the evolution of $\mathrm{\omega_{r}}(u_\mathrm{t})$ is dominated by the intrinsic nonlinear coupling, and follows the evolution given in equation \ref{eq:omegat}. In Fig.\ref{fig2}.c, the two contributions are comparable, and the non-monotonic dependence of $ \mathrm{\omega_{r}}(u_\mathrm{t})$ is due to the non-negligible electromechanical coupling. More example of two-tone spectroscopy are given in Supplementary Section \ref{SI-alpha3t-structuredataset}.

\begin{figure*}[h!]
\centering
\includegraphics[width=1\linewidth]{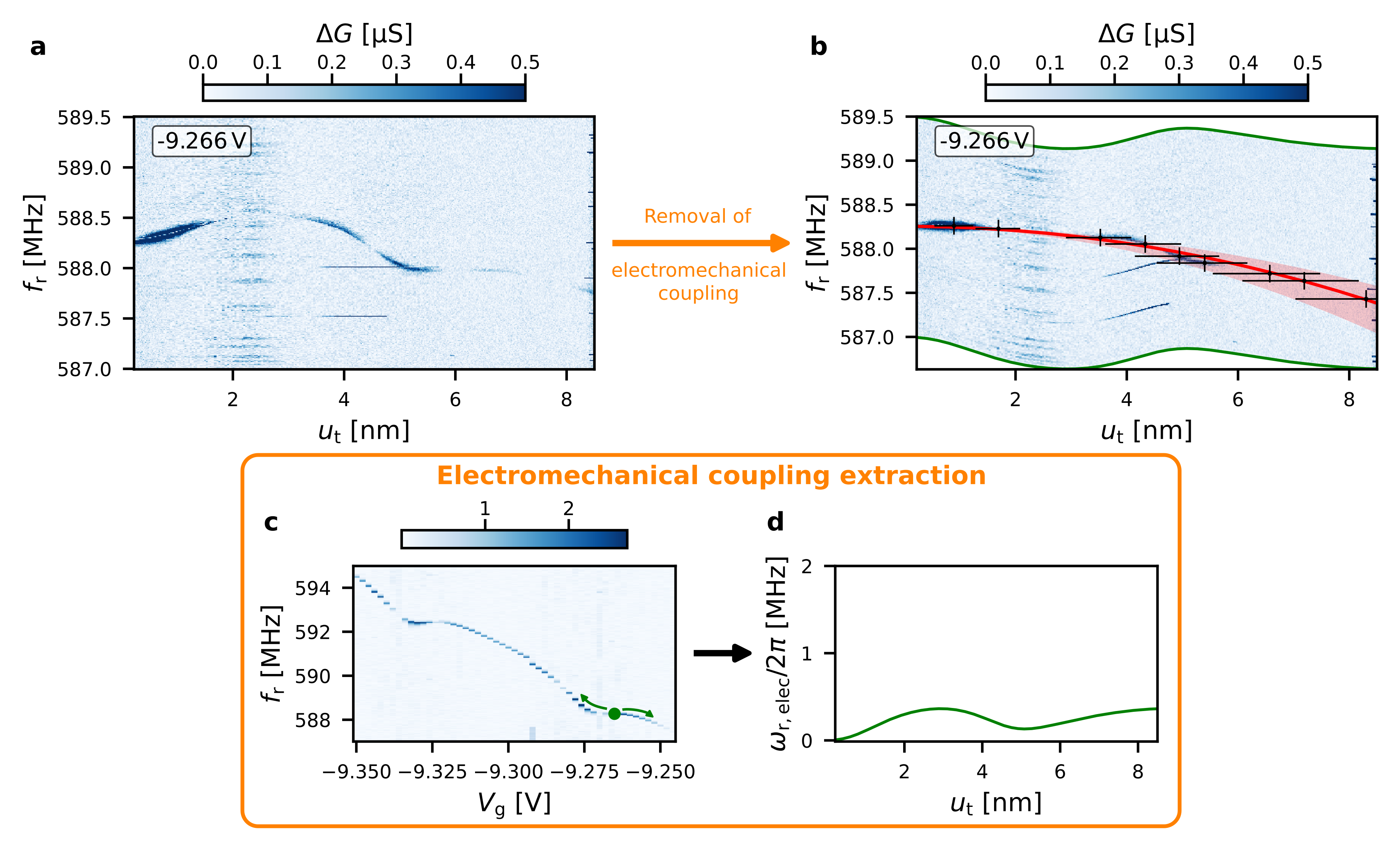}
\caption{
\textbf{Extraction of the intrinsic Duffing nonlinearity.} 
The method to extract the intrinsic Duffing nonlinearity is illustrated  here  on the dataset around $V_\mathrm{g}=-9.2\,$V.
\textbf{a} $f_\mathrm{r} - u_\mathrm{t}$ two-tone spectroscopy taken at $V_\mathrm{g}=-9.266\,$V ($P_\mathrm{r}=-60\,$dBm and $P_\mathrm{t}=-37\,$dBm).  The visible dark blue line is the resonance frequency of the tangential mode $\omega_\mathrm{r}/2\pi$. 
\textbf{b} Same two-tone spectroscopy as in \textbf{a} with  the removal of the electromechanical contribution to the variation of $\omega_\mathrm{r}/2\pi$ .  The variation of $\omega_\mathrm{r}/2\pi$ as a function of $u_\mathrm{t}$ is therefore only due to the intrinsic nonlinearity. $\omega_{\mathrm{r,int}}(u_\mathrm{t})$ is fitted using equation \ref{eq:omegat} to obtain the intrinsic nonlinearity $\alpha_{3,\mathrm{r}}$. Details on the fitting procedure and the uncertainty estimation are given in Supplementary Section \ref{SI-alpha3t}.
\textbf{c} Resonance frequency of the radial mode  as a function of the gate voltage at low driving power ($P_\mathrm{r}=-60\,$dBm). The green  dot indicates the gate voltage position $V_\mathrm{g} = -9.266\,\mathrm{V}$.  The green arrows represent the averaging of the $\omega_{\mathrm{r}}(V_\mathrm{g})$ around $V_\mathrm{g}\,=\,-9.266V$. 
\textbf{d} Electromechanical contribution to the variations of the  radial  mode resonance frequency  $\omega_{\mathrm{r,elec}}(u_\mathrm{t})$ as a function of the  tangential  mode displacement $u_\mathrm{t}$.  Details about  the method to extract $\omega_{\mathrm{r,elec}}(u_\mathrm{t})$ from the data presented in \textbf{c}  are given  in Supplementary Section \ref{SI-alpha3t-extractionalpha3t-determinationalpha3t}. }

\label{fig3}
\end{figure*}

In order to extract solely the intrinsic nonlinearity, the removal of the electromechanical contribution is therefore necessary. The strength of electromechanical coupling varies strongly in the vicinity of a Coulomb resonance and can be accurately predicted based on the exact knowledge of $ \mathrm{\omega_{r}}(V_\mathrm{g})$. In Fig.\ref{fig3}.c, a one-tone spectroscopy of the radial mode directly provides its resonance frequency $ \mathrm{\omega_{r}}(V_\mathrm{g})$ as a function of gate voltage. To extract the electromechanical contribution, we model the effective gate-voltage modulation $\mathrm{\Delta V_g}$ induced by the mechanical oscillation of the tangential mode. This modulation produces a corresponding variation of $ \mathrm{\omega_{r}}$ through its dependence on $V_\mathrm{g}$. We identify this variation as the electromechanical contribution $\mathrm{\omega_{r,elec}}$, shown in Fig.\ref{fig3}.d.  A detailed description of the extraction method of the electromechanical contribution is given in Supplementary Section \ref{SI-alpha3t-extractionalpha3t-determinationalpha3t}.

The conversion and calibration steps presented in Fig.\ref{fig2} and Fig.\ref{fig3} need to be performed for each gate voltage $V_\mathrm{g}$ at which the intrinsic nonlinearity is extracted. The necessary measurements are regrouped in datasets, the structure of which is detailed in Supplementary Section \ref{SI-alpha3t-structuredataset}. An example of a fit of $\mathrm{\omega_{r}}(u_\mathrm{t})$ used for the extraction of the intrinsic nonlinearity $\alpha_\mathrm{3, r}$ is shown in Fig.\ref{fig3}.b.




\begin{figure*}[h!]
\centering
\includegraphics[width=1\linewidth]{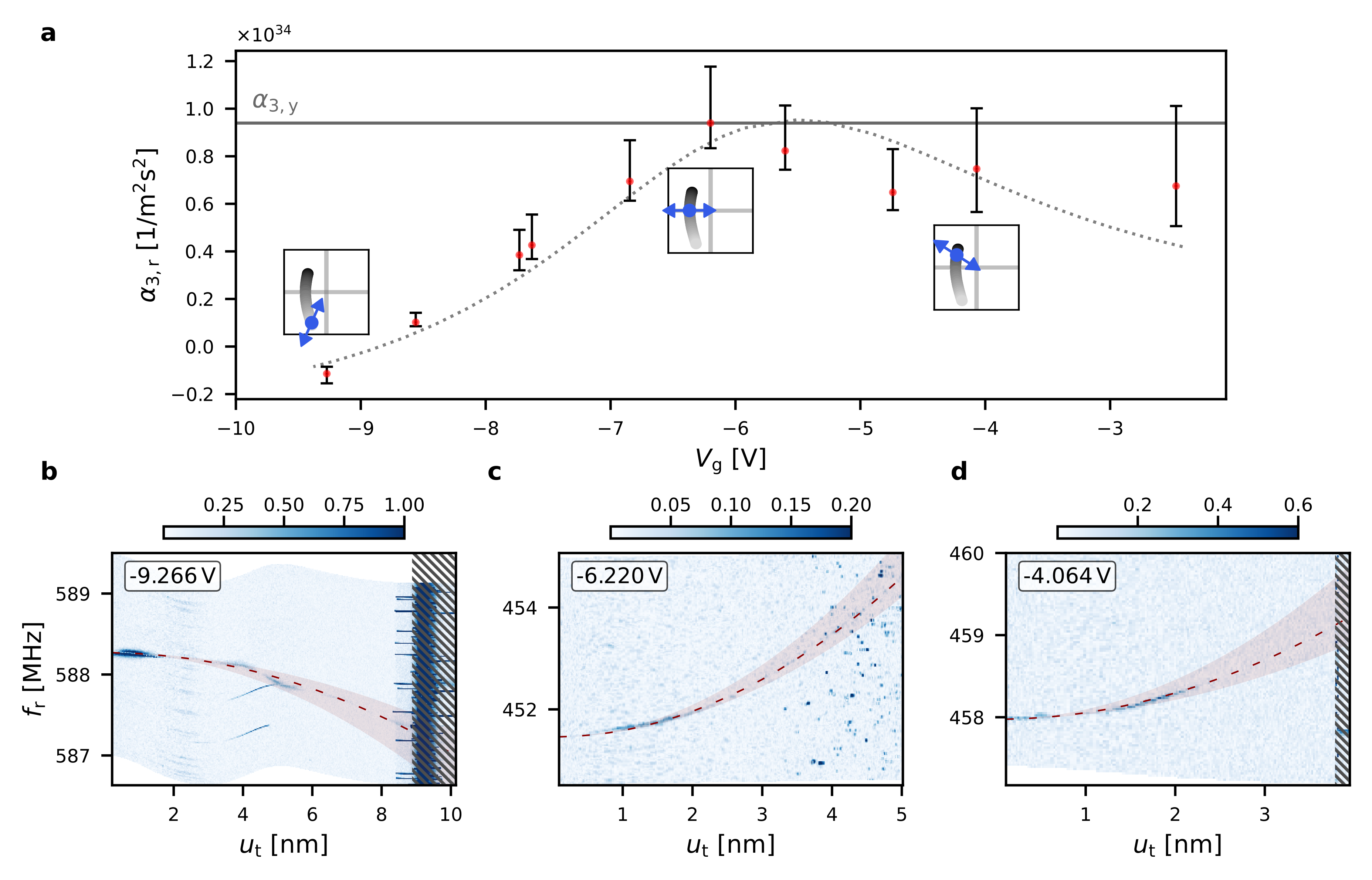}
\caption{\textbf{Control of Duffing intrinsic nonlinearity via polarization rotation.}
\textbf{a} Intrinsic mechanical nonlinearity of the  radial  mode $\mathrm{\alpha_{3,r}}$ as a function of gate voltage $V_\mathrm{g}$. The calculation of the error bars is detailed in Supplementary Section \ref{SI-estimationuncertainty}. Insets: oscillation direction of the  radial  mode. Dotted line: fit with $\alpha_{3,y}$ and $\alpha_{3,z}$ as fitting parameters. At $V_\mathrm{g} \simeq -6.2\,$V, the measured intrinsic nonlinearity is maximal. Here, the oscillation direction of the  radial  mode is exclusively in the $y$-direction (see central inset), which leads to $\mathrm{\alpha_{3,r}}\,=\, \alpha_{3,y}$, represented by a black line.
\textbf{b-d} Two-tone spectroscopy $f_\mathrm{r}$-$u_\mathrm{t}$ maps with subtracted electromechanical contribution from which the $\alpha_\mathrm{3,r}$-values shown in \textbf{a} are extracted. The variation of $\omega_\mathrm{r}/2\pi$ appears in $|\Delta G| = | \frac{\partial I}{\partial V_\text{sd}} - \langle \frac{\partial I}{\partial V_\text{sd}} \rangle_\text{sweep}|$ and is fitted by a red dotted line with only $\alpha_\mathrm{3,r}$ as a free parameter.
}
\label{fig4}
\end{figure*}

\subsection*{Intrinsic Duffing nonlinearity}
The intrinsic Duffing parameter is expected to be independent of the gate voltage \cite{castellanos2012strong} (Supplementary Section \ref{SI-th}). Nevertheless, by assuming an anisotropy in the $y$-$z$ plane (as introduced in equation \ref{eq:alpha23}), the observed gate dependence of $\mathrm{\alpha_{3,r}}$ can be convincingly reproduced (Fig.\ref{fig4}a). One probable origin for this anisotropy is the presence of an inhomogeneous electrostatic gradient field between the nanotube and the gate electrode, which can contribute to $\alpha_{3,z}$. Such a mechanism was already proposed to enhance the intrinsic mechanical nonlinearity \cite{rips2013quantum, rips2012steady, rips2014nonlinear}. Another possibility is an asymmetry in the clamping direction of the nanotube on the contact electrodes leading to different modeshapes in the $y$- and $z$-direction. This directly translates into an anisotropy of the intrinsic nonlinearity (Supplementary Section \ref{SI-th-eigenvalueproblem}). 

One key advantage of using an upward-buckled nanotube is the ability to tune the mechanical resonator through a "sweet spot" (around $V_\mathrm{g}=-6\,$V) via the change of the polarization angle. At this gate voltage, a minimum of the resonance frequency $\mathrm{\omega_r}$ coincides with the maximum of the Duffing parameter. The maximum nonlinearity is $ \mathrm{\alpha_{3,r} = \alpha_{3,y} = (9.4^{+2.3}_{-1.1})\times10^{33} \, m^{-2}s^{-2}}$ for a polarization angle $\theta =0\degree$ corresponding to a purely intrinsic nonlinearity, since the inhomogeneous electrostatic gradient field has no effect along the $y$-direction.


In order to compare with other works, one considers the ratio $u_\mathrm{nl}/u_\mathrm{zp}$ \cite{samanta2023nonlinear}, where $u_\mathrm{nl}$ is the oscillation amplitude at which mechanical bistability emerges \cite{Lif08}, and $u_\mathrm{zp}$ is the zero-point amplitude. This ratio can be interpreted as the phonon number at the onset of mechanical bistability. At the "sweet spot", we obtain an upper-bound of $u_\mathrm{nl}/u_\mathrm{zp}=34.0\pm 9.7 $ (Supplementary Section \ref{SI-alpha3t-nonlinearity-unluzp}) which is two orders of magnitude lower than previous works using intrinsic nonlinearity \cite{Eichler12}, and comparable to very recent experiments based on electromechanical nonlinearity \cite{samanta2023nonlinear}. This shows that only $34.0\pm 9.7 $ phonons in the resonator are sufficient to lead to mechanical bistability, thus demonstrating strong intrinsic nonlinearity in our system.

By scaling down the resonator to realistic dimensions, such as a length of $L=100\,$nm and a nanotube diameter of $d=2\,$nm, one estimates a Kerr parameter of $K = \mathrm{2\pi \times  913 \,Hz}$ (Supplementary Section \ref{SI-alpha3t-nonlinearity-Kerr}). Comparing this to the loss rate $\kappa$ reported for state-of-the-art carbon nanotube resonators ($\kappa=9\,$Hz) \cite{moser2014nanotube}, one obtains the ratio $K/\kappa = \mathrm{101 \gg 1}$, which clearly demonstrates that the single-phonon Kerr regime is well within reach for a purely mechanical system. For slightly more challenging dimensions such as $L=60\,$nm and $d=1.5\,$nm, $K/\kappa =  626$, thus even approaching the level of anharmonicity of superconducting qubits \cite{krantz2019quantum}.



\subsection*{Conclusions}
We investigated the intrinsic mechanical nonlinearities in an upward-buckled carbon nanotube resonator. By performing two-tone spectroscopy and exploiting nonlinear mode-mode coupling, the intrinsic Duffing parameter was selectively measured. We demonstrate a novel tuning mechanism of the nonlinearity relying on the control of a polarization angle offered by the upward-buckled nature of the nanotube. This allowed us to reach a "sweet spot" with the highest measured intrinsic nonlinearity corresponding to an onset of bistability for only  $34.0\pm 9.7$ phonons, which is about two orders of magnitude lower than similar approaches \cite{Eichler12}. 

Extrapolation to a realistic reduced resonator size shows that a purely mechanical system can operate deep within the single-phonon Kerr regime. This opens the way to intrinsic mechanical qubits and to the generation of long-lived mechanical cat states, the later being particularly crucial for exploring gravitational effects on quantum mechanical objects. \cite{gely2021superconducting, bassi2013models}.

\subsection*{Methods}

\paragraph{Device fabrication}
The device is fabricated using our custom nanoassembly technique detailed in \cite{althuon2024nano, cubaynes2020nanoassembly}, where a carbon nanotube grown with chemical vapor deposition (CVD) is mechanically transferred onto the electrodes of a circuit chip. The circuit electrode material consists of Mo (15$\,$nm), MoRe (41$\,$nm) and Pt (11$\,$nm) and is sputtered onto a p-doped SiO$_2$ (300$\,$nm) / Si (500$\,\mu$m) substrate and patterned using electron-beam and optical lithography. After removing the Pt with argon milling, reactive-ion etching (RIE) with SF$_6$/Ar/O$_2$ is used to etch Mo/MoRe and CHF$_3$/Ar/O$_2$ to form self-aligned trenches in the silicon oxide. Gate electrodes made from Ti (20$\,$nm) and Pt (25$\,$nm) are deposited in these trenches in a subsequent lift-off process. Lastly, as a requirement for the carbon nanotube transfer, 12$\,\mu$m deep trenches are patterned with optical lithography and reactive-ion etching using a combination of CHF$_3$ (SiO$_2$ etching) and SF$_6$/C$_4$F$_8$ (Si etching).


\bibliographystyle{unsrt}
\bibliography{list_paper}

\onecolumn



\onehalfspacing
\setcounter{figure}{0}
\setcounter{table}{0}
\setcounter{section}{0}
\setcounter{equation}{0}
\renewcommand{\thefigure}{S\arabic{figure}}
\renewcommand{\thetable}{S\arabic{table}}
\renewcommand{\theHfigure}{supp.\arabic{figure}}
\renewcommand{\theHtable}{supp.\arabic{table}}
\renewcommand{\theHsection}{supp.\arabic{section}}
\renewcommand{\theHequation}{supp.\arabic{equation}}

\begin{center}
  {\LARGE\bfseries Supplemental Material\par}
  \medskip
  {\Large Intrinsically nonlinear nanomechanical resonator toward the single-phonon Kerr regime\par}
  \medskip
  {\normalsize S. Klingel, T. Althuon, C. Wassmer, C. Sürgers, T. Cubaynes, and W. Wernsdorfer\par}
\end{center}

\startcontents[supp]
\section*{Contents}
\printcontents[supp]{}{1}{2}{}

\section{Theoretical model}
\label{SI-th}
In this section we describe the calculation and the model used to fit the mechanical resonance frequency in figure \ref{fig1} and to estimate the DC displacement of the carbon nanotube (CNT) which is coming along. This model also provides a rough estimation of the nonlinearity terms allowing to extrapolate the dependence as a function of CNT length. These calculations closely follow the results of Rechnitz et al.  \cite{rechnitz2022mode, rechnitz2023theoretical}.

\subsection{Euler-Bernoulli equations of motion}
\label{SI-th-EulerBernoullieom}
To describe the dynamics of the carbon nanotube flexural modes we use the Euler-Bernoulli beam equations in 2D, thus leading to a system of two equations:

\begin{equation}
\begin{split}
    \rho S \,\ddot{Z}(x,t) + E I \,Z''''(x,t) - T \,Z''(x,t) & = F\subr{DC}(x) + F_{\mathrm{AC,}z} \\
    \rho S \,\ddot{Y}(x,t) + E I\, Y''''(x,t) - T\, Y''(x,t) & = F_{\mathrm{AC,}y} \\
    \label{eq:EB_1}
\end{split}
\end{equation}

Where $\rho$ is the graphene density, $S$ the cross-section area ($S = \pi \big[ (d+ \Delta d)^2 - (d - \Delta d)^2 \big]$, with $d$: diameter of the nanotube, $\Delta d$: nanotube wall thickness), $E$ the Young modulus, and $I$ the second moment of inertia of the carbon nanotube ($I = \frac{\pi}{4} \big[ (\frac{d}{2} + \frac{\Delta d}{2})^4 - (\frac{d}{2} - \frac{\Delta d}{2})^4 \big]$). $Z'$ and $\dot Z$ are the derivatives with respect to the coordinate $x$ and the time $t$, respectively. Here, dissipation is neglected. $T$ is the axial tension in the CNT. On the right-hand side of equations \ref{eq:EB_1} one finds the different external drives. The DC part is solely due to the capacitive force between the gate electrode and the carbon nanotube. Due to the geometry of the circuit, it has no component along the $y$-direction. The AC part of the external force is the sum of the oscillatory motion of the nanotube in the electric field produced by the gate and a RF-field produced by a nearby antenna. Because the antenna has no particular alignment with the carbon nanotube, the two components '$y$' and '$z$' are non-zero. 

The axial tension is the source of nonlinearity of the flexural mode and thus plays a key role in this model. The tension reads:

\begin{equation}
\begin{split}
    T & = T_0 +\frac{ES}{L} \int_0^L \epsilon(x,t)  \mathrm{d}x \\
    \label{eq:tension}
\end{split}
\end{equation}
Where $T_0$ is the initial axial tension in the nanotube, $L$ is the total length of the suspended section of the nanotube, and $\epsilon(x,t)$ is the axial strain in the CNT due to the bending of the nanotube:
\begin{equation}
\begin{split}
    \epsilon(x,t) & = \frac{\mathrm{d}l(x)-\mathrm{d}x}{\mathrm{d}x} = \sqrt{1 + |\vec{U}(x,t)|'^{2}} -1 \\
                  & \simeq \frac{1}{2} |\vec{U}(x,t)|'^{2} - \frac{1}{8}|\vec{U}(x,t)|'^{4} + O\big( |\vec{U}(x,t)|'^{4} \big)\\
                  & \simeq \frac{1}{2} \big( Y'^2 + Z'^2 \big) - \frac{1}{8} \big( Y'^4 + Z'^4 \big)
    + O\big( Y'^4, Z'^4 \big)
    \label{eq:strain}
\end{split}
\end{equation}

Where $\mathrm{d}l$ is the length of the CNT corresponding to a segment $\mathrm{d}x$, and $\vec{U}(x,t) = Y(x,t) \vec{e}_y + Z(x,t) \vec{e}_z$ is the transverse displacement of the nanotube. To model the data presented in the main text, only the first order of the Taylor expansion was considered. 
Finally, the tension has the following form:
\begin{equation}
\begin{split}
    T & \simeq T_0 +\frac{ES}{2L} \int_0^L  (Y'^2 + Z'^2 )\, \mathrm{d}x \\
    \label{eq:tension2}
\end{split}
\end{equation}
In equations \ref{eq:EB_1}, $F\subr{DC}(x)$ is the static capacitive force per unit length exerted by the gate electrode onto the CNT. Assuming the potential of the quantum dot $V\subr{dot} \ll V\subr{g}$, the static capacitive force reads: 
\begin{equation}
\begin{split}
    F\subr{DC}(x) & = \frac{1}{2}\frac{\partial \delta C\subr{g}(x)}{\partial z\subr{CNT}} V\subr{g}^2 \\
    \label{eq:forceDC}
\end{split}
\end{equation}
where $\delta C\subr{g}(x)$ is the capacitance per unit length between the gate electrode and the CNT, $z\subr{CNT}$ is the distance between the CNT and the gate electrode, and $V\subr{g}$ is the gate voltage.

\subsection{Non-dimensional parameters}
\label{SI-th-nondimensionalparams}
To simplify formulas and facilitate numerical calculations, non-dimensional parameter are used, and their definitions can be found in Table \ref{table:1}. 

\begin{table}[ht]
\centering
\begin{tabular}{c c l} 
 \hline \hline
 $\tilde{x}$ & = & $\frac{x}{L}$  \\ [1ex] 
 $\tilde{Y}$ & = & $\frac{Y}{g_0}$  \\[1ex] 
 $\tilde{Z}$ & = & $\frac{Z}{g_0}$  \\  [1ex] 
 $P$ & = & $\frac{T_0 L^2}{E I}$  \\ [1ex] 
 $\alpha$ & = & $\frac{S g_0^2}{2 I}$  \\ [1ex] 
 $\tilde{t}$ & = & $ t \sqrt{\frac{E I}{\rho S L^4}}$  \\ [1ex]
 $\tilde{F}\subr{DC}$ & = & $F\subr{DC} \frac{L^4}{E I g_0}$  \\ [1ex]
 $\tilde{F}_{\mathrm{AC},i}$ & = & $F_{\mathrm{AC},i} \frac{L^4}{E I g_0}$  \\ [1ex] 
 \hline
\end{tabular}
\caption{\textbf{List of non-dimensional parameters.} $g_0$ is the height difference between the contacts and the gate electrodes ($g_0 = 65\,$nm).}
\label{table:1}
\end{table}

The Euler-Bernoulli beam equations \ref{eq:EB_1} thus leads to the following system of coupled equations:

\begin{equation}
\begin{split}
    \ddot{\tilde{Z}}(\tilde{x},\tilde{t}) 
    + \tilde{Z}''''(\tilde{x},\tilde{t}) 
    - \Bigg( \alpha \int_0^1 (\tilde{Y}'^2 + \tilde{Z}'^2 )\, \mathrm{d}\tilde{x} + P \Bigg) \tilde{Z}''(\tilde{x},\tilde{t}) & = 
    \tilde{F}\subr{DC}(\tilde{x}) + \tilde{F}_{z\mathrm{,AC}} \\
    \ddot{\tilde{Y}}(\tilde{x},\tilde{t}) 
    + \tilde{Y}''''(\tilde{x},\tilde{t}) 
    - \Bigg( \alpha \int_0^1 (\tilde{Y}'^2 + \tilde{Z}'^2 )\, \mathrm{d}\tilde{x} + P \Bigg) \tilde{Y}''(\tilde{x},\tilde{t}) & = 
    \tilde{F}_{y\mathrm{,AC}} \\
    \label{eq:EB_2}
\end{split}
\end{equation}

In the following we will omit the $\tilde{\square}$ symbol. Using a similar decomposition as introduced in ref. \cite{rechnitz2023theoretical}, the solutions ($Y(x,t)$, $Z(x,t)$) to the equation \ref{eq:EB_2} are assumed to have the form:

\begin{equation}
\begin{split}
    Z(x,t) & = Z_0(x) + Z\subr{g}(x) + Z\subr{d}(x,t) \\
    Y(x,t) & = Y_0(x) + Y\subr{g}(x) + Y\subr{d}(x,t) \\
    \label{eq:yz_form}
\end{split}
\end{equation}

Where the solutions $Y(x,t)$ and $Z(x,t)$ are the superposition of three different contributions. 
The subscript $()_0$ refers to the static displacement in absence of any gate voltage, thus including the initial buckling. $(Y_0, \,Z_0)$ is the solution to the time-independent version of the set of equation \ref{eq:EB_2} with $F\subr{DC}(x) = 0$. The subscript $()\subr{g}$ refers to the static displacement due to the electrostatic gate force. $(Y_0 + Y\subr{g},\, Z_0 + Z\subr{g})$ is the solution to the time-independent version of the set of equation \ref{eq:EB_2} with $F\subr{DC}(x) \neq 0$. The subscript $()\subr{d}$ refers to the dynamic displacement. This leads to the three following pairs of equations:
\paragraph{Initial displacement - Buckling}
\begin{equation}
\begin{split}
    Z_{0}'''' - \Bigg( \alpha\int_0^1  (Y_{0}'^2 + Z_{0}'^2 ) \mathrm{d}x  + P \Bigg) Z_{0}'' & = 0 \\
    Y_{0}'''' - \Bigg( \alpha\int_0^1  (Y_{0}'^2 + Z_{0}'^2 ) \mathrm{d}x  + P \Bigg) Y_{0}'' & = 0 \\
    \label{eq:EB_yz_0}
\end{split}
\end{equation}
\paragraph{Static displacement}
\begin{equation}
\begin{split}
    Z\subr{g}''''  - Z\subr{g}'' \Bigg( \alpha\int_0^1  \Big[  (Y_{0}' + Y\subr{g}')^2 + (Z_{0}' + Z\subr{g}')^2 \Big] \mathrm{d}x  + P \Bigg) \\ 
    - Z_{0}'' \Bigg( \alpha\int_0^1  \Big[   2 (Z_0' Z\subr{g}' + Y_0' Y\subr{g}') + Z\subr{g}'^2 + Y\subr{g}'^2  \Big] \mathrm{d}x \Bigg)   & = F\subr{DC} \\
    Y\subr{g}''''  - Y\subr{g}'' \Bigg( \alpha\int_0^1  \Big[  (Y_{0}' + Y\subr{g}')^2 + (Z_{0}' + Z\subr{g}')^2 \Big] \mathrm{d}x  + P \Bigg) \\
    - Y_{0}'' \Bigg( \alpha\int_0^1  \Big[   2 (Z_0' Z\subr{g}' + Y_0' Y\subr{g}') + Z\subr{g}'^2 + Y\subr{g}'^2  \Big] \mathrm{d}x \Bigg)   & = 0 \\
    \label{eq:EB_yz_st}
\end{split}
\end{equation}
\paragraph{Dynamic displacement}
\begin{equation}
\begin{split}
    Z\subr{d}'''' + \ddot{Z\subr{d}} - Z\subr{d}'' \Bigg( \alpha\int_0^1  \Big[  (Y_{0}' + Y\subr{g}' + Y\subr{d}')^2 + (Z_{0}' + Z\subr{g}' + Z\subr{d}')^2 \Big] \mathrm{d}x  + P \Bigg) \\ 
    - (Z_{0}'' + Z\subr{g}'') \Bigg( \alpha\int_0^1  \Big[  Y\subr{d}'^2 + Z\subr{d}'^2 + 2 Z\subr{d}'(Z_0' +  Z\subr{g}') + 2 Y\subr{d}'( Y_0' +  Y\subr{g}') + Y\subr{g}'^2 + Z\subr{g}'^2  \Big] \mathrm{d}x \Bigg)   & = F_{\mathrm{AC},z} \\
    Y\subr{d}'''' + \ddot{Y\subr{d}} - Y\subr{d}'' \Bigg( \alpha\int_0^1  \Big[  (Y_{0}' + Y\subr{g}' + Y\subr{d}')^2 + (Z_{0}' + Z\subr{g}' + Z\subr{d}')^2 \Big] \mathrm{d}x  + P \Bigg) \\ 
    - (Y_{0}'' + Y\subr{g}'') \Bigg( \alpha\int_0^1  \Big[  Y\subr{d}'^2 + Z\subr{d}'^2 + 2 Z\subr{d}'(Z_0' +  Z\subr{g}') + 2 Y\subr{d}'( Y_0' +  Y\subr{g}') + Y\subr{g}'^2 + Z\subr{g}'^2  \Big] \mathrm{d}x \Bigg)   & = F_{\mathrm{AC},y} \\
    \label{eq:EB_yz_d}
\end{split}
\end{equation}

\subsection{Galerkin decomposition}
\label{SI-th-Galerkindecomp}
In order to numerically solve these three sets of equations \ref{eq:EB_yz_0}, \ref{eq:EB_yz_st}, \ref{eq:EB_yz_d}, the Galerkin method is commonly used, allowing us to reduce them to algebraic equations. The idea is to discretize the problem by projecting these equations onto a finite number of modes which is sufficient to capture the dynamics of the system. Here, only the modeshapes $\phi(x) = \frac{1}{2} \big(1-\cos(2 \pi x) \big)$ of the first mode in the bending regime are considered:

\begin{align*}
Y_0(x)&= y_{0} \phi(x)    &  Y\subr{g}(x) &=y\subr{g}(V\subr{g}) \phi(x)     &  Y\subr{d}(x,t)&=y\subr{d}(t) \phi(x)\\
Z_0(x)&= z_{0} \phi(x)    &  Z\subr{g}(x) &=z\subr{g}(V\subr{g}) \phi(x)     &  Z\subr{d}(x,t)&=z\subr{d}(t) \phi(x)
\end{align*}

The projection operation corresponds to taking the integral of the product with $\phi(x)$ ($\int_0^1 ... \phi(x)  \mathrm{d}x $). It is useful to define geometric factors which are specific to the modeshape $\phi(x)$ used in this model. These factors read:
\begin{align*}
a_1&= \int_0^1 \phi(x) \mathrm{d}x  & a_2 &= \int_0^1 \phi^{2}(x) \mathrm{d}x     &  b_2&= \int_0^1 \phi^{'2}(x) \mathrm{d}x  &  c_2&= \int_0^1 \phi^{''2}(x) \mathrm{d}x
\end{align*}

The projection operation transforms the three sets of equations \ref{eq:EB_yz_0}, \ref{eq:EB_yz_st}, \ref{eq:EB_yz_d} into:

\paragraph{Initial displacement - Buckling}
\begin{equation}
\begin{split}
    c_2 + b_2\Big( \alpha b_2 \big(y_{0}^2 + 
    z_{0}^2 \big)  + P \Big)  & = 0 \\
    y_{0}^2 + z_{0}^2 & = -\frac{\frac{c_2}{b_2} + P}{\alpha b_2}
    \label{eq:GD_yz_0}
\end{split}
\end{equation}
The initial displacement ($y_0,z_0$)=($R \cos(\theta), R \sin(\theta)$) with the radius $R=\sqrt{-\frac{\frac{c_2}{b_2} + P}{\alpha b_2}}$ and the angle $\theta$ which emerges from the symmetry breaking due to asymmetric clamping of the CNT. For the radius $R$ to be real, the condition $ T_0 < -\frac{c_2 E I}{b_2 L^2} $ has to be fulfilled. This is known as the buckling criterion, defining the amount of axial compression ($T_0 <0$: compression) required for the beam to buckle.

\paragraph{Static displacement}
\begin{equation}
\begin{aligned}
     & z\subr{g} c_2 + z\subr{g}b_2 \bigg( \alpha b_2  \Big[ \big(z_0 + z\subr{g} \big)^2 + \big(y_0 + y\subr{g} \big)^2 \Big] + P \bigg)\\ 
     &+ \alpha b_2^2 \bigg(  2 \Big[ z_0 z\subr{g} + y_0 y\subr{g} \Big] + z\subr{g}^2 + y\subr{g}^2 \bigg) z_0 = \int_0^1 F\subr{DC} \phi(x)\mathrm{d}x\\ 
     & y\subr{g} c_2 + y\subr{g}b_2 \bigg( \alpha b_2  \Big[ \big(z_0 + z\subr{g} \big)^2 + \big(y_0 + y\subr{g} \big)^2 \Big] + P \bigg)\\
     & + \alpha b_2^2 \bigg(  2 \Big[ z_0 z\subr{g} + y_0 y\subr{g} \Big] + z\subr{g}^2 + y\subr{g}^2 \bigg) y_0 = 0
    \label{eq:GD_yz_st}
\end{aligned}
\end{equation}

The gate dependent part of the static displacement ($y\subr{g}$, $z\subr{g}$) is solution to the pair of equations \ref{eq:GD_yz_st}. These two algebraic equations are solved numerically. An estimation of the static force integral ($\int_0^1 F\subr{DC} \phi(x)\mathrm{d}x$) is found using a finite-element simulation (COMSOL \cite{comsol}) of the capacitance between the buckled nanotube and the gate electrode for various $z\subr{st}$ displacements (more details in section \ref{SI-th-implementation}).

\paragraph{Dynamic displacement}
\begin{equation}
\begin{split}
    & z\subr{d} c_2 + \ddot{z}\subr{d} a_{2} + z\subr{d}b_2 \bigg( \alpha b_2  \Big[ \big(z_0 + z\subr{g} + z\subr{d} \big)^2 + \big(y_0 + y\subr{g} + y\subr{d} \big)^2 \Big] + P \bigg)\\ 
    & + \alpha b_2^2 \bigg( 2 z\subr{d} \big[ z_0 + z\subr{g} \big] + 2 y\subr{d} \big[ y_0 + y\subr{g} \big] + z\subr{d}^2 + y\subr{d}^2 \bigg) (z_0 + z\subr{g}) = \int_0^1 F_{\mathrm{AC,}z} \phi(x)\mathrm{d}x\\
    & y\subr{d} c_2 + \ddot{y}\subr{d} a_{2} + y\subr{d}b_2 \bigg( \alpha b_2  \Big[ \big(z_0 + z\subr{g} + z\subr{d} \big)^2 + \big(y_0 + y\subr{g} + y\subr{d} \big)^2 \Big] + P \bigg)\\ 
    & + \alpha b_2^2 \bigg( 2 z\subr{d} \big[ z_0 + z\subr{g} \big] + 2 y\subr{d} \big[ y_0 + y\subr{g} \big] + z\subr{d}^2 + y\subr{d}^2 \bigg) (y_0 + y\subr{g}) = \int_0^1 F_{\mathrm{AC,}y} \phi(x)\mathrm {d}x\\ 
    \label{eq:GD_yz_d}
\end{split}
\end{equation}

\subsection{Eigenvalue problem}
\label{SI-th-eigenvalueproblem}
To find the eigenvalues ($\omega\subr{r}$ and $\omega\subr{t}$) and the eigenvector (oscillation direction of the modes) of the set of equations \ref{eq:GD_yz_d}, one takes $ \ddot{z}\subr{d} = -\omega^2 z\subr{d} $, and $ \ddot{y}\subr{d} = -\omega^2 y\subr{d} $, and linearizes the equations (remove all nonlinear coupling terms). The set of dynamics equations \ref{eq:GD_yz_d} can then be written in matrix form:

\begin{equation}
\begin{pmatrix}
H_{11} & H_{12}\\
H_{21} & H_{22} 
\end{pmatrix}
\begin{pmatrix}
z\subr{d}\\
y\subr{d}
\end{pmatrix}
= \omega^2
\begin{pmatrix}
z\subr{d}\\
y\subr{d}
\end{pmatrix}
\label{eq:matrix_form}
\end{equation}

with:
\begin{equation}
\begin{split}
    H_{11} & = \frac{c_2}{a_{2}} + \frac{b_2}{a_{2}} \bigg( \alpha b_2  \Big[ 3 \big(z_0 + z\subr{g} \big)^2 + \big(y_0 + y\subr{g} \big)^2 \Big] + P \bigg) \\
    H_{22} & = \frac{c_2}{a_{2}} + \frac{b_2}{a_{2}} \bigg( \alpha b_2  \Big[ \big(z_0 + z\subr{g} \big)^2 + 3 \big(y_0 + y\subr{g} \big)^2 \Big] + P \bigg) \\
    H_{12} & = H_{21} = \frac{2 \alpha b_2^2}{a_{2}} \Big(z_0 + z\subr{g} \Big) \Big(y_0 + y\subr{g} \Big) \\
    \label{eq:matrixElements}
\end{split}
\end{equation}

\paragraph{Eigenfrequencies}
The two eigenfrequencies are obtained from $\mathrm{det}\begin{pmatrix}
H_{11}-\omega^2 & H_{12}\\
H_{21} & H_{22}-\omega^2 
\end{pmatrix} =0 $, and redimensionalizing $\omega_{i} = \sqrt{\frac{E I}{\rho A L }} \tilde{\omega}_{i} $ where the $\tilde{\square}$ denotes the dimensionless quantity, and $i \in{[\mathrm{t,t}]}$.

\paragraph{Eigenvectors}
The eigenvectors determine the direction of oscillation of the modes  tangential  and  radial , which are fully captured by the polarization angle $\theta$ introduced in the main text. In the basis ($ \vec{e}_z, \vec{e}_y $), the eigenvectors read:

\begin{align*}
\vec{e}\subr{t} &= 
\begin{pmatrix}
v_{z,\mathrm{t}}\\
v_{y,\mathrm{t}}
\end{pmatrix}
& \vec{e}\subr{r} &=
\begin{pmatrix}
v_{z,\mathrm{r}}\\
v_{y,\mathrm{r}}
\end{pmatrix}
\end{align*}
with:
\begin{equation}
\begin{split}
    v_{z,\mathrm{r}} &=  v_{y,\mathrm{t}} = \sin(\theta)  =  \frac{1}{\sqrt{1+ \Big( \frac{\omega\subr{r}^2- H_{11}}{H_{12}} \Big)^2}}\\
    v_{y,\mathrm{r}} &=  -v_{z,\mathrm{t}} = \cos(\theta)  = \frac{1}{\sqrt{1+ \Big( \frac{H_{12}}{\omega\subr{r}^2- H_{11}} \Big)^2}}
    \label{eq:eigenvectors}
\end{split}
\end{equation}

To write the set of equations \ref{eq:GD_yz_d} in the rotated frame of ($ \vec{e}\subr{t}, \vec{e}\subr{r} $), we perform the following transformations:

\begin{align*}
\begin{pmatrix}
z\subr{d}\\
y\subr{d}
\end{pmatrix}
&= 
\begin{pmatrix}
-\cos{\theta} & \sin{\theta}\\
\sin{\theta} & \cos{\theta} 
\end{pmatrix}
\begin{pmatrix}
u\subr{t}\\
u\subr{r}
\end{pmatrix}
& \begin{pmatrix}
\ddot{z\subr{d}}\\
\ddot{y\subr{d}}
\end{pmatrix}
&=
\begin{pmatrix}
\omega\subr{t}^2 \cos{\theta} & -\omega\subr{r}^2 \sin{\theta}\\
-\omega\subr{t}^2 \sin{\theta} &  -\omega\subr{r}^2 \cos{\theta} 
\end{pmatrix}
\begin{pmatrix}
u\subr{t}\\
u\subr{r}
\end{pmatrix}
\end{align*}

To take into account an anisotropy in the nonlinearity, we consider the geometric factors to be different in the $y$- and the $z$-direction, such that: $a_{1,z} \neq a_{1,y}$, $a_{2,z} \neq a_{2,y}$, $b_{2,z} \neq b_{2,y}$, and $c_{2,z} \neq c_{2,y}$. Physically, this corresponds to different modeshapes in the $y$- and $z$-direction. In the rotated frame ($ \vec{e}\subr{t}, \vec{e}\subr{r} $), the set of equations \ref{eq:GD_yz_d} becomes:

{
\setlength{\abovedisplayskip}{0pt}%
\setlength{\belowdisplayskip}{0pt}%
\begin{align}
 & -c_{2,z}\cos(\theta)u\subr{t}
   + c_{2,z}\sin(\theta)u\subr{r}
   + a_{2,z}\omega\subr{t}^2\cos(\theta)u\subr{t}
   + a_{2,z}\omega\subr{r}^2\sin(\theta)u\subr{r} \nonumber \\[2pt]
 & + [-b_{2,z}\cos(\theta)u\subr{t} + b_{2,z}\sin(\theta)u\subr{r}]\,
   \Big[
      \alpha b_{2,z}(
         [z_0 + z\subr{g} - \cos(\theta)u\subr{t} + \sin(\theta)u\subr{r}]^2  \nonumber \\[2pt]
 &\qquad + [y_0 + y\subr{g} + \sin(\theta)u\subr{t} + \cos(\theta)u\subr{r}]^2
      ) + P
   \Big] \label{eq:rotated_z} \\[2pt]
 & + \alpha b_{2,z}^2\Big(
        [-\cos(\theta)u\subr{t} + \sin(\theta)u\subr{r}]^2 
      + [\sin(\theta)u\subr{t} + \cos(\theta)u\subr{r}]^2 \nonumber \\[2pt]
 &\qquad
      + 2[z_0 + z\subr{g}]\,[ -\cos(\theta)u\subr{t} + \sin(\theta)u\subr{r}]
      + 2[y_0 + y\subr{g}]\,[\sin(\theta)u\subr{t} + \cos(\theta)u\subr{r}]
   \Big)(z_0 + z\subr{g}) \nonumber \\[2pt]
 & = \int_0^1 F_{\mathrm{AC},z}\,\phi(x)\,\mathrm{d}x \nonumber
\end{align}
}
\vspace{1cm}
{
\setlength{\abovedisplayskip}{0pt}%
\setlength{\belowdisplayskip}{0pt}%
\begin{align}
 & c_{2,y}\sin(\theta)u\subr{t}
   + c_{2,y}\cos(\theta)u\subr{r}
   - a_{2,y}\omega\subr{t}^2\sin(\theta)u\subr{t}
   - a_{2,y}\omega\subr{r}^2\cos(\theta)u\subr{r} \nonumber \\[2pt]
 & + [b_{2,y}\sin(\theta)u\subr{t} + b_{2,y}\cos(\theta)u\subr{r}]\,
   \Big[
      \alpha b_{2,y}(
         [z_0 + z\subr{g} - \cos(\theta)u\subr{t} + \sin(\theta)u\subr{r}]^2 \nonumber \\[2pt]
 &\qquad + [y_0 + y\subr{g} + \sin(\theta)u\subr{t} + \cos(\theta)u\subr{r}]^2
      ) + P
   \Big]  \label{eq:rotated_y} \\[2pt]
 & + \alpha b_{2,y}^2\Big(
        [-\cos(\theta)u\subr{t} + \sin(\theta)u\subr{r}]^2
      + [\sin(\theta)u\subr{t} + \cos(\theta)u\subr{r}]^2 \nonumber \\[2pt]
 &\qquad
      + 2[z_0 + z\subr{g}]\,[-\cos(\theta)u\subr{t} + \sin(\theta)u\subr{r}]
      + 2[y_0 + y\subr{g}]\,[\sin(\theta)u\subr{t} + \cos(\theta)u\subr{r}]
   \Big)(y_0 + y\subr{g}) \nonumber \\[2pt]
 & = \int_0^1 F_{\mathrm{AC},y}\,\phi(x)\,\mathrm{d}x \nonumber
\end{align}
}

\vspace{1cm}

In order to obtain the equation of motion of the  radial  mode, we take the linear combination $\sin{\theta} \times$ eq.\ref{eq:rotated_z} $+ \cos{\theta} \times$ eq.\ref{eq:rotated_y}. This leads to an equation of the form:
\begin{equation}
\begin{aligned}
\ddot{u}\subr{r} & + \omega\subr{r}^2 u\subr{r} + \lambda\subr{t} u\subr{t} + \alpha\subr{2,rr} u\subr{r}^2 + \alpha\subr{2,tt} u\subr{t}^2 + \alpha\subr{2,tr} u\subr{t} u\subr{r}\\
& + \alpha\subr{3,ttt} u\subr{t}^3 + \alpha\subr{3,ttr} u\subr{t}^2 u\subr{r} + \alpha\subr{3,trr} u\subr{t} u\subr{r}^2 + \alpha\subr{3,rrr} u\subr{r}^3 = F\subr{r}
\label{eq:final}
\end{aligned}
\end{equation}
with $F_\mathrm{r}$ the external driving force on the  radial  mode and:
\allowdisplaybreaks

\begin{align}
\omega\subr{r}^2 = & \frac{c_{2,z}}{a_{2,z}} + \frac{P}{a_{2,z}} \Big( b_{2,z} \sin^2(\theta) + b_{2,y} \cos^2(\theta) \Big) \\
& + \frac{\alpha}{a_{2,z}} \Big( b_{2,z}^2 \sin^2(\theta) + b_{2,y}^2 \cos^2(\theta) \Big) \Big( \big[ z_0 + z\subr{g} \big]^2 + \big[ y_0 + y\subr{g} \big]^2 \Big) \notag \\
& + \frac{2 \alpha}{a_{2,z}} \Big( b_{2,z}^2 \sin(\theta) \big[ z_0 + z\subr{g} \big] + b_{2,y}^2 \cos(\theta) \big[ y_0 + y\subr{g} \big] \Big) \notag \\ 
& \hspace{1.2cm} \Big( \sin(\theta) \big[ z_0 + z\subr{g} \big] +  \cos(\theta) \big[ y_0 + y\subr{g} \big] \Big) \notag \\
\vspace{0.5cm} \notag \\
\lambda\subr{t} = & \frac{P}{a_{2,z}} \sin(\theta) \cos(\theta) \big( b_{2,y} - b_{2,z} \big)\\
& + \frac{\alpha}{a_{2,z}} \sin(\theta) \cos(\theta) \big( b_{2,y}^2 - b_{2,z}^2 \big) \Big( \big[ z_0 + z\subr{g} \big]^2 + \big[ y_0 + y\subr{g} \big]^2 \Big) \notag \\
& + \frac{2 \alpha}{a_{2,z}} \Big( b_{2,z}^2 \sin(\theta) \big[ z_0 + z\subr{g} \big] + b_{2,y}^2 \cos(\theta) \big[ y_0 + y\subr{g} \big] \Big) \notag \\
&  \hspace{1.2cm} \Big( \sin(\theta) \big[ y_0 + y\subr{g} \big] - \cos(\theta) \big[ z_0 + z\subr{g} \big] \Big) \notag \\
\vspace{0.5cm} \notag \\
\alpha\subr{2,rr} = & \frac{2 \alpha}{a_{2,z}} \Big( b_{2,z}^2 \sin^2(\theta) + b_{2,y}^2 \cos^2(\theta) \Big) \Big( \sin(\theta)\big[ z_0 + z\subr{g} \big] + \cos(\theta) \big[ y_0 + y\subr{g} \big] \Big) \\
& + \frac{\alpha}{a_{2,z}} \Big( b_{2,z}^2 \sin(\theta) \big[ z_0 + z\subr{g} \big] + b_{2,y}^2 \cos(\theta) \big[ y_0 + y\subr{g} \big] \Big) \notag \\
\vspace{0.5cm} \notag\\
\alpha\subr{2,tt} = & \frac{2 \alpha}{a_{2,z}} \sin(\theta) \cos(\theta) \Big( b_{2,y}^2 - b_{2,z}^2 \Big) \Big( \sin(\theta)\big[ y_0 + y\subr{g} \big] - \cos(\theta) \big[ z_0 + z\subr{g} \big] \Big)  \\
& + \frac{\alpha}{a_{2,z}} \Big( b_{2,z}^2 \sin(\theta) \big[ z_0 + z\subr{g} \big] + b_{2,y}^2 \cos(\theta) \big[ y_0 + y\subr{g} \big] \Big) \notag \\
\vspace{0.5cm} \notag \\
\alpha\subr{2,tr} = & \frac{2 \alpha}{a_{2,z}} \sin(\theta) \cos(\theta) \Big( b_{2,y}^2 - b_{2,z}^2 \Big) \Big( \sin(\theta)\big[ z_0 + z\subr{g} \big] + \cos(\theta) \big[ y_0 + y\subr{g} \big] \Big)  \\
& + \frac{2\alpha}{a_{2,z}} \Big( b_{2,z}^2 \sin^2(\theta) + b_{2,y}^2 \cos^2(\theta) \Big) \Big( \sin(\theta) \big[ y_0 + y\subr{g} \big] - \cos(\theta) \big[ z_0 + z\subr{g} \big] \Big) \notag \\
\vspace{0.5cm} \notag \\
\alpha\subr{3,ttt} = & \frac{ \alpha}{a_{2,z}} \sin(\theta) \cos(\theta) \Big( b_{2,y}^2 - b_{2,z}^2 \Big)\\
\vspace{0.5cm} \notag \\
\alpha\subr{3,ttr} = & \frac{ \alpha}{a_{2,z}} \Big( b_{2,y}^2 \cos^2(\theta) + b_{2,z}^2 \sin^2(\theta) \Big) \\
\vspace{0.5cm} \notag \\
\alpha\subr{3,trr} = & \frac{ \alpha}{a_{2,z}} \sin(\theta) \cos(\theta) \Big( b_{2,y}^2 - b_{2,z}^2 \Big) \\
\vspace{0.5cm} \notag \\
\alpha\subr{3,rrr} = & \frac{ \alpha}{a_{2,z}} \Big( b_{2,y}^2 \cos^2(\theta) + b_{2,z}^2 \sin^2(\theta) \Big)\\
\notag
\label{eq:final_parameter}
\end{align}
\subsection{Derivation of equations (1) and (4) of the main text}
\label{SI-th-derivationequations}
 In the data presented in figure \ref{fig2} of the main text, only the  radial  mode is driven, therefore the equations \ref{eq:EOM1}, and \ref{eq:alpha23} of the main text are:
 
 \begin{equation}
\ddot{u}\subr{r} + \omega\subr{r}^2 u\subr{r} + \alpha\subr{2,r} u\subr{r}^2  + \alpha\subr{3,r} u\subr{r}^3 = F\subr{r}
\label{eq:final_v1}
\end{equation}
where $\alpha\subr{2,r}$ and $\alpha\subr{3,r}$: 
\begin{align}
\alpha\subr{2,r} & = \alpha\subr{2,rr} = c_0(\theta)y\subr{st} + c_1(\theta) z\subr{st} \\
\alpha\subr{3,r} & = \alpha\subr{3,rrr} =  \alpha_{3,y} \cos^2(\theta) + \alpha_{3,z} \sin^2(\theta)
\end{align}
 with:
\begin{align}
c_0(\theta) & = \frac{E I}{\rho S L^4 g_0} \frac{2\alpha}{a_{2,z}} \sin(\theta) \bigg(  \sin^2(\theta) b_{2,z}^2 + \Big[ \frac{1}{2} + \cos^2(\theta) \Big] b_{2,y}^2 \bigg) \\
c_1(\theta) & = \frac{E I}{\rho S L^4 g_0} \frac{2\alpha}{a_{2,z}} \sin(\theta) \bigg( \Big[ \frac{1}{2} + \sin^2(\theta) \Big] b_{2,z}^2 + \cos^2(\theta) b_{2,y}^2 \bigg) \\
\alpha_{3,y}& = \frac{E I}{\rho S L^4 g_0^2} \frac{\alpha b_{2,y}^2}{a_{2,z}} \\
\alpha_{3,z}& = \frac{E I}{\rho S L^4 g_0^2} \frac{\alpha b_{2,z}^2}{a_{2,z}}
\end{align}

which are the expressions of the dimensional quantities presented in the main text, with $z\subr{st} = z_0 + z\subr{g}$, and $y\subr{st} = y_0 + y\subr{g}$.

In the data presented in figure \ref{fig3} and \ref{fig4} of the main text, the  radial  and the  tangential  modes are driven. Because the  radial  mode is only weakly driven, the terms $u\subr{r}^2$, $u\subr{r}^3$, $u\subr{r}^2u\subr{t}$ are negligible, and the equation \ref{eq:final} reduces to:
\begin{equation}
\ddot{u}\subr{r} + \omega\subr{r}^2 u\subr{r} + \alpha\subr{2,tt} u\subr{t}^2 + \alpha\subr{2,tr} u\subr{r} u\subr{t}  + \alpha\subr{3,ttt} u\subr{t}^3 + \alpha\subr{3,ttr} u\subr{t}^2 u\subr{r}  = F\subr{r}
\label{eq:final_v2a}
\end{equation}
Moreover, since the measurement of the differential conductance requires integration time way longer than the characteristic time of the mechanical resonator, we are only sensitive to the effect of the RMS amplitude of the  tangential  mode \cite{ari2018intermodal}. This leads to $\langle u\subr{t} \rangle = 0$, $\langle u\subr{t}^2 \rangle = \frac{u\subr{t}^2}{2}$, and $\langle u\subr{t}^3 \rangle = 0$ where $\langle ... \rangle$ denotes the time average. Therefore, we obtain the following equation of motion:

\begin{equation}
\ddot{u}\subr{r} + (\omega\subr{r}^2 + \alpha\subr{3,ttr} \frac{u\subr{t}^2}{2})u\subr{r} + \alpha\subr{2,tt} \frac{u\subr{t}^2}{2}  = F\subr{r}
\label{eq:final_v2b}
\end{equation}
from which the equation \ref{eq:omegat} of the main text is derived.

\subsection{Force integral estimation}
\label{SI-th-implementation}
Because of the complex geometry of our device, an analytical expression of the force integral term in the equation \ref{eq:GD_yz_st} ($\int_0^1 F\subr{DC}(x) \phi(x)\mathrm{d}x$) cannot be found. Therefore, we rely on a polynomial expression as a function of $z_\mathrm{st}$ the static displacement of the nanotube with respect to the horizontal situation ($z\subr{st}=z_0 +z\subr{g}$):

\begin{equation}
\begin{split}
\int_0^1 F\subr{DC}(x) \phi(x)\mathrm{d}x & = \frac{V\subr{g}^2}{2} \int_0^1 \frac{\partial \delta C\subr{g}(x,z\subr{st})}{\partial z\subr{st}}\phi(x)\mathrm{d}x\\
 & = \big( A_{0} + A_{1} z_\mathrm{st}  \big)V\subr{g}^2
\label{eq:forceDC2}
\end{split}
\end{equation}
$\delta C\subr{g}(x,z\subr{st})$ is the capacitance between the nanotube and the gate per unit length for a static displacement amplitude $z\subr{st}$. $A_0$ and $A_1$ are coefficients determined based on AFM (Atomic Force Microscope) and SEM (Scanning Electron Microscope) images (see figure \ref{fig:AFM-SEM}) of the device post-measurement, as well as on the finite-element simulation (COMSOL \cite{comsol}) of the total capacitance between the carbon nanotube and the gate electrode for different displacements $C\subr{tot}(z\subr{st})$. In practice, the simulation of $C\subr{tot}(z\subr{st})$ only allows us to find an upper bound of the force integral:

\begin{equation}
\begin{split}
\frac{\partial C\subr{tot}}{\partial z\subr{st}}(z\subr{st}) =  \int_0^1 \frac{\partial \delta C\subr{g}(x,z\subr{st})}{\partial z\subr{st}} \mathrm{d}x > \int_0^1 \frac{\partial \delta C\subr{g}(x,z\subr{st})}{\partial z\subr{st}} \phi(x) \mathrm{d}x = 2 \big( A_{0} + A_{1} z\subr{st}  \big)
\label{eq:total-capacitance-gradient-bound}
\end{split}
\end{equation}
By calculating $\frac{\partial C\subr{tot}}{\partial z\subr{st}}(z\subr{st})$ and then fitting this function with a second-order polynomial, we obtain the following upper bounds: 
\begin{equation}
\begin{split}
A\subr{0,up} &= 3.69 \times10^{-12} \, \text{N.V}^{-2} \\
A\subr{1,up} &= 1.05 \times10^{-4} \, \text{N.m}^{-1}\text{V}^{-2}
\label{eq:dimensional-force-upper-bounds}
\end{split}
\end{equation}
which correspond to the following dimensionless quantities:
\begin{equation}
\begin{split}
\tilde{A}\subr{0,up} &= \frac{L^4}{E I g_0^2} A\subr{0,up} = 1.64 \\
\tilde{A}\subr{1,up} &= \frac{L^4}{E I g_0} A\subr{1,up} = 5.55  
\label{eq:dimensionless-force-upper-bounds}
\end{split}
\end{equation}

These upper-bounds are important to constrain as many parameters as possible in the fit of the  radial  resonance frequency (see figure \ref{fig:omega_fit}).

\subsection{Extraction of $z_\mathrm{st}(V_\mathrm{g})$, $y_\mathrm{st}(V_\mathrm{g})$ and $\theta(V_\text{g})$ - Fitting procedure}
\label{SI-th-fitting}
\begin{figure}[H]
  \centering
  \includegraphics[width=0.75\linewidth]{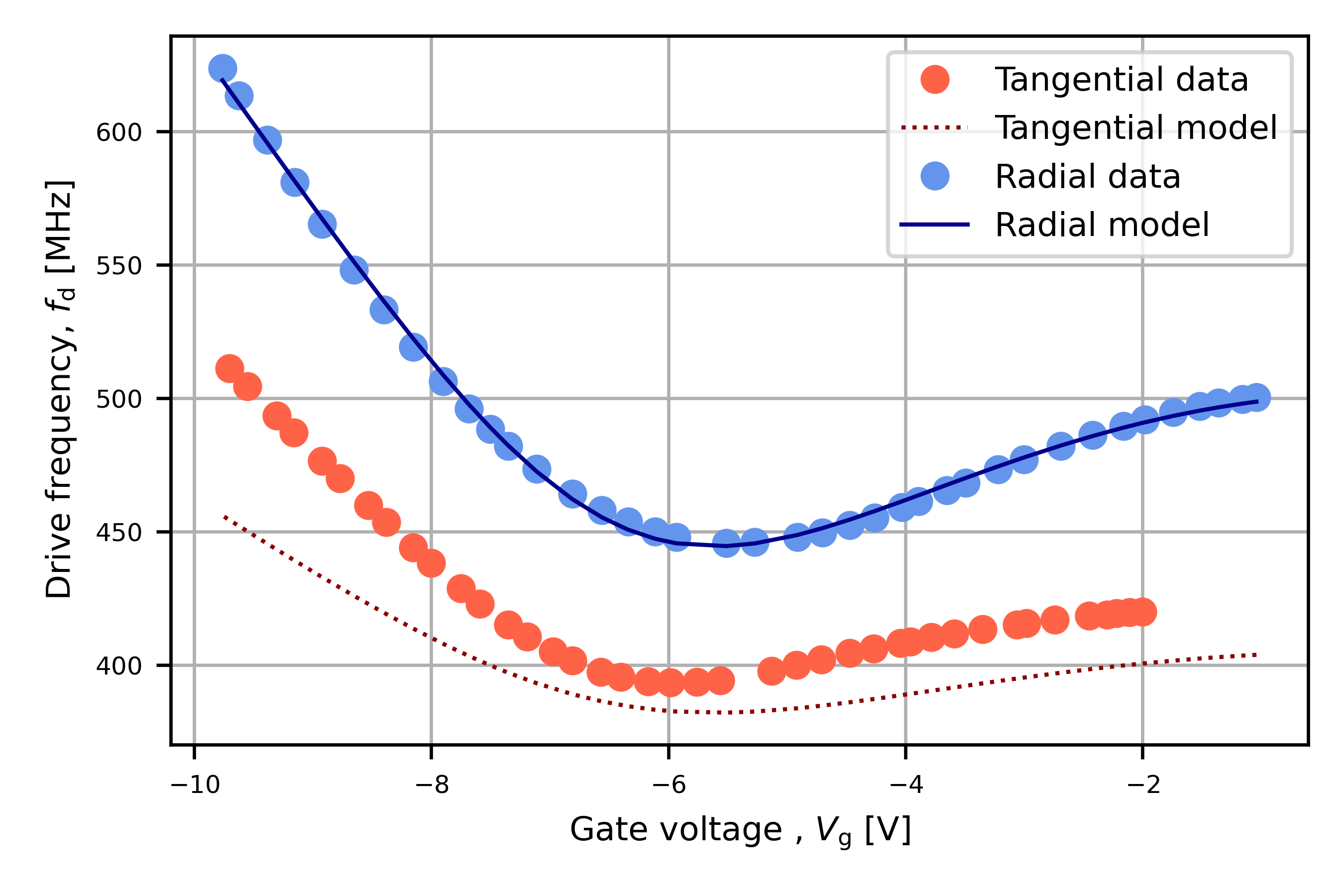}
  \caption{\textbf{Fit of the  radial  mode resonance frequency.} The fit (dark blue line) is only performed over the data points from the  radial  mode (light blue dots).}
  \label{fig:omega_fit}
\end{figure}

To obtain the static displacement ($y_\mathrm{st}(V_\mathrm{g}), z_\mathrm{st}(V_\mathrm{g})$) shown in the figure \ref{fig1}c of the main text, we perform a fit of the  radial  mode resonance frequency (figure \ref{fig:omega_fit}). The discrepancy between the measured  tangential  mode resonance frequencies (light red dots) and the predicted resonance frequencies (dark red dotted-line) could be explained by torsional effect which were not taken into account in our model  \cite{rechnitz2022mode, rechnitz2023theoretical}.  Nevertheless this discrepancy is comprised between $5-10\,\%$ of the tangential mode resonance frequency. This is relatively weak considering the simplicity of the model.

The higher energy solution of the eigenvalues system of equations \ref{eq:matrix_form} is used to fit the measured resonance frequency of the  radial  mode. The fitting procedure requires to solve the systems of algebraic equations \ref{eq:GD_yz_0} and \ref{eq:GD_yz_st} at each iteration. The fixed and free parameters are given in Table \ref{table:fixparam} and Table \ref{table:freeparam} respectively.

\begin{table}[ht]
\centering
{\begin{tabular}{l  l } 
 \hline \hline
 Parameter & value  \\ [1ex]
 \hline
 Graphene density, $\rho$  &  $\rho = 2300\,$kg.m$^{-3}$  \\ [1ex]
 Integral mode factor, $a_2$  &  $a_2 = 0.375$  \\ [1ex]
 Integral mode factor, $b_2$  &  $b_2 = 4.93$  \\ [1ex]
 Integral mode factor, $c_2$  &  $c_2 = 194.8$  \\ [1ex]
 Nanotube resonator length, $L$  &  $L = 385\,$nm  \\ [1ex]
 Nanotube-gate distance (for: $z\subr{st}=0$), $g_0$  &  $g_0 = 65\,$nm  \\ [1ex]
 \hline
\end{tabular}}
\caption{{Fixed parameters used in the Euler-Bernoulli model for the fitting of the  radial  mode resonance frequency. The integral mode factors are defined in section \ref{SI-th-Galerkindecomp}}.}
\label{table:fixparam}
\end{table}

\begin{table}[ht]
\centering
{\begin{tabular}{l | c | c | c  } 
 \hline \hline
 Parameter & Initial values & Boundaries & Best fit value  \\ [1ex]
 \hline
 Initial axial tension, $T_0$ &  $-4\,$nN.m$^{-1}$ & Not constrained & $-4.74\,$nN.m$^{-1}$   \\ [1ex]
 Force parameter, $A_0$  & 1 & Upper bound: 1.64 & 1.02   \\ [1ex]
 Force parameter, $A_1$  & 4 & Upper bound: 5.55 & 1.19  \\ [1ex]
 Initial angle $\theta_0$ & $\frac{3 \pi}{4}$ & Not constrained & $133^{\circ} $ \\ [1ex]
 Young modulus, $E$ & $0.3\,$TPa & Not constrained &  $0.23\,$TPa \\ [1ex]
 CNT diameter, $d$ & $4\,$nm & 2 - 4.5 nm & $4.3 \,$nm  \\ [1ex]
 CNT wall thickness, $dr$ & $1\,$nm & 0.34 - 1.36 nm & $1.02\,$nm  \\ [1ex]
 \hline
\end{tabular}}
\caption{{Free parameters used in the Euler-Bernoulli model for the fitting of the  radial  mode resonance frequency.}}
\label{table:freeparam}
\end{table}

\begin{figure}[H]
  \centering
  \includegraphics[width=0.7\linewidth]{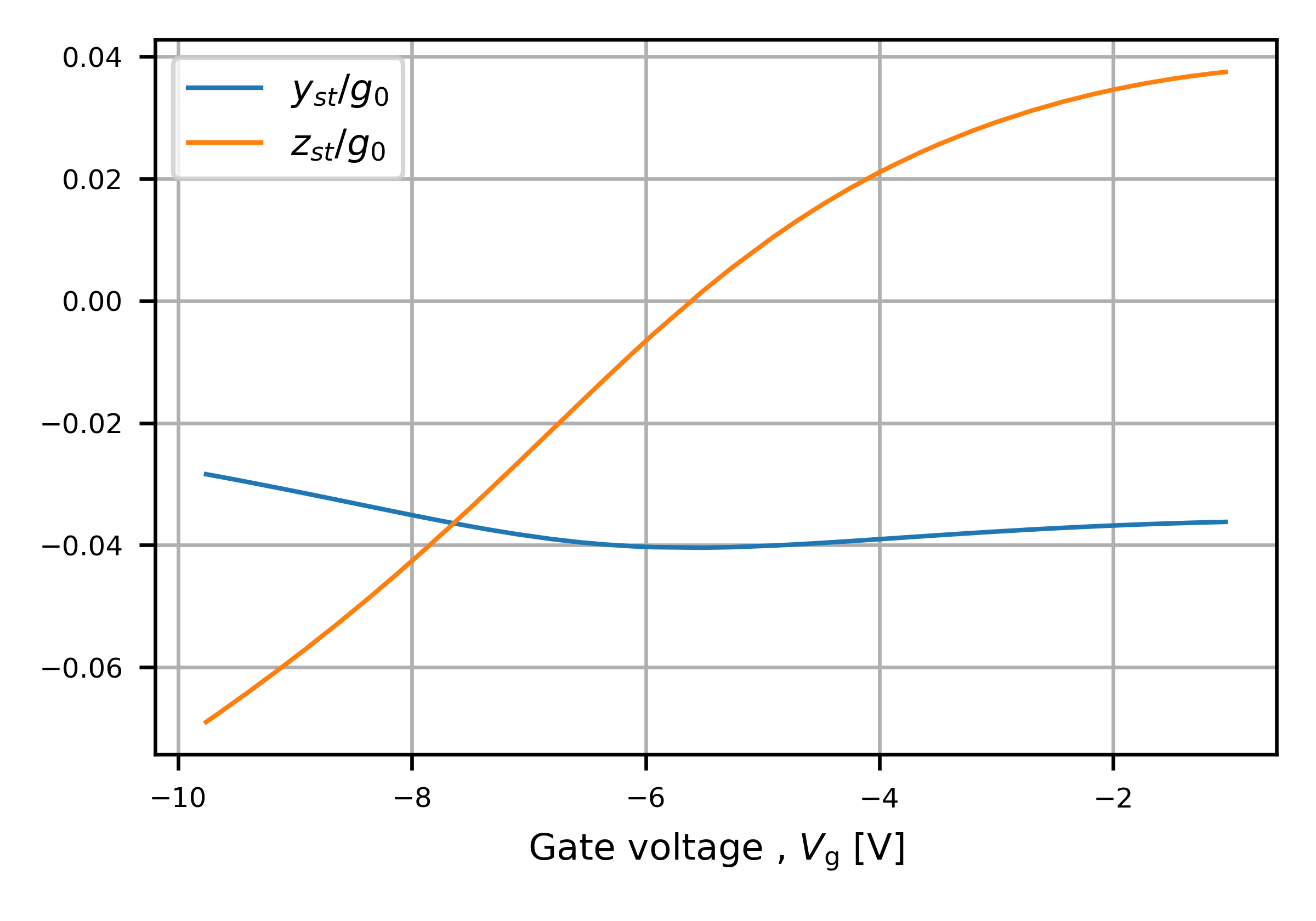}
  \caption{{Static displacement of the midpoint of the carbon nanotube as a function of the gate voltage $V_\text{g}$.}}
  \label{fig:ys-zs}
\end{figure}

The upper bounds on $A_0$, and $A_1$  come from the finite-element simulations and are derived in section \ref{SI-th-implementation}. From a statistical characterization of our nanotube growth process using a Transmission Electron Microscope, we can conservatively limit the nanotube diameter to the range ($2\,$nm - $4.5\,$nm) as well as the nanotube wall thickness to ($0.34\,$nm - $1.36\,$nm) which corresponds to 1 to 4 walls. The best fit value $dr = 1.0\,$nm corresponds to a multi-walled carbon nanotube, which is consistent with the low Young's modulus we obtained (for SWCNT: $E \simeq 0.5-3\,$TPa \cite{rechnitz2022mode, Eichler12}, for MWCNT: $E \simeq 0.2-1\,$TPa \cite{yu2000strength}).
The resulting coordinates $y_\text{st}$ and $z_\text{st}$ of the static displacement of the midpoint of the carbon nanotube are shown in figure \ref{fig:ys-zs}, as well as in figure \ref{fig1}.c of the maintext.

\subsection{Sensitivity analysis of EB-model}
\label{SI-th-sensitivity-analysis}

In order to quantify the uncertainty of the fitting parameters due to a wrong evaluation of the dimensions of the device, we performed a sensitivity analysis. To do so, we tested the influence of the carbon nanotube length which is the only dimensional parameter which directly plays a role in the Euler-Bernoulli model. The uncertainties on the five other dimensional parameters (four parameters describing the gate electrode dimensions and one parameter for the contact height asymmetry, see Supplementary Note \ref{SI-estimationuncertainty-us}) act on the force parameters $A_0$ and $A_1$ which were taken as free parameters.

We consider three different carbon nanotube lengths $L = 375\,$nm$,\, 385\,$nm$, \,395\,$nm where $L = 385\,$nm is found from SEM images including the visible tilt of the nanotube (see Supplementary Note \ref{SI-alpha3t-DeltaVgus-geometriccharac}). In table \ref{table:sensitivity} the dispersion of each parameter is presented, where the $\delta$ are defined as the difference between the maximum and minimum values found for the three tested nanotube lengths.

\begin{table}[ht]
\centering
{\begin{tabular}{l | c } 
 \hline \hline
 Parameter & Geometric dimensions sensitivity  \\ [1ex]
 \hline
 Initial axial tension, $ T_0$ &  $ \delta T_0 = 1.53 \,$ N.m$^{-1}$  \\ [1ex]
 Force parameter, $A_0$  &  $\delta A_0 = 0.22 $   \\ [1ex]
 Force parameter, $A_1$  &  $\delta A_1 = 0.08$  \\ [1ex]
 Initial buckling angle $\theta_0$ & $\delta \theta_0 = 3.4^{\circ}$ \\ [1ex]
 Young modulus, $E$ & $\delta E = 0.1\,$TPa \\ [1ex]
 CNT diameter, $d$ & $\delta d = 0.4\,$nm \\ [1ex]
 CNT wall thickness,$dr$ & $\delta dr = 0.3 \,$nm \\ [1ex]
 \hline
\end{tabular}}
\caption{{Sensitivity analysis with respect to the nanotube length $L$.}}
\label{table:sensitivity}
\end{table}

The uncertainty on the initial axial tension $T_0$, on the Young modulus $E$ and on  the initial buckling angle $\theta_0$, can be propagated to the initial nanotube buckling ($y_0, z_0$). For this purpose we use the expression $y_0 = R\cos(\theta_0)$ and $z_0 = R\sin(\theta_0)$ with $R=\sqrt{-\frac{\frac{c_2}{b_2} + P}{\alpha b_2}}$ and $P = \frac{T_0 L^2}{E I}$ as previously defined in Supplementary Note \ref{SI-th-nondimensionalparams} and \ref{SI-th-Galerkindecomp}. 
This results in $y_0/g_0 = -0.036 \pm 0.018 $ and $z_0/g_0 = 0.039 \pm 0.019 $.

\begin{figure}[H]
  \centering
  \includegraphics[width=1.0\linewidth]{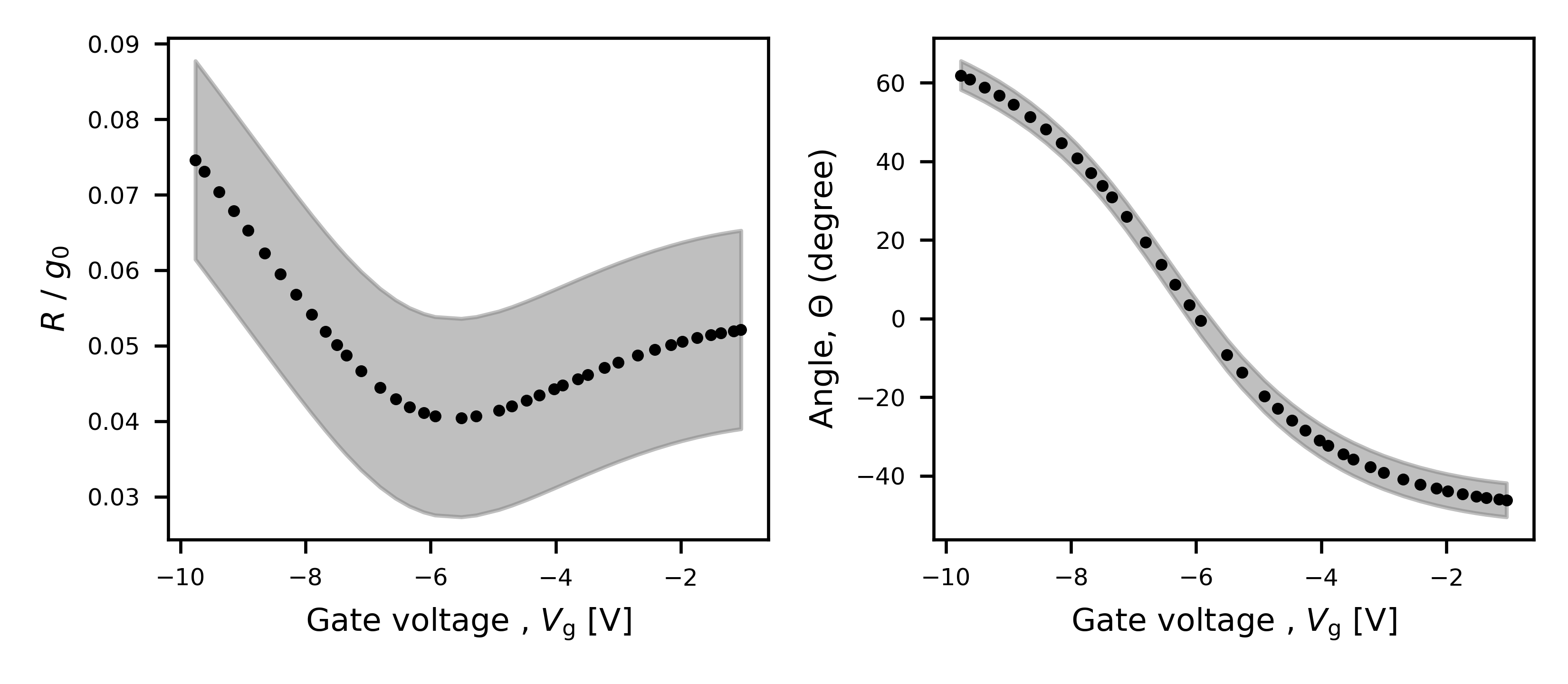}
  \caption{{Radius and angle describing the static displacement of the midpoint of the carbon nanotube such that $y_0 = R\cos(\theta_0)$ and $z_0 = R\sin(\theta_0)$. The light gray areas correspond to the uncertainty estimated from a sensitivity analysis of the Euler-Bernoulli model. }}
  \label{fig:R-and-theta-error}
\end{figure}

The uncertainty on $R(V_\text{g})$ and $\theta(V_\text{g})$ are also computed, and the result is shown in figure \ref{fig:R-and-theta-error}, and as the light yellow area shown in figure \ref{fig1}.c of the maintext. The uncertainty on the CNT diameter $d$ is used in Supplementary Note \ref{SI-alpha3t-nonlinearity-unluzp-uncertainty} to estimate an uncertainty on the phonon number of the bistability onset.

\section{Electrical and mechanical characterization of the device}
\label{SI-EMcharac}
\subsection{Single quantum dot charge stability diagram}
\label{SI-EMcharac-QDchargestability}
\begin{figure}[H]
  \centering
  \includegraphics[width=1\linewidth]{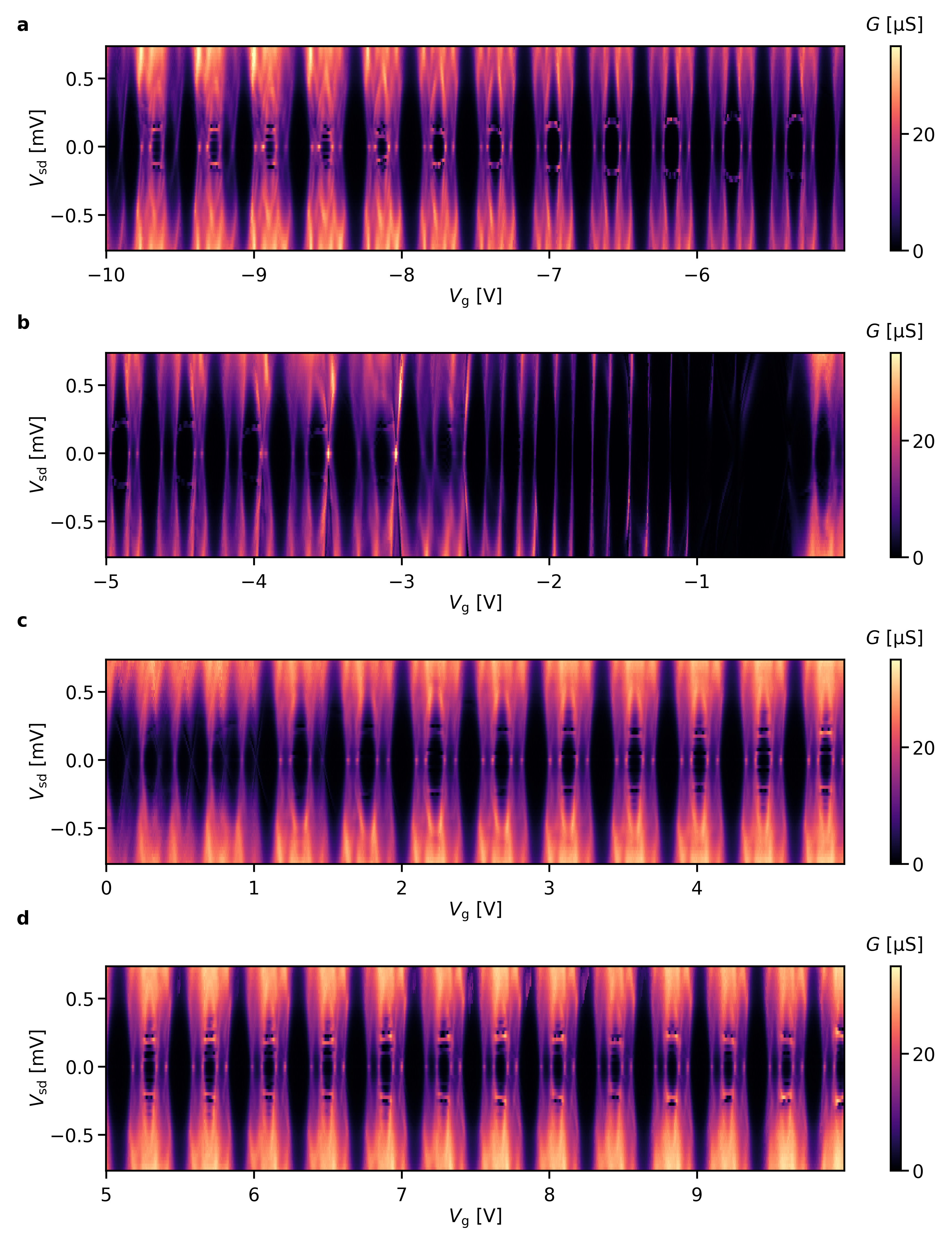}
  \caption{\textbf{Single quantum dot charge stability diagram of the device.} Differential conductance $G$ measured as a function of gate voltage $V_\mathrm{g}$ and source-drain bias voltage $V_\mathrm{sd}$. \textbf{a} $-10\,\text{V} < V_\text{g} < -5\,\text{V}$, \textbf{b} $-5\,\text{V} < V_\text{g} < 0\,\text{V}$, \textbf{c} $0\,\text{V} < V_\text{g} < 5\,\text{V}$, \textbf{d} $5\,\text{V} < V_\text{g} < 10\,\text{V}$.}
  \label{fig:vsd-vg}
\end{figure}

\begin{figure}[H]
  \centering
  \includegraphics[width=1\linewidth]{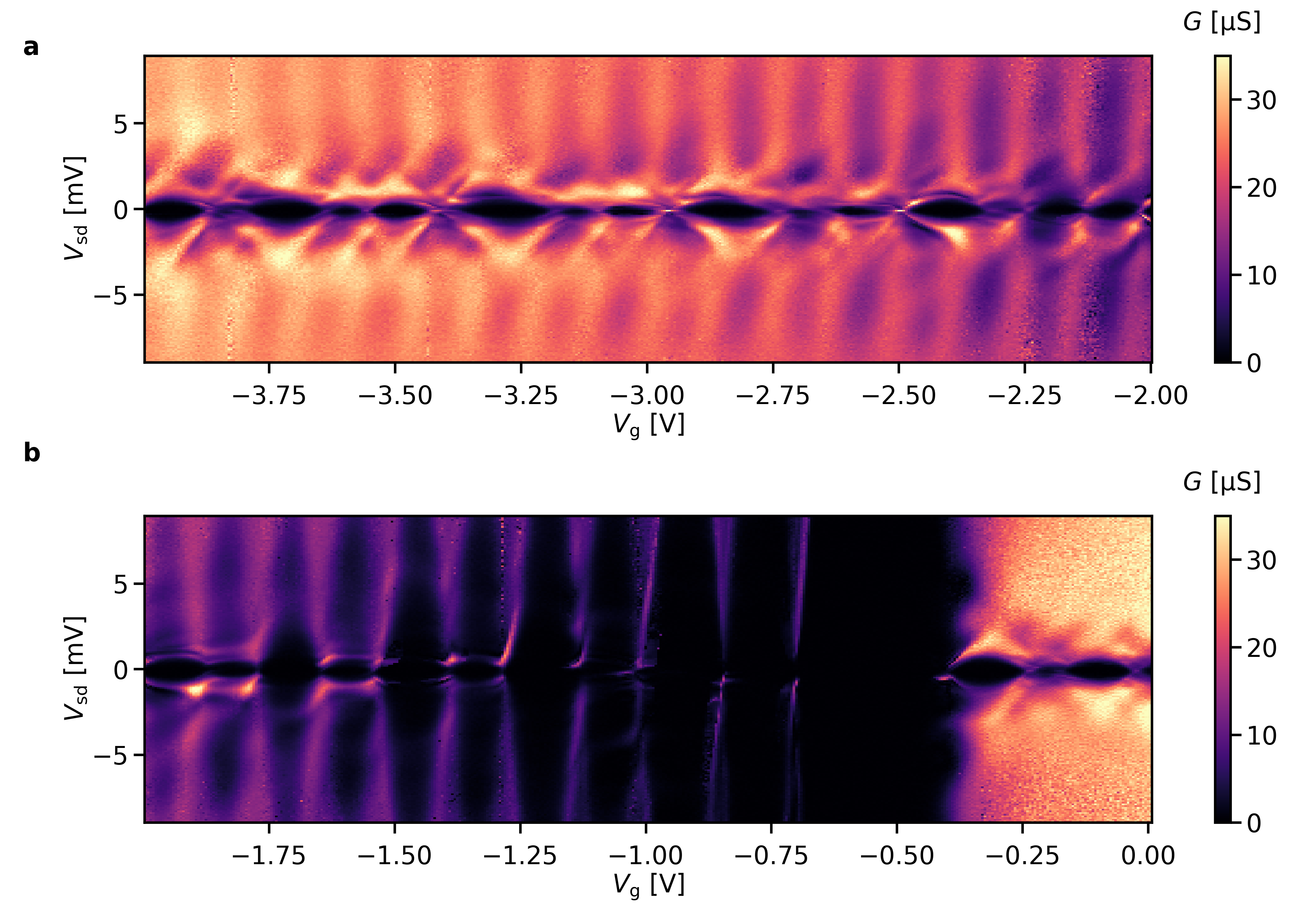}
  \caption{\textbf{Single quantum dot charge stability diagram of the device with an extended range of source-drain voltage.} \textbf{a} $-4\,\text{V} < V_\text{g} < -2\,\text{V}$, \textbf{b} $-2\,\text{V} < V_\text{g} < 0\,\text{V}$. The charging energy of the quantum dot $E_\text{C}$ can be derived from this measurement and is estimated to $E_\text{C} \approx 25-30\,\text{mV}$.}
  \label{fig:vsd-vg-big}
\end{figure}

\subsection{Frequency-gate voltage maps}
\label{SI-EMcharac-Frequencygatemaps}

\begin{figure}[H]
  \centering
  \includegraphics[width=1\linewidth]{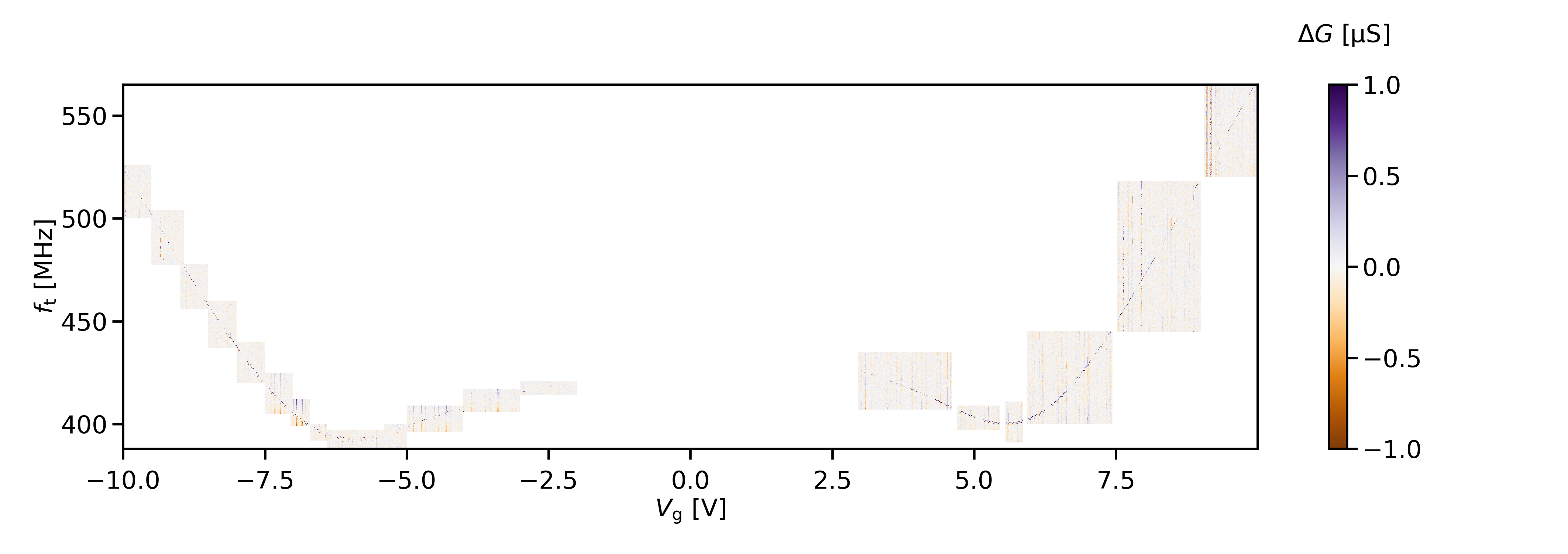}
  \caption{\textbf{Broad range frequency-gate voltage map of the  tangential  mode.} The variation of differential conductance $\Delta G$ is plotted as a function of gate voltage $V_\mathrm{g}$ and drive frequency around the resonance frequency of the  tangential  mode $f_\mathrm{t}$. The resonance frequency of the  tangential  mode follows a w-shape which is expected for upward-buckled carbon nanotubes \cite{rechnitz2022mode}. The slight shift of the mechanical resonance frequency between positive and negative gate voltages is due to a current-induced annealing of the carbon nanotube between the different measurements.}
  \label{fig:freq-vg-swing}
\end{figure}

\subsection{Frequency sweep}
\label{SI-EMcharac-Frequencysweep}
\begin{figure}[H]
  \centering
  \includegraphics[width=1\linewidth]{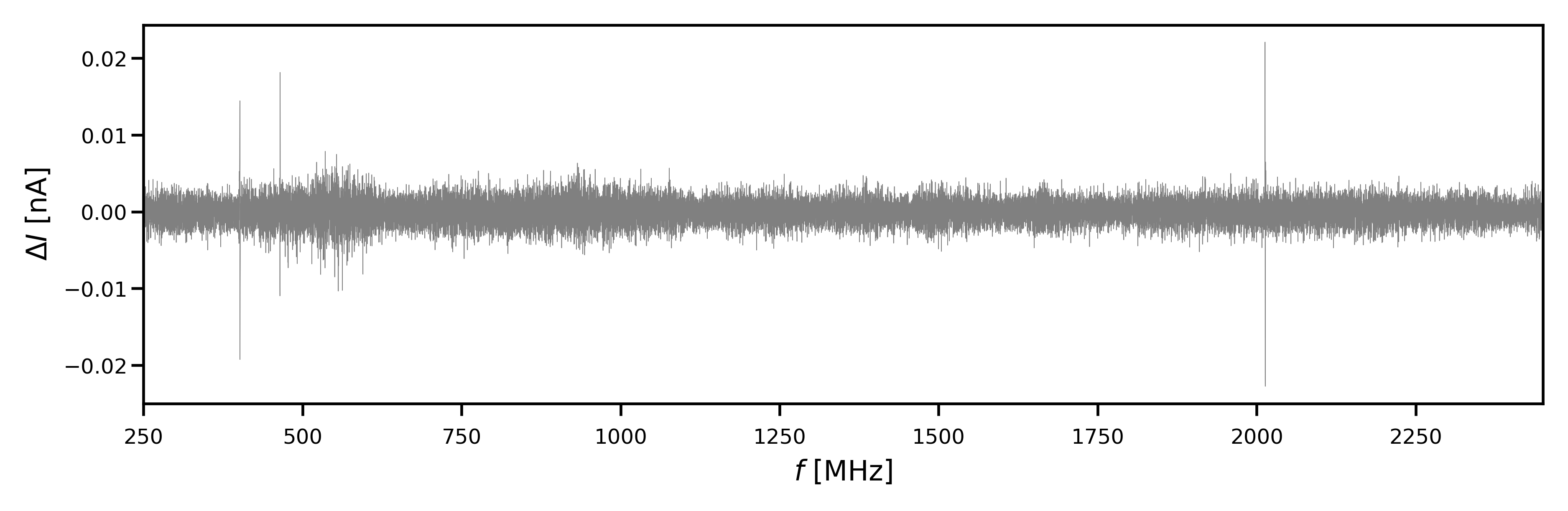}
  \caption{\textbf{Broad-range frequency-sweep.} The change in current is taken at $V_\text{g}\,=\,-6.844\,\text{V}$ as a function of drive frequency $f$ at a power of $-45\,\text{dBm}$. The background signal has been removed. Three mechanical resonances of the carbon nanotube can be distinguished as sharp peaks at $402\,\text{MHz}$, $465\,\text{MHz}$ and $2013\,\text{MHz}$. }
  \label{fig:freq-sweep}
\end{figure}

In figure \ref{fig:freq-sweep}, the frequency is swept across a wide range of frequencies to detect mechanical resonances, which appear as a change in the time-averaged current. Three mechanical resonance frequencies can be distinguished, two with frequencies between $400$ and $500\,\text{MHz}$ and one with a frequency above $2\,\text{GHz}$.
The resonances at lower frequencies can be attributed to the transverse bending modes of the carbon nanotube ( radial  and  tangential  modes). The third resonance around $2\,$GHz is associated to the 3rd higher-order of the lowest frequency mode.

\section{FEM simulation benchmarking - Gate lever arm estimation}
\label{SI-FEM_benchmarking}

\begin{figure}[H]
  \centering
  \includegraphics[width=1\linewidth]{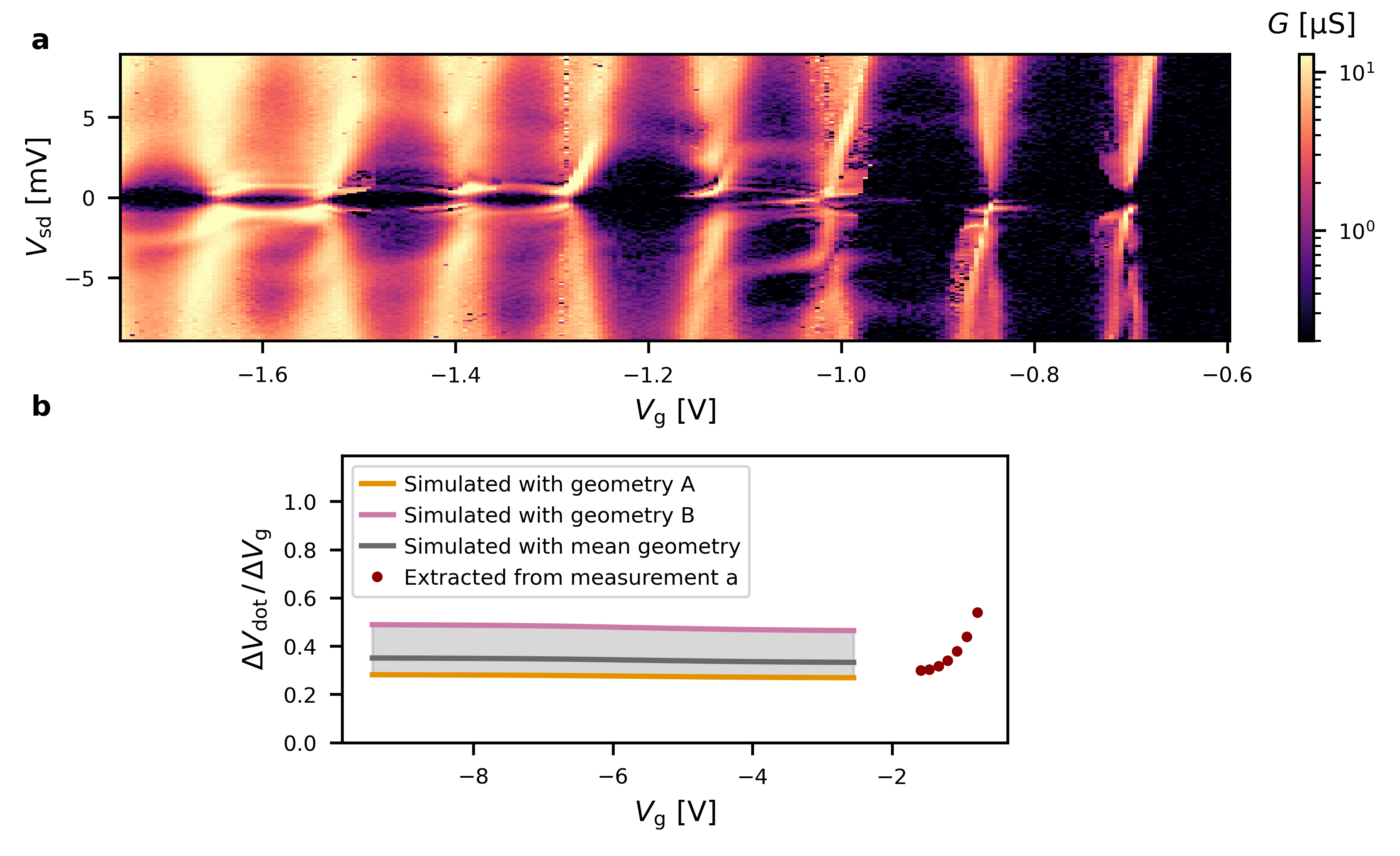}
  \caption{\textbf{Gate lever arm estimation.}
\textbf{a} Single quantum dot charge stability diagram of the device. The lever arm values of the first seven Coulomb diamonds are shown as dark red dots in panel \textbf{b}.
\textbf{b} Lever arm values plotted as a function of gate voltage. The dark gray line represents the lever arm calculated from the FEM simulation with the mean geometry, while the pink and orange lines indicate the values obtained for the two boundary geometries A and B presented in Supplementary Section \ref{SI-estimationuncertainty-us}.}

  \label{fig:FEMbenchmark}
\end{figure}

The conversion $\Delta V_\text{g}-u_\text{t}$ is a crucial step in the extraction method of the intrinsic nonlinearity presented in Supplementary section \ref{SI-alpha3t}. To validate the finite-element model used in this conversion, we benchmark its predictions against experimental measurements. One way to do so is by extracting the lever arm from the $V_\text{sd}-V_\text{g}$ map shown in Figure \ref{fig:FEMbenchmark}a, which is defined as
\begin{equation}
    \text{lever arm}
    \,=\,
    \frac{\Delta V_\text{dot}}{\Delta V_\text{g}}.
\end{equation}
Indeed, the lever arm is directly proportional to the amplitude of motion of the  tangential  mode, as defined in Supplementary Section \ref{SI-alpha3t-DeltaVgus} it holds
\begin{equation}
    u_\text{t}\,=\,\frac{1}{\cos(\theta)}\cdot \frac{\text{lever arm}}{E(z_\text{st}(V_\text{g}^\text{DC}), V_\text{g}^\text{DC})} \cdot \Delta V_\text{g}.
\end{equation}

The lever arm used in the $\Delta V_\text{g}-u_\text{t}$ conversion presented in Supplementary Section \ref{SI-alpha3t-DeltaVgus} is extracted from the finite-element model used to simulate the electrostatic potential around the carbon nanotube. It describes the change in dot potential corresponding to a given change in gate voltage and can be extracted from the slopes of the Coulomb-diamond edges. The lever arm values extracted from the measurement shown in Figure \ref{fig:FEMbenchmark}a are shown as dark red dots in Figure \ref{fig:FEMbenchmark}b.\\
From the finite-element model used in our analysis, the lever arm $\Delta V_\text{dot}/\Delta V_\text{g}$ can be calculated, corresponding to the dark gray line shown in Figure \ref{fig:FEMbenchmark}b. The pink and orange lines show the corresponding lever arms obtained from simulations using the two boundary geometries considered in the uncertainty analysis in Supplementary Section \ref{SI-estimationuncertainty-us} (illustrated in Figure \ref{fig:error_geometry} using the same color code). Together, these two curves define the enclosed light-gray uncertainty band for $\Delta V_\text{dot}/\Delta V_\text{g}$. Importantly, the experimentally extracted values (dark red dots) lie within this uncertainty band. 
\\ \\
For more negative gate voltages, the presence of subgap conductance prevents a reliable extraction of the lever arm from the $V_\text{sd}$-$V_\text{g}$ measurements (see, e.g., Figure \ref{fig:vsd-vg-big}a). Nevertheless, the trend of the experimentally extracted lever arm values suggests that it approaches a constant value as the occupation number of the quantum dot increases. \\
This behavior is physically expected: as the occupation number increases, Coulomb repulsion leads to an expansion of the quantum dot size, but only up to the total length of the carbon nanotube. Beyond this point, the dot size saturates and the lever arm is expected to remain approximately constant, consistent with the stabilization observed in the lever arm values for the final measured Coulomb diamonds. \\
Consequently, for even more negative gate voltages, similar lever arm values are expected, which would remain within the uncertainty band of the simulation. This agreement provides strong support for the validity of the finite-element model used in our analysis.

\section{Validity of the approximation $u_\text{t} \gg u_\text{r}$}
\label{SI-approx_validity}

\begin{figure}[H]
  \centering
  \includegraphics[width=1\linewidth]{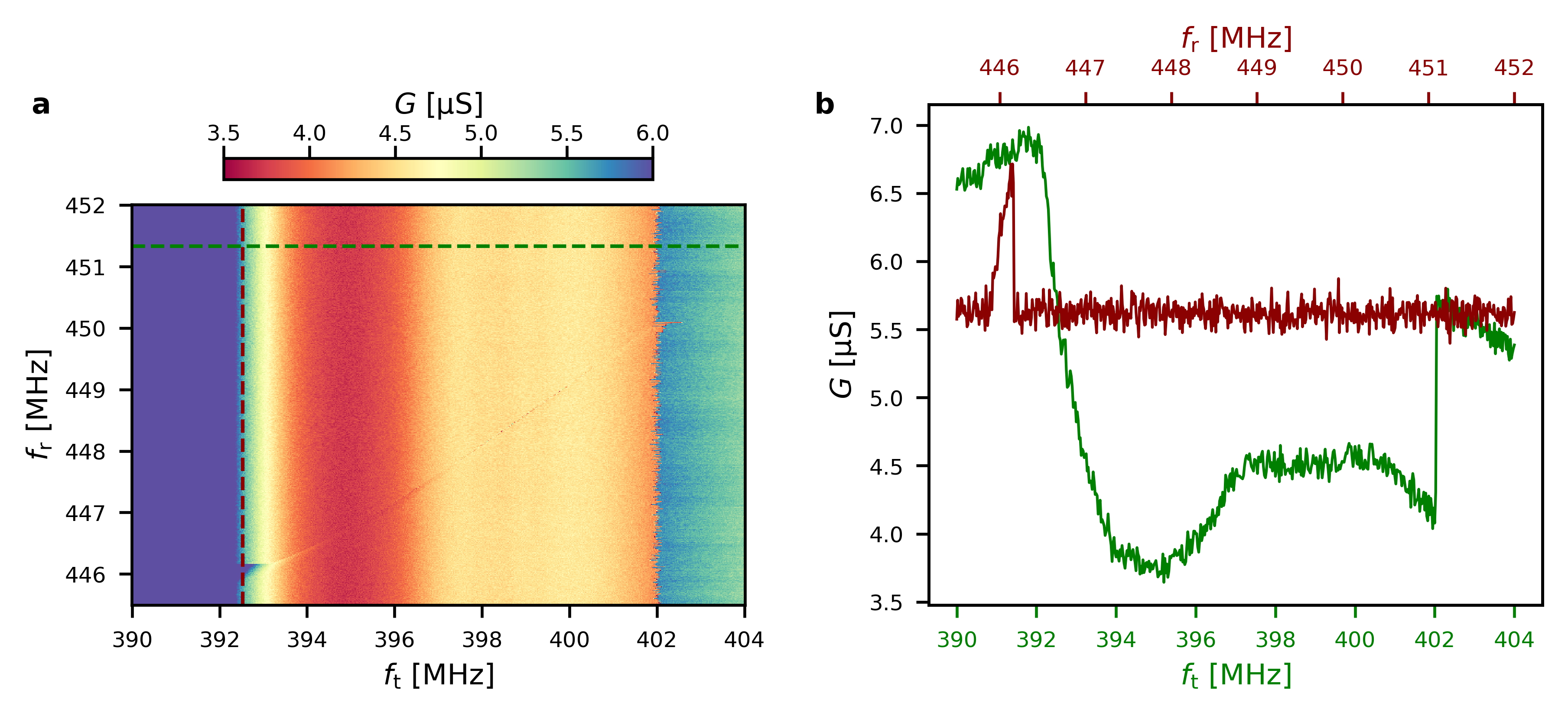}
  \caption{\textbf{Validity of assumption $u_\mathrm{t} \gg u_\mathrm{r}$.}
\textbf{a} Two-tone spectroscopy as the one presented in Figure \ref{fig2} of the main text ($V_\text{g}\,=\,-5.612\,\text{V}$) with $P_\text{r} = -40\,\mathrm{dBm}$ and $P_\text{t} = -38.5\,\mathrm{dBm}$, with the exception that the conductance $\frac{\partial I}{\partial V_\text{sd}}$ is plotted and not the variation of the conductance $|\Delta G|$ as it is the case in figure \ref{fig2}. \textbf{b} Line-cuts of the mechanical resonance for the  tangential  mode (green curve) and the  radial  mode (red curve) taken at the locations highlighted in \textbf{a}.
  }
  \label{fig:approxvalidity}
\end{figure}

In Supplementary Section \ref{SI-th}, the equation \ref{eq:omegat} of the main-text is derived by using the approximation that the amplitude of the  tangential  mode is significantly higher than the amplitude of the  radial  mode $u_\text{t} \gg u_\text{r}$. While we cannot directly measure the amplitude of the displacement associated to each modes we can assess the validity of this approximation based on the bifurcation frequency extent of the mechanical resonance. For example, as can be seen in Figure \ref{fig:approxvalidity}b, the  tangential  mode's resonance extends over more than $10\,\mathrm{MHz}$, whereas the  radial  mode's resonance spans only about $200\,\mathrm{kHz}$. This indicates a significantly stronger bifurcation of the  tangential  mode and, consequently, a much larger oscillation amplitude compared to the  radial  mode. 
\\
It is important to note that the oscillation amplitude depends on the effective driving power exciting the modes, which in general is different from the applied RF power at the antenna. This is due to the fact that the oscillation directions of the mechanical modes are, in most cases, misaligned with respect to the radiation field of the RF antenna, which is located approximately $1\,\mathrm{mm}$ above the circuit. 
This explains why the antenna driving power of the  tangential  mode and the  radial  are so similar in Fig.~\ref{fig:approxvalidity}a. Indeed, at $V_\text{g} \approx -5.6\,\mathrm{V}$, the  radial  mode oscillates predominantly parallel to the device plane, resulting in inefficient coupling to the antenna RF field. Consequently, similar antenna driving powers of the two modes result in a weak driving of the  radial  mode and a strong driving of the  tangential  mode.
\\
For this reason, irrespective of the RF power applied at the antenna, we ensure in all two-tone spectroscopy measurements that the bifurcation of the  tangential  mode is larger than the bifurcation of the  radial  mode, validating the approximation $u_\text{t} \gg u_\text{r}$.

The list of all driving powers used for the frequency-frequency maps of each dataset is presented in table \ref{table:list_powers}. The power applied to drive the  tangential  mode is usually much larger than the driving power of the  radial  mode, except for the dataset at $V_\text{g}\approx-5.6\,\text{V}$ due to the reason given above.

\begin{table}[ht]
\centering
\begin{tabular}{c c c} 
\hline \hline
$V_\text{g}$ (V) & $P_\text{t}$ (dBm) & $P_\text{r}$ (dBm)\\
 \hline
 -2.4 & -40 & -70  \\ [1ex] 
 -4.0 & -37 & -60  \\[1ex] 
 -4.7 & -37 & -55  \\  [1ex] 
 -5.6 & -38.5 & -40  \\ [1ex] 
 -6.2 & -37 & -55  \\ [1ex] 
 -6.8 & -40 & -50  \\ [1ex]
 -7.6 & -37 & -65 \\ [1ex]
 -7.7 & -37 & -65  \\ [1ex]
 -8.5 & -37 & -65  \\ [1ex]
 -9.2 & -37 & -60  \\
 \hline
\end{tabular}
\caption{\textbf{List of driving powers used for the frequency-frequency maps for each dataset.} $P_\text{t}$ is the driving power of the  tangential  mode, and $P_\text{r}$ is the driving power of the  radial  mode.}
\label{table:list_powers}
\end{table}

\section{Extraction of the intrinsic Duffing nonlinearity $\alpha_\text{3,r}$} \label{SI-alpha3t}
\subsection{Extraction procedure}
\label{SI-alpha3t-outline}
\begin{figure}[H]
  \centering
  \includegraphics[width=0.5\linewidth]{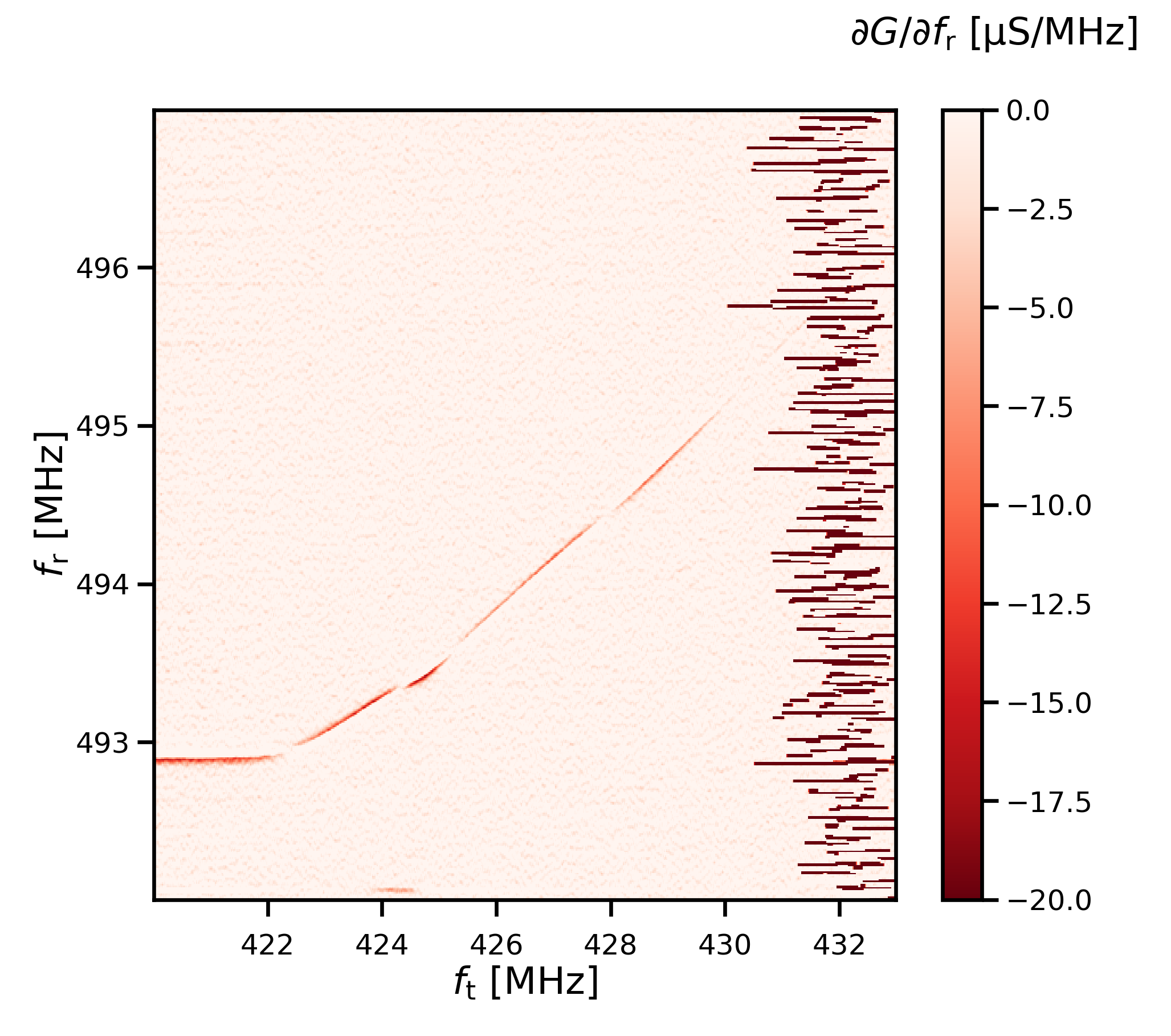}
  \caption{\textbf{Frequency-frequency map.} Two drive frequencies $f_\mathrm{t}$ and $f_\mathrm{r}$ are applied simultaneously around the resonance frequencies of the  tangential  and the  radial  mode, respectively (at $V_\text{g}\,=\,-7.624\,\text{V}$).}
  \label{fig:freq-freq-raw}
\end{figure}

\begin{figure}[H]
  \centering
  \includegraphics[width=1\linewidth]{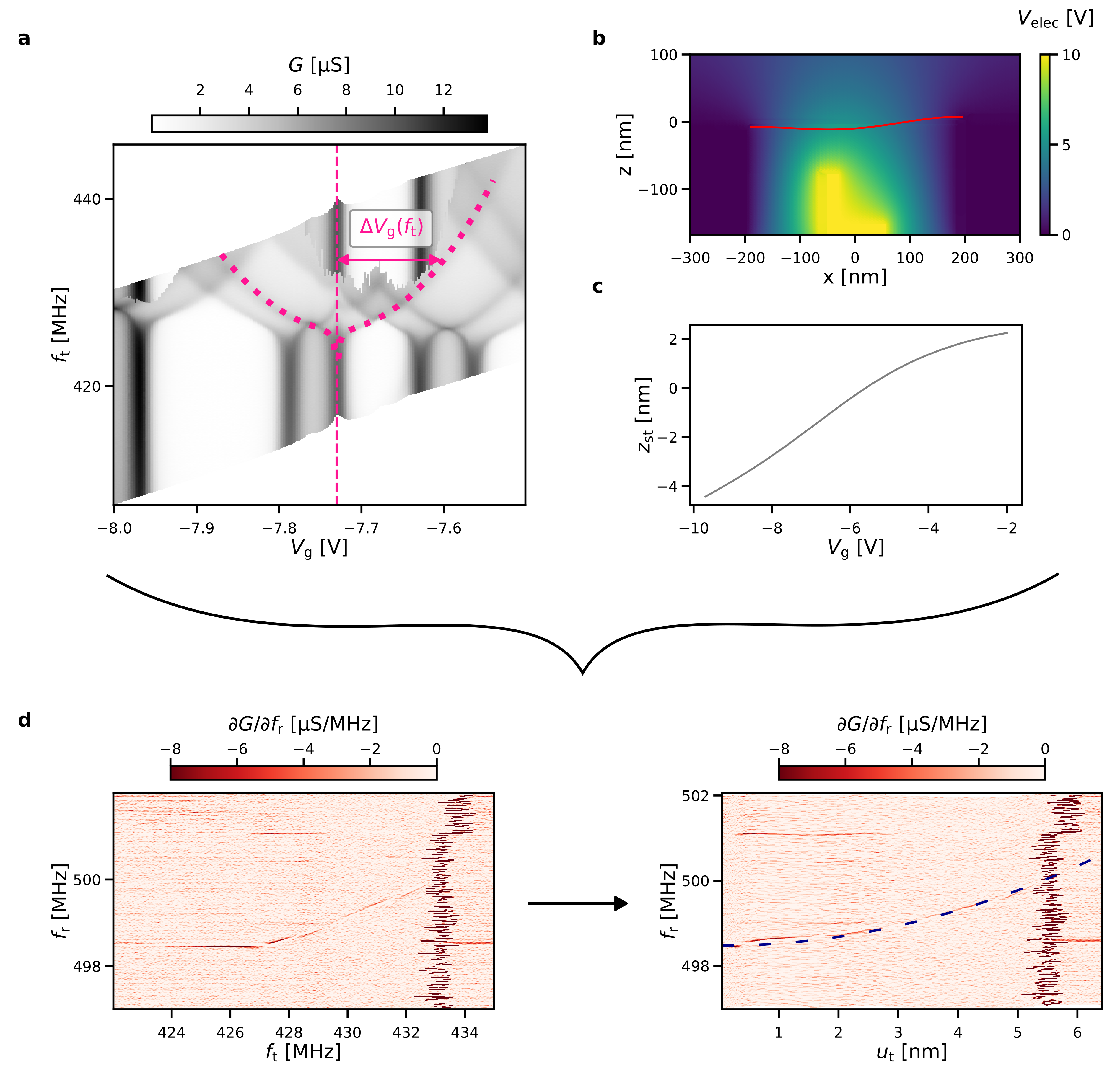}
  \caption{\textbf{Summary of the extraction procedure of the intrinsic cubic nonlinearity of the  radial  mode illustrated on dataset $V_\mathrm{g} = -7.75\,$V.} \textbf{a} Frequency-gate voltage map of the  tangential  mode taken at strong drive ($P\,=\,-37\,\text{dBm}$). The shift of the Coulomb peak is fitted (pink dotted line) and enables us to find the relation $f_\text{t} \leftrightarrow\Delta V_\text{g}\,$. \textbf{b} Finite-element simulation of the electrostatic potential around the carbon nanotube created by a voltage of $V_\text{g}\,=\,10\,\text{V}$ applied to the gate electrode. \textbf{c} Static displacement $z_\mathrm{st}$ of the carbon nanotube as a function of the applied gate voltage. \textbf{d} Conversion of a frequency-frequency map to an amplitude-frequency map. The resonance frequency of the  radial  mode with respect to the amplitude of the  tangential  mode is then fitted with equation \ref{eq:freq_freq_fit}.}
  \label{fig:summary_procedure}
\end{figure}

In the following, we summarize the procedure to extract the value of the intrinsic cubic nonlinearity $\alpha_\text{3,r}$ (cf. figure \ref{fig4} of the main text) from two-tone spectroscopy such as the one shown in figure \ref{fig:freq-freq-raw}. The individual steps are described in the following subsections. \\ \\
Actuating the carbon nanotube with two drive frequencies allows us to follow the change in frequency of one mode as a function of the amplitude of the other mode (see figure \ref{fig:freq-freq-raw}).\\
The variation of the resonance frequency of the  radial  mode $\omega_{\text{r,int}}$ depending on the oscillation amplitude of the  tangential  mode $u_{\text{t}}$ via the intrinsic cubic nonlinearity of the  radial  mode $\alpha_{3\text{,r}}$ in the carbon nanotube (see section \ref{SI-th-derivationequations}) is given by:
\begin{equation}
\omega_{\text{r,int}}(u_{\text{t}}) \,=\,  \omega_{\text{r},0} \sqrt{1 + \frac{\alpha_{3\text{, r}}}{2} \frac{u_{\text{t}}^2}{\omega^2_{\text{r},0}}} \,,
\label{eq:freq_freq_fit}
\end{equation}
with $\omega_{\text{r},0}$ the natural resonance frequency of the  radial  mode. One can then extract the intrinsic Duffing nonlinearity $\alpha_{3\text{,r}}$ by measuring the variation of $\omega_{\text{r}}$ as a function of the  tangential  mode amplitude $u_\text{t}$. In practice, in a two-tone spectroscopy measurement, we do not directly change the  tangential  mode amplitude $u_\text{t}$ but only the driving frequency $f_\text{t}$. For this reason, we first developed a procedure allowing us to convert the frequency axis $f_\text{t}$ into  tangential  mode amplitude $u_\text{t}$. This conversion is split in two steps:
\begin{itemize}
    \item $\mathbf{f_t \rightarrow \Delta V_g:}$ When one strongly drives the  tangential  mode, its mechanical oscillation will cause the dot formed in the carbon nanotube to sense the gradient of electric potential produced by the gate electrode. The displacement of the dot will therefore result in a splitting of the Coulomb peaks that we call in the following: $2\Delta V_\text{g}$. An example of such a measurement is presented in Fig.\ref{fig:summary_procedure}.a. This conversion step of the procedure is described in section \ref{SI-alpha3t-fsDeltaVg}.
    \item $\mathbf{\Delta V_g \rightarrow u_t:}$ By a precise knowledge of the electric field produced by the gate electrode, one can then relate the Coulomb peak shift $\Delta V_\text{g}$ to the carbon nanotube displacement of the  tangential  mode $u_\text{t}$. This step relies on finite element simulations and is described in section \ref{SI-alpha3t-DeltaVgus}.
\end{itemize}

Before fitting the evolution of $\omega_{\text{r,int}}(u_{\text{t}})$ using the expression (\ref{eq:freq_freq_fit}), one has to remove the contribution of the electromechanical nonlinear coupling $\omega_{\text{r,elec}}(u_{\text{t}})$ from $\omega_{\text{r}}(u_{\text{t}})$. It holds

\begin{equation}
    \omega_{\text{r}}(u_{\text{t}})\,=\,\omega_{\text{r,int}}(u_{\text{t}})+\omega_{\text{r,elec}}(u_{\text{t}}).
\end{equation}
The electromechanical contribution $\omega_{\text{r,elec}}(u_{\text{t}})$ arises from the strong electron-phonon interactions in the carbon nanotube \cite{Steele09, lassagne2009coupling} and results in large variations of $\omega_{\text{r}}$ in the vicinity of Coulomb peaks. A measurement of the dependence of $\omega_{\text{r}}$ as a function of the gate voltage $V_\text{g}$ in the vicinity of a Coulomb peak allows us to infer this contribution.
After estimating and removing the electromechanical contribution from the $f_\text{t}  - u_\text{t}$ two-tone spectroscopy maps, one can then fit the evolution of $\omega_{\text{r,int}}(u_{\text{t}})$ using the expression (\ref{eq:freq_freq_fit}) and extract the intrinsic Duffing nonlinearity of the  radial  mode $\alpha_{3\text{,r}}$. The details about this fit and the removal of the electromechanical contribution are discussed in section \ref{SI-alpha3t-extractionalpha3t}.
\\
Figure \ref{fig:summary_procedure} provides an overview of the main steps of the procedure. 

\subsection{Structure of a dataset}
\label{SI-alpha3t-structuredataset}

\begin{figure}[H]
  \centering
  \includegraphics[width=1\linewidth]{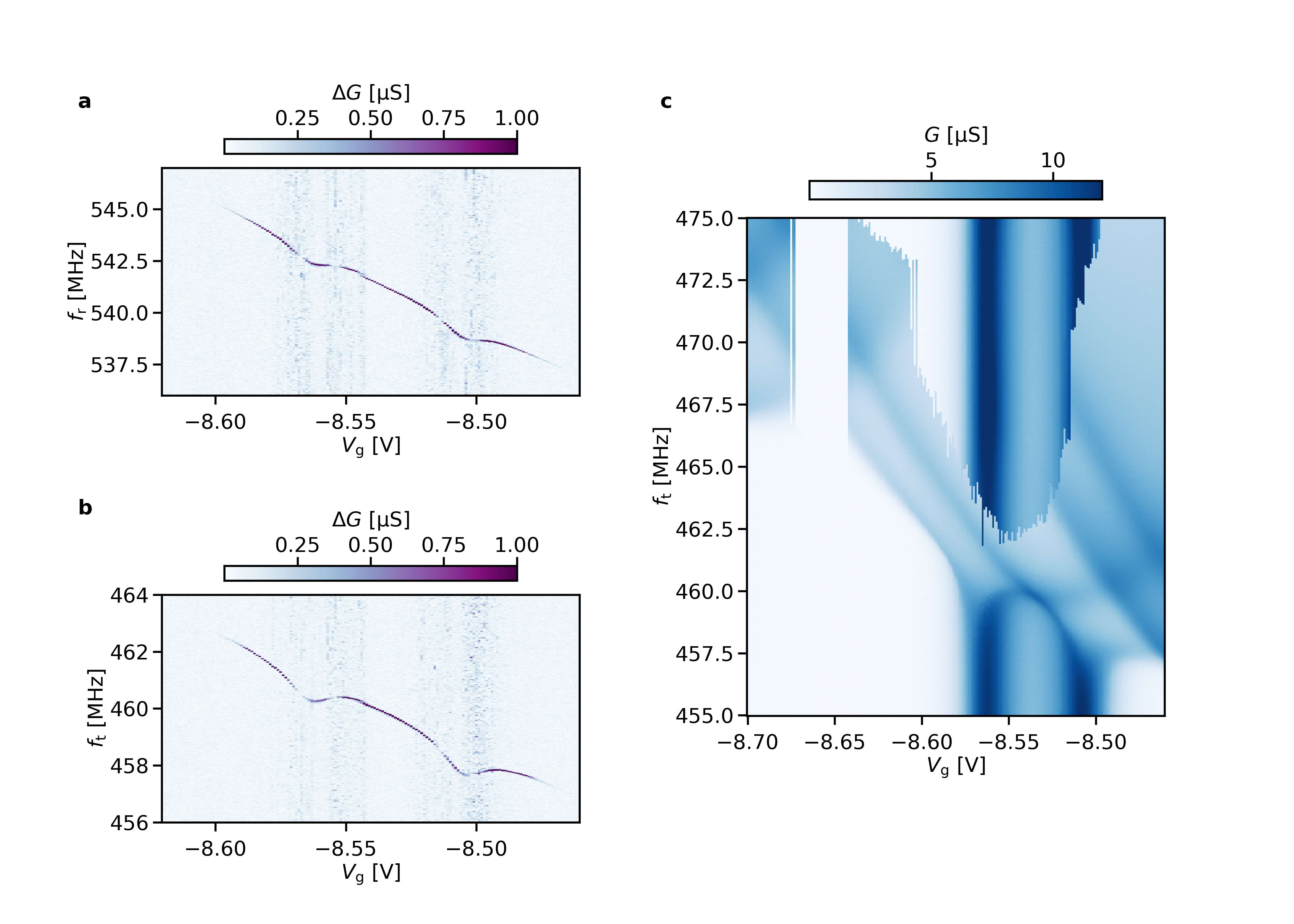}
  \caption{\textbf{Frequency–gate voltage maps from the dataset with $V_\text{g}\approx-8.5\,\text{V}$.} \textbf{a} Gate-dependence of the resonance frequency of the  radial  mode $\omega_\text{r,0}(V_\text{g}^\text{DC})$ with $P_\text{r}\,=\,-65\,\text{dBm}$. \textbf{b} Gate-dependence of the resonance frequency of the  tangential  mode $\omega_\text{t,0}(V_\text{g}^\text{DC})$ with $P_\text{t}\,=\,-75\,\text{dBm}$. \textbf{c} Frequency-gate voltage map of the  tangential  mode with $P_\text{t}\,=\,-37\,\text{dBm}$ used for the $f_\text{t}\rightarrow \Delta V_\text{g}$-conversion.}
  \label{fig:freq-vg-dataset}
\end{figure}

\begin{figure}[H]
  \centering
  \includegraphics[width=1\linewidth]{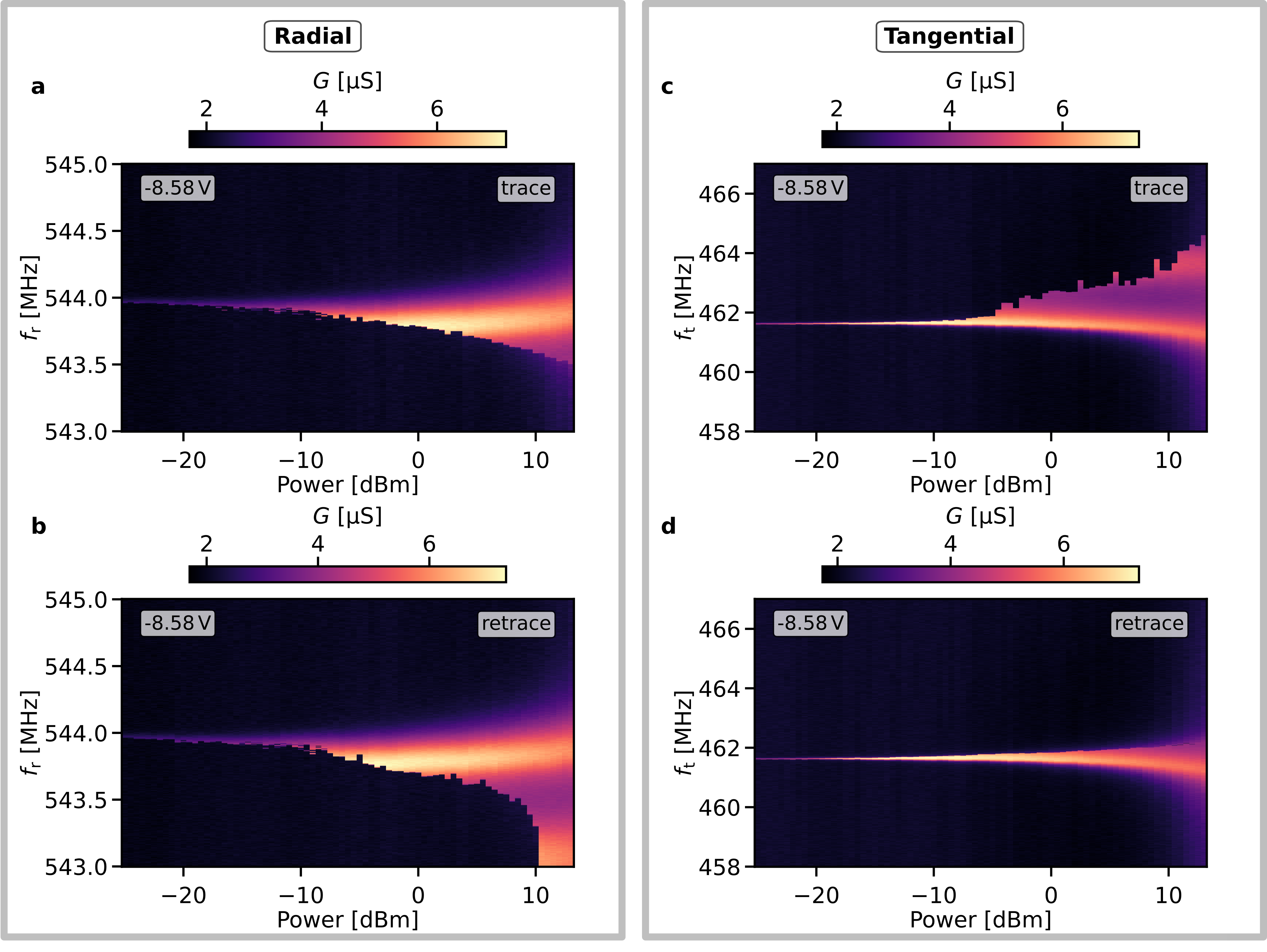}
  \caption{\textbf{Frequency–power maps from the dataset with $V_\text{g}\approx-8.5\,\text{V}$.} \textbf{a, b} Trace and retrace of the frequency-power map of the  radial  mode. \textbf{c, d} Trace and retrace of the frequency-power map of the  tangential  mode.}
  \label{fig:freq-power-dataset}
\end{figure}

\begin{figure}[H]
  \centering
  \includegraphics[width=0.95\linewidth]{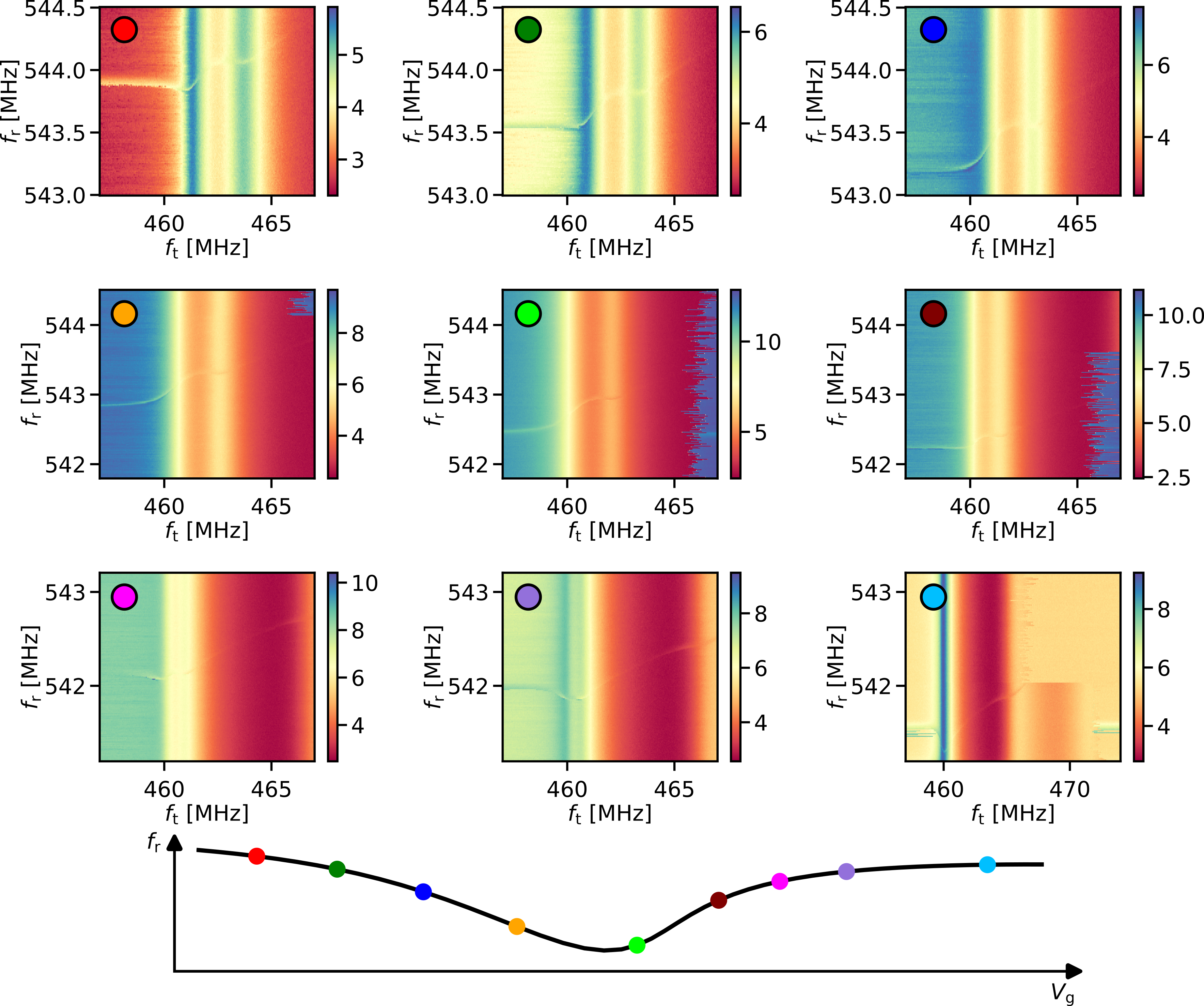}
  \caption{\textbf{Frequency–frequency maps from the dataset with $V_\text{g}\approx-8.5\,\text{V}$.} Two drive frequencies are applied around the resonance frequencies of the  tangential  and the  radial  mode and the differential conductance $G$ is measured in \unit{\micro \siemens} (colorbar). At the bottom, the position of the gate voltage of each frequency-frequency map is sketched on the resonance frequency curve of the  radial  mode.}
  \label{fig:freq-freq-dataset}
\end{figure}

The procedure introduced in the previous section is repeated for each data point shown in Fig.\ref{fig4}.a of the main-text. We define a dataset as the ensemble of measurements taken around a single Coulomb peak required for the application of the procedure (ie: the extraction of the intrinsic cubic nonlinearity $\alpha_\text{3,r}$). Such a dataset is composed of:

\begin{itemize}
    \item a frequency-gate voltage map of the  radial  mode at low driving power, allowing for the extraction of $\omega_\text{r,0}(V_\text{g})$ (see Fig.\ref{fig:freq-vg-dataset}a for an example).
    \item a frequency-gate voltage map of the  tangential  mode at low driving power, allowing for the extraction of $\omega_\text{t,0}(V_\text{g})$ (see Fig.\ref{fig:freq-vg-dataset}b for an example).
    \item a frequency-gate voltage map of the  tangential  mode at high driving power, used for the conversion $f_\text{t}\rightarrow \Delta V_\text{g}$ (see Fig.\ref{fig:freq-vg-dataset}c for an example).
    \item a frequency-power map of the  radial  mode, used to determine the low driving power to be used in the two-tone spectroscopy as well as in the previous frequency gate map (see Fig.\ref{fig:freq-power-dataset}a and b for an example).
    \item a frequency-power map of the  tangential  mode, used to determine the low driving power to be used in frequency-gate voltage map, as well as the high power to be used in the two-tone spectroscopy (see Fig.\ref{fig:freq-power-dataset}c and d for an example).
    
    \item 8 to 16 two-tone spectroscopy maps with the  tangential  mode strongly driven and the  radial  mode weakly driven (see Fig.\ref{fig:freq-freq-dataset} for an example). Each of the two-tone spectroscopy is taken at a different gate voltage value in the vicinity of the Coulomb peak of interest. Because the intrinsic nonlinearity is not expected to vary over such a narrow gate voltage range ($20-30\,$mV), the repetition of the full extraction procedure on each of these two-tone maps provides a sanity check of the method. In practice, small variations of $\alpha_{3,\text{r}}$ are observed (more details in Supplementary Note \ref{SI-alpha3t-bias}).
\end{itemize}

\subsection{Conversion: $f_\text{t} \rightarrow\Delta V_\text{g}$}
\label{SI-alpha3t-fsDeltaVg}

\begin{figure}[H]
  \centering
  \includegraphics[width=0.55\linewidth]{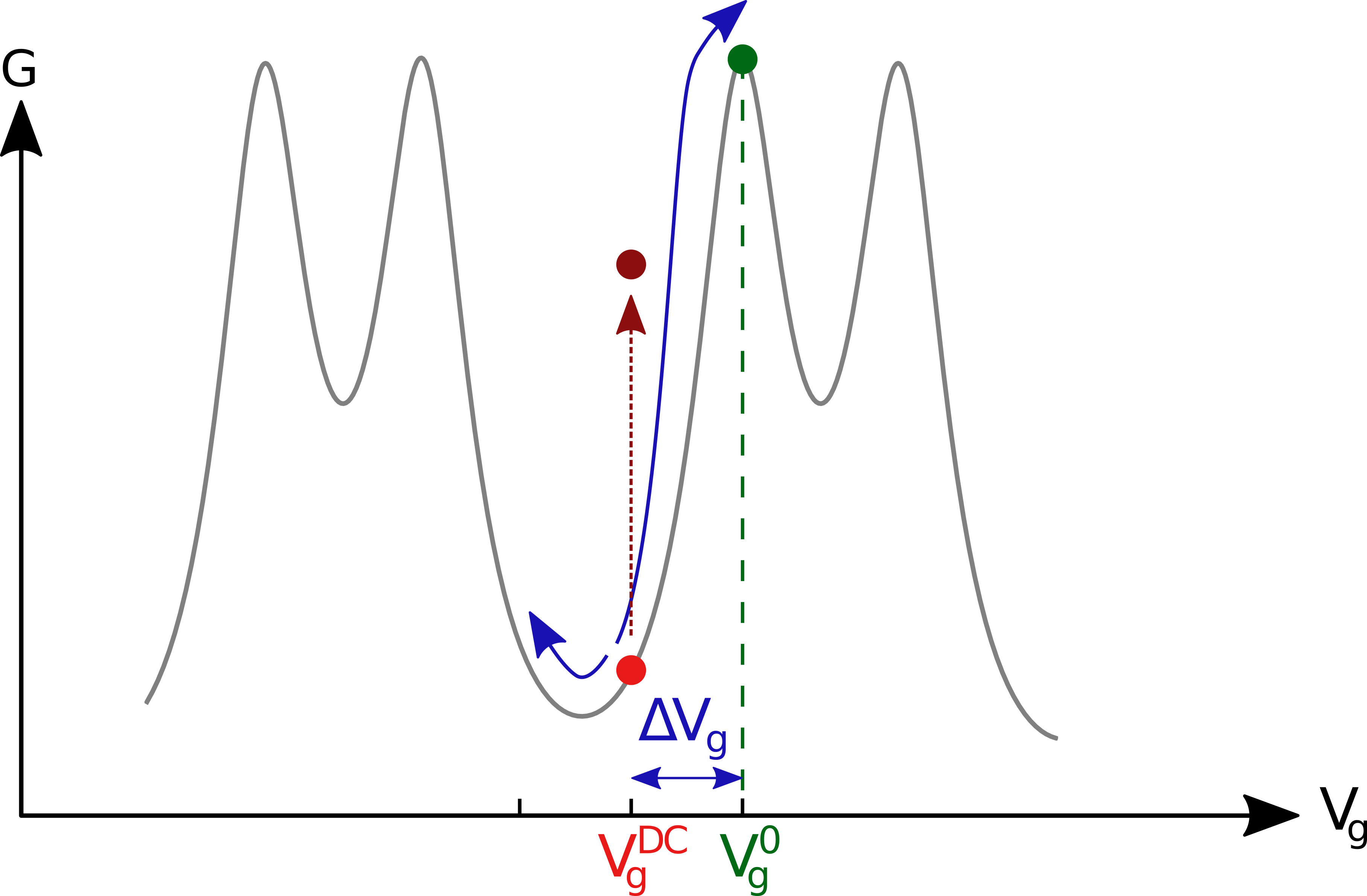}
  \caption{\textbf{Schematic representation of the shift in the Coulomb peak position in $V_\text{g}$ under actuation.} For an oscillation amplitude which corresponds to an effective gate voltage change of $\Delta V_\text{g}\,=\,|V_\text{g}^\text{DC}-V_\text{g}^0|$ (blue arrows), the top of the Coulomb peak is found at $V_\text{g}^\text{DC}$ (red dot) instead of $V_\text{g}^0$ (green dot). The differential conductance measured at $V_\text{g}^\text{DC}$ increases (dark red) in this case.}
  \label{fig:coulomb_peak_transport}
\end{figure}

In figure \ref{fig:summary_procedure}a, the differential conductance is measured as a function of gate voltage and drive frequency around the  tangential  mode resonance, using a strong RF-drive. For small drive frequencies, vertical blue lines can be seen, which represent the Coulomb peaks we expect when sweeping the gate voltage. \\
For intermediate drive frequencies, a more complex pattern is observed, due to a growth in the motion amplitude of the  tangential  mode $u_\mathrm{t}$ as the drive frequency $f_\mathrm{t}$ increases. The Coulomb peak splits from its usual gate voltage value $V_\text{g}^0$ into an arch-like shape with increasing drive frequency. This pattern overlaps with the neighboring peaks at high oscillation amplitudes. \\
The splitting can be explained by an effective shift in gate voltage felt by the carbon nanotube $\Delta V_\text{g}$, which increases with the oscillation amplitude, as schematized in figure \ref{fig:coulomb_peak_transport}. The Coulomb peak is shifted by $V_\text{g}^\text{DC}\,=\,V_\text{g}^0 \pm \Delta V_\text{g}$.

\begin{figure}[H]
  \centering
  \includegraphics[width=1\linewidth]{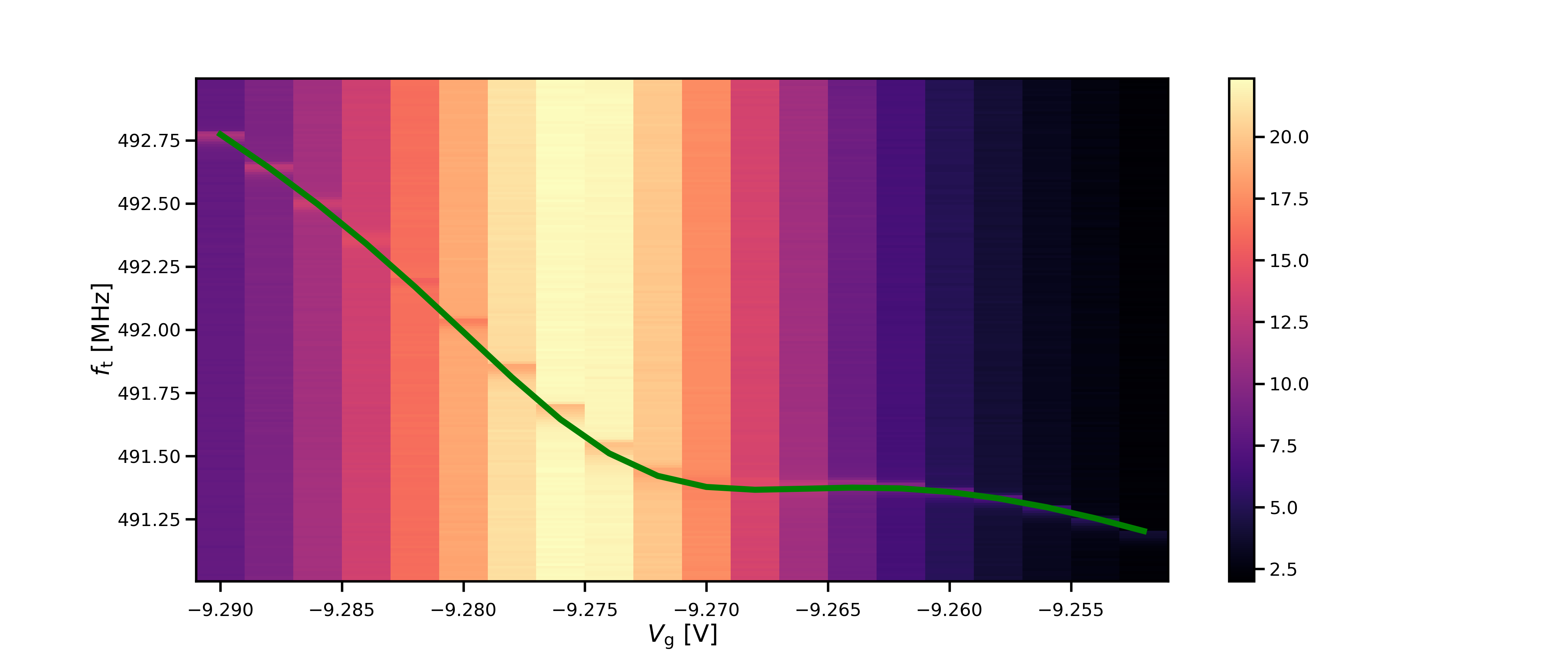}
  \caption{\textbf{Fit of the resonance frequency of the  tangential  mode at low RF-driving power.} The gate voltage is swept around the Coulomb peak at $V_\text{g}^0\,=\,-9.275\,\text{V}$. The differential conductance $G$ is measured in \unit{\micro\siemens} (colorbar). }
  \label{fig:ss_fit}
\end{figure}

As shown in figure \ref{fig:summary_procedure}a, the shift in gate voltage of the Coulomb peak $\Delta V_\text{g}$ depends on the drive frequency $f_\text{t}$. Therefore, the splitting of a Coulomb peak at strong drive enables us to extract the amplitude–frequency response curve of the  tangential  mode by exploiting the relation between $\Delta V_\text{g}$ and $u_\mathrm{t}$ derived in the next section. The upper branch of the frequency response of the nonlinear Duffing oscillator \cite{Nay08} is used to fit the shift of an individual Coulomb peak:
\begin{equation}
    f_\text{t}-\frac{\omega_{\text{t},0}(V_\text{g}^0)}{2\pi} \,=\,A \cdot u_\text{t}(|\Delta V_\text{g}|)^2-\frac{\mu}{2\pi}\sqrt{\frac{u_{\text{pt}}^2}{u_\text{t}(|\Delta V_\text{g}|)^2}-1}.
    \label{eq:amp-f-curve-for-fit}
\end{equation}
It is important to note that, while this expression is used to fit the $f_\text{t}(V_\text{g})$ dependence, other polynomial forms could equally well have been employed. Equation \ref{eq:amp-f-curve-for-fit} was selected for convenience, as it provides a good fit to the data, but no assumptions about the underlying nonlinearity of the system are required for the fitting procedure.  The correspondence between $f_\mathrm{t}$ and $\Delta V_\mathrm{g}$ in equation \ref{eq:amp-f-curve-for-fit} is the frequency response for the fixed DC-working point on the Coulomb peak at $V_\text{g}^0$. \\ \\
The frequency response for other gate voltages changes due to the dependence between $\omega_\text{t,0}$ and $V_\text{g}$, as shown in figure \ref{fig:ss_fit}. The complicated dependence arises from two different contributions. While the global slope arises from the increase in tension in the carbon nanotube with $|V_\mathrm{g}|$, the local dip in the resonance frequency at the position of each Coulomb peak is due to the softening of the spring constant \cite{Steele09, lassagne2009coupling}. When the gate voltage is swept across a Coulomb peak, single electrons can tunnel through the quantum dot. The total charge on the dot is consequently able to temporarily screen the potential induced by the gate electrode. This reduces the overall tension in the nanotube, which produces the local dips observed in the resonance-frequency dependence.\\ \\
Therefore, for a working point at a gate voltage $V_\text{g}^\text{DC}$, $\omega_{\text{t},0}(V_\text{g}^0)$ should be replaced by $\omega_{\text{t},0}(V_\text{g}^\text{DC})$ in equation \ref{eq:amp-f-curve-for-fit}. \\
However, for a stronger RF-drive, the value of $\omega_{\text{t},0}(V_\text{g}^\text{DC})$ is unknown, since the splitting of the Coulomb peak appears and smears the dependence $\omega_{\text{t},0}(V_\text{g}^\text{DC})$. As a reasonable approximation, the $\omega_{\text{t},0}(V_\text{g}^\text{DC})$-dependence obtained from a low-power measurement (as shown in figure \ref{fig:ss_fit}) is taken instead. How this approximation affects the extraction of $\alpha_\mathrm{3,r}$ is described in Section \ref{SI-alpha3t-bias}.

\subsubsection*{Removal of the spring softening effect}
\begin{figure}[H]
  \centering
  \includegraphics[width=0.65\linewidth]{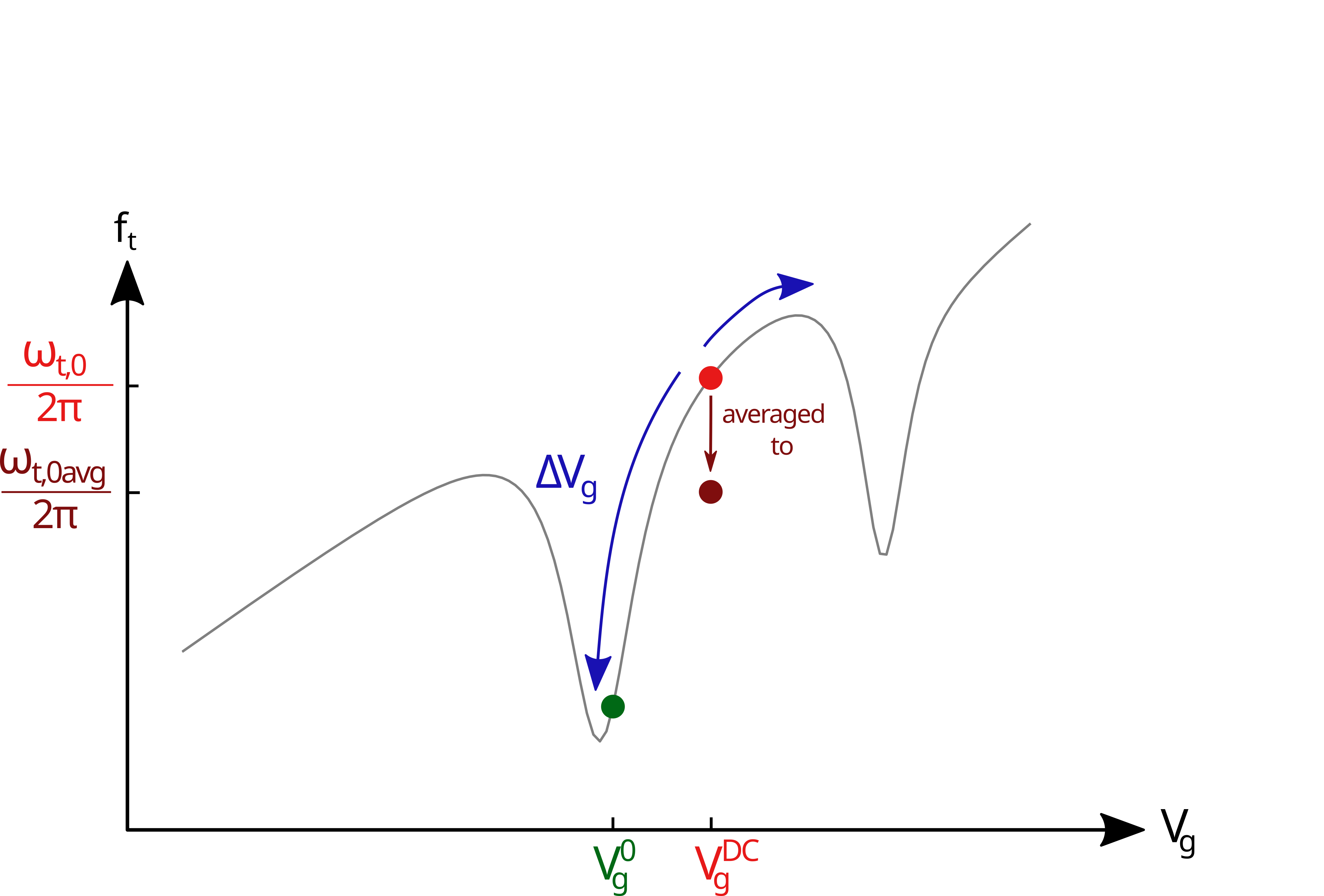}
  \caption{\textbf{Schematic representation of the averaging of the mechanical resonance frequency of the  tangential  mode.} For an effective change in gate voltage $\Delta V_\text{g} \,=\, |V_\text{g}^\text{DC}-V_\text{g}^0|$, the resonance frequency changes from $\omega_\mathrm{t,0}$ to $\omega_\mathrm{t,0avg}$.}
  \label{fig:ss_swing}
\end{figure}

The equation \ref{eq:amp-f-curve-for-fit} to fit the splitting of a Coulomb peak depends only on intrinsic nonlinearities. Therefore, the spring softening contribution of the  tangential -mode needs to be removed from the $f_\mathrm{t}\text{-}V_\mathrm{g}$ map before the fitting procedure. \\
Figure \ref{fig:summary_procedure}a is shifted to compensate for the change in mechanical resonance caused by the gate-induced tension and the spring softening effect. These effects can be removed by shifting the entire frequency-gate voltage map with $\Delta f_\text{t} = 1/2\pi \cdot (\omega_\text{t,0}(V_\text{g}^0) - \omega_\text{t,0avg}(V_\text{g}^\text{DC}, \Delta V_\text{g}))$. The value of $\omega_\text{t,0avg}(V_\text{g}^\text{DC}, \Delta V_\text{g})$ (see figure \ref{fig:ss_swing}) is computed with the sine-weighted average of the gate-dependent mechanical resonance frequency of the  tangential  mode at low power (usually around $-65\,\text{dBm}$) around a gate voltage $V_\text{g}^\text{DC}$ with amplitude $\Delta V_\text{g} \,=\, |V_\text{g}^\text{DC}-V_\text{g}^0|$.

\subsection{Conversion: $\Delta V_\mathrm{g}\rightarrow u_\mathrm{t}$}
\label{SI-alpha3t-DeltaVgus}
\subsubsection{Geometric modelization and characterization of the device}
\label{SI-alpha3t-DeltaVgus-geometriccharac}
\begin{figure}[H]
  \centering
  \includegraphics[width=0.95\linewidth]{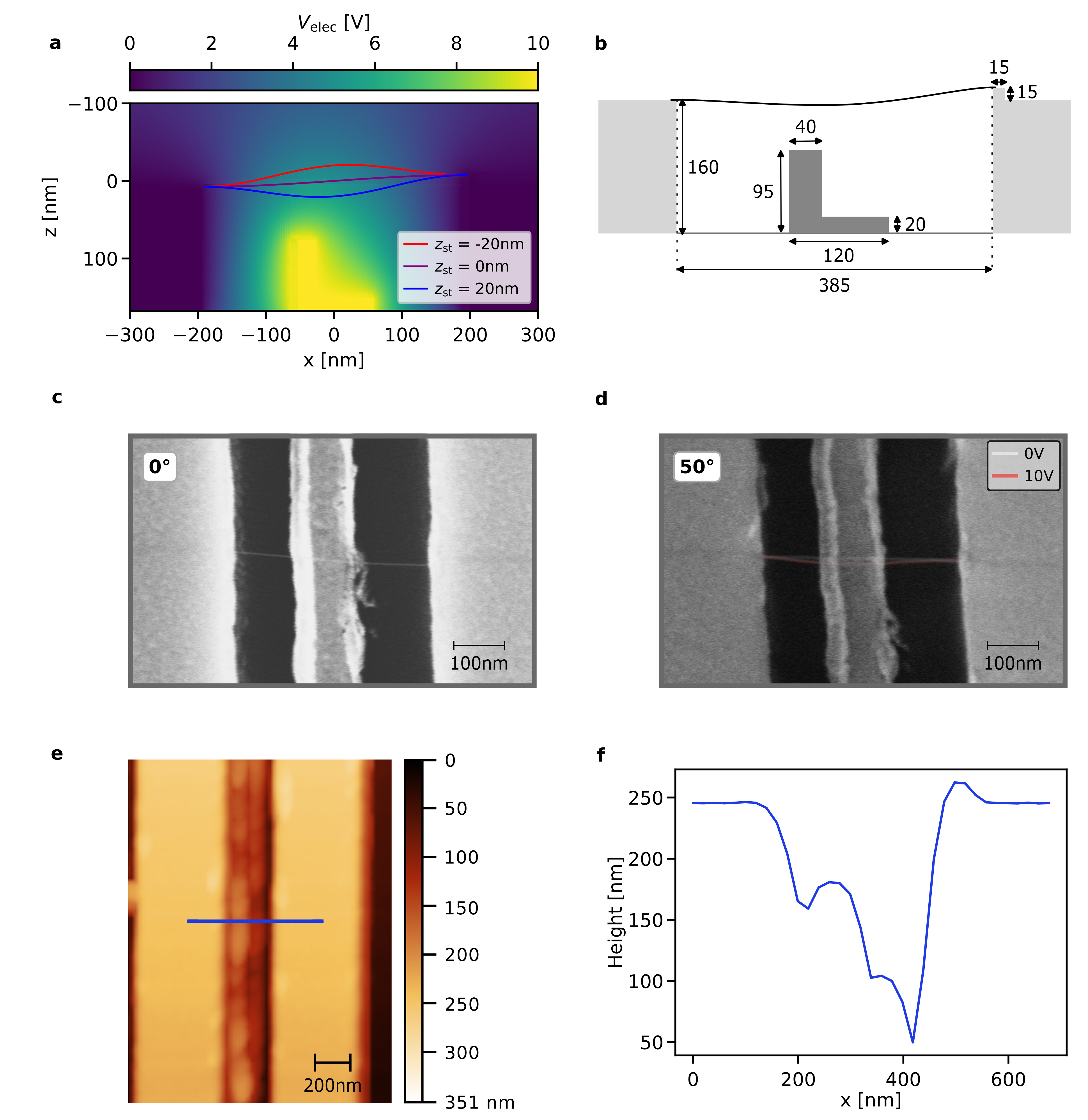}
  \caption{\textbf{Geometric modelization and characterization of the device.} \textbf{a} Simulation showing the electrostatic potential around the carbon nanotube with an applied gate voltage of $V_\text{g}\,=\,10\,\text{V}$ for three different values of static displacement $z_\mathrm{st}$. \textbf{b} Schematic of the device with its real dimensions (in nm) extracted from the AFM and SEM characterization. \textbf{c} Top‐view SEM image with no gate voltage applied. \textbf{d} Static displacement of the carbon nanotube of the same device under an applied gate voltage taken under a 50$^{\circ}$ angle with respect to the normal. Two images are superposed: one at $V_\text{g}\,=\,0\,\text{V}$ and one at $V_\text{g}\,=\,10\,\text{V}$. The carbon nanotube with no gate voltage applied is represented by a thin white line. It moves closer to the gate electrode when a voltage is applied (thin red line). \textbf{e} AFM-image of the surface profile of the contact and gate electrodes above which the carbon nanotube is suspended. The position of the carbon nanotube is represented by a blue line. \textbf{f} Height profile across the electrodes extracted from the blue line in e}
  \label{fig:AFM-SEM}
\end{figure}

To be able to convert the gate voltage into oscillation amplitude of the  tangential  mode, the relation between the change in potential of the quantum dot $\Delta V_\text{dot}$ and the static displacement of the carbon nanotube $z_\mathrm{st}$ needs to be extracted. For this purpose, the electrostatic potential arising from the gate electrode is simulated. The geometry of the system is characterized beforehand using SEM for lateral dimensions and AFM for vertical dimensions (see figure \ref{fig:AFM-SEM}).

The geometric characteristics of the carbon nanotube used for the calculation of $\alpha_\text{3,r}$ are presented in table \ref{table:fixparam}.

\subsubsection{$\Delta V_\text{g}$-$u_\text{t}$ conversion}
\label{SI-alpha3t-DeltaVgus-DeltaVgus}
The oscillation amplitude $u_{\text{t,}z}$ in $z$-direction leads to an effective gate voltage change $\Delta V_\mathrm{g}$ sensed by the carbon nanotube dependent on the gate voltage working point $V_\text{g}^\text{DC}$. The relation between $u_{\text{t,}z}$ and $\Delta V_\mathrm{g}$ is given by
\begin{equation}
    u_{\text{t,}z} \,=\, \left|\frac{1}{E(z_\mathrm{st}(V_\text{g}^\text{DC}),\,V_\text{g}^\text{DC})}\cdot \frac{ V_\text{dot, 10}(z_\mathrm{st}(V_\text{g}^\text{DC}))}{10} \cdot \Delta V_\text{g}\right|.
    \label{eq:us_to_DeltaVg}
\end{equation}
In the following section, the steps needed to derive equation \ref{eq:us_to_DeltaVg} are listed. For simplicity and for clearer visualization in Figure \ref{fig:summary_procedure}a, the explanation focuses on the case of the Coulomb-peak shift.
\\ \\
As described in section \ref{SI-alpha3t-fsDeltaVg}, the Coulomb peak is shifted by $\Delta V_\text{g}\,=\,|V_\text{g}^\text{DC}-V_\text{g}^0|$, depending on the drive frequency. \\
For a carbon nanotube with an oscillation amplitude in $z$-direction $u_{\text{t,}z}$, the quantum dot potential at the top of the Coulomb peak $V_\text{dot}$ must be identical to that of an undriven carbon nanotube, hence
\begin{equation}
    V_\text{dot}(z_\mathrm{st}(V_\text{g}^0)) \overset{!}{=} V_\text{dot}(z_\mathrm{st}(V_\text{g}^\text{DC})) \pm u_{\text{t,}z} \cdot E(z_\mathrm{st}(V_\text{g}^\text{DC}), V_\text{g}^\text{DC}).
    \label{eq:vdotequal}
\end{equation}
The components of equation \ref{eq:vdotequal} are derived below.

\begin{figure}[H]
  \centering
  \includegraphics[width=0.65\linewidth]{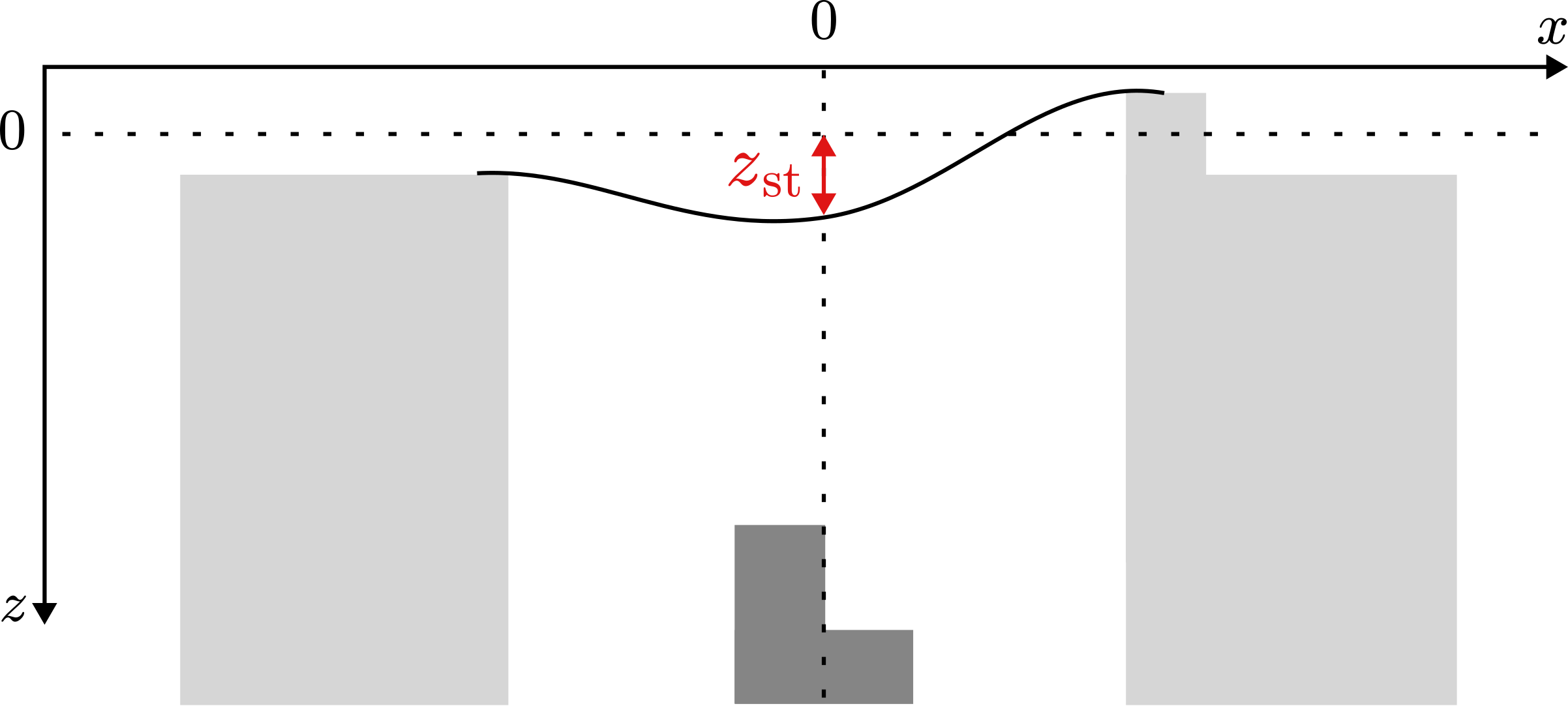}
  \caption{Representation of the suspended carbon nanotube and definition of $z_\text{st}$.}
  \label{fig:zdc0}
\end{figure}

\begin{itemize}
    \item \textbf{Initialization:} The relation between the applied gate voltage and the static displacement of the carbon nanotube $z_\mathrm{st}$ (see figure \ref{fig:zdc0}) is extracted from finite-element simulations using COMSOL \cite{comsol}, as well as the value of the electrostatic potential arising from an applied gate voltage of $V_\text{g}^\text{DC}\,=\,10\,\text{V}$, $V_{10}(x, \,z)$ (see figures \ref{fig:summary_procedure}b and \ref{fig:summary_procedure}c).
    
    \item \textbf{Calculation of the mean electrostatic potential at the quantum dot:} The mean electrostatic potential at the quantum dot along the $x$-direction for a given static displacement of the carbon nanotube $z_\mathrm{st}$ and an applied gate voltage of $V_\text{g}^\text{DC}\,=\,10\,\text{V}$ can be calculated with 
    \begin{equation}
        V_\text{dot, 10}(z_\mathrm{st})\,=\, \int_{x\,=\,-l/2}^{x\,=\,l/2}V_{10}(x,\,z_\text{st}(x))\,\mathrm{d}x,
    \end{equation}
    with $l$ the extent of the quantum dot inside the carbon nanotube. We assume hard confinement boundaries at the carbon nanotube-metal contact, thus $l\,=\,385\,\text{nm}$. $z_\text{st}(x)$ is the position of the carbon nanotube in the $z$-direction along $x$ with a static displacement of $z_\mathrm{st}$. The carbon nanotube operates in the bending limit (the applied tension is much smaller than the flexural rigidity) and is modeled by the shape     
    \begin{equation}
        z_\text{st}(x)\,=\,7.5 \cdot\cos\left(\frac{\pi}{d}x + \frac{3\pi}{2}\right)-\frac{1}{2}\cdot z_\mathrm{st}\cdot \left( 1+ \cos\left( \frac{2\pi x}{d}\right)\right),
    \end{equation}
    with $d\,=\,385\,\text{nm}$ the distance between the two contact electrodes and taking into account the different heights of the contact electrodes. \\
    $V_\text{dot, 10}(z_\mathrm{st})$ is calculated for each simulated $z_\mathrm{st}(V_\text{g}^\text{DC})$ in the gate voltage range $[-9.55; -2]\,\text{V}$ in $10\,\text{mV}$ steps.
    \item \textbf{Rescaling of the mean voltage of the quantum dot:} The mean voltage of the quantum dot $V_\text{dot, 10}(z_\mathrm{st})$, obtained for $V_\text{g}^\text{DC}\,=\,10\,\text{V}$, is then rescaled for an arbitrary gate voltage $V_\text{g}^\text{DC}$. We use
    \begin{equation}
        V_\text{dot}(z_\mathrm{st})\,=\,\frac{V_\text{dot, 10}(z_\mathrm{st})}{10} \cdot V_\text{g}^\text{DC},
        \label{eq:vdot_rescaling}
    \end{equation}
    assuming a linear dependence of the quantum dot potential on the applied gate voltage.
    \item \textbf{Calculation of the electric field:} The electric field in the $z$-direction is defined as
    \begin{equation}
        E(z_\mathrm{st},\,V_\text{g}^\text{DC})\,=\,\frac{\partial V_\text{dot}(z_\mathrm{st})}{\partial z_\mathrm{st}}.
    \end{equation}
    It is calculated for a fixed gate voltage value $V_\text{g}^\text{DC}$. The electric field with respect to the static displacement is approximated with a linear fit.

    \item \textbf{Calculation of the oscillation amplitude in $z$-direction:} The  tangential  mode amplitude in $z$-direction as shown in equation \ref{eq:us_to_DeltaVg} is obtained by combining equation \ref{eq:vdotequal} with equation \ref{eq:vdot_rescaling}. The electrostatic potential $V_\mathrm{dot,\,10}$ felt by the quantum dot for a fixed gate voltage $V_\mathrm{g}\,=\,10\,\text{V}$ can be assumed constant: $V_\text{dot, 10}(z_\mathrm{st}(V_\text{g}^\text{DC})) \approx V_\text{dot, 10}(z_\mathrm{st}(V_\text{g}^0))$ (according to the values of the electrostatic potential and static displacement from the finite-element simulation). 
    \item \textbf{Derivation of the oscillation amplitude:} The  tangential  mode amplitude is finally obtained by dividing the oscillation amplitude in $z$-direction with the cosine of the polarization angle $\theta$ derived with the Euler-Bernoulli model,
    \begin{equation}
        u_\text{t} \,=\, \frac{u_{\text{t,}z}}{\cos (\theta)}\, .
    \end{equation}
\end{itemize}

\subsection{Fitting of $\omega_{\text{r}}(u_{\text{t}})$}
\label{SI-alpha3t-extractionalpha3t}

To extract the value of the intrinsic Duffing nonlinearity of the  radial  mode $\alpha_\mathrm{3,r}$ for a particular gate voltage value, we convert a frequency–frequency map as the one on the left of figure \ref{fig:summary_procedure}d into the frequency-amplitude map displayed on the right (see section \ref{SI-alpha3t-fsDeltaVg}). The electromechanical coupling is removed and the shift in $\omega_\text{r}$ is fitted with equation \ref{eq:freq_freq_fit}. In the following, we detail the steps required to fit equation \ref{eq:freq_freq_fit}.

\begin{figure}[H]
  \centering
  \includegraphics[width=0.7\linewidth]{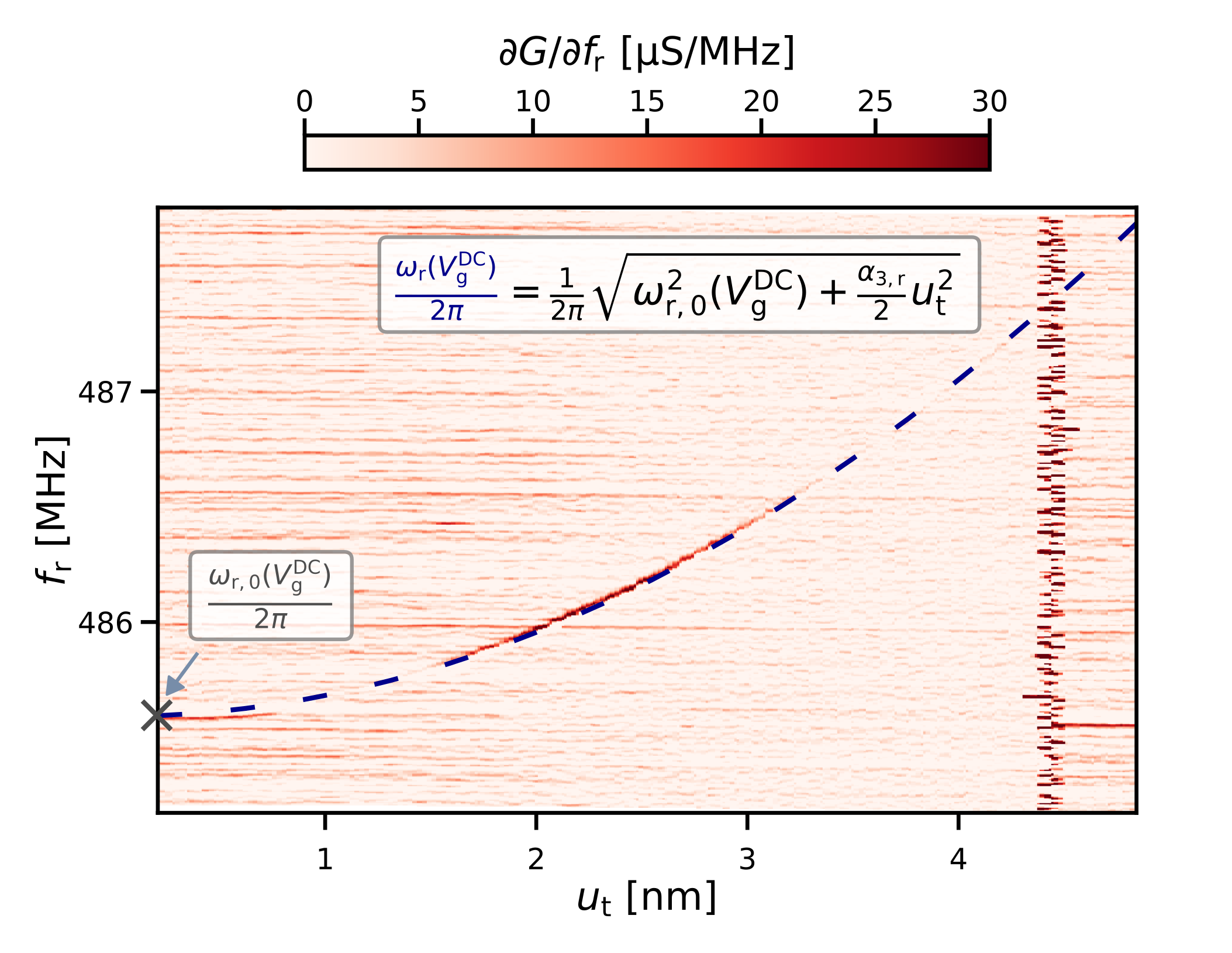}
  \caption{\textbf{Fit of $\omega_{\text{r}}/2\pi$ in a $f_\mathrm{r}$-$u_\mathrm{t}$ map.} The measurement is taken at $V_\text{g}^\mathrm{DC}\,=\,-2.454\,\text{V}$. The value $\omega_{\text{r},0}/2\pi$ is taken as the $f_\mathrm{r}$-intercept and is represented by the dark gray cross. $\omega_{\text{r}}/2\pi$ is fitted with Eq. \ref{eq:freq_freq_fit} (dark-blue dashed line), using $\alpha_{\mathrm{3,r}}$ as the only free fit parameter.}
  \label{fig:ft_us_fit}
\end{figure}

\subsubsection{Extraction of $\omega_{\text{r,0}}(V_\text{g}^\text{DC})$}
\label{SI-alpha3t-extractionalpha3t-omegat0}
\begin{figure}[H]
  \centering
  \includegraphics[width=1\linewidth]{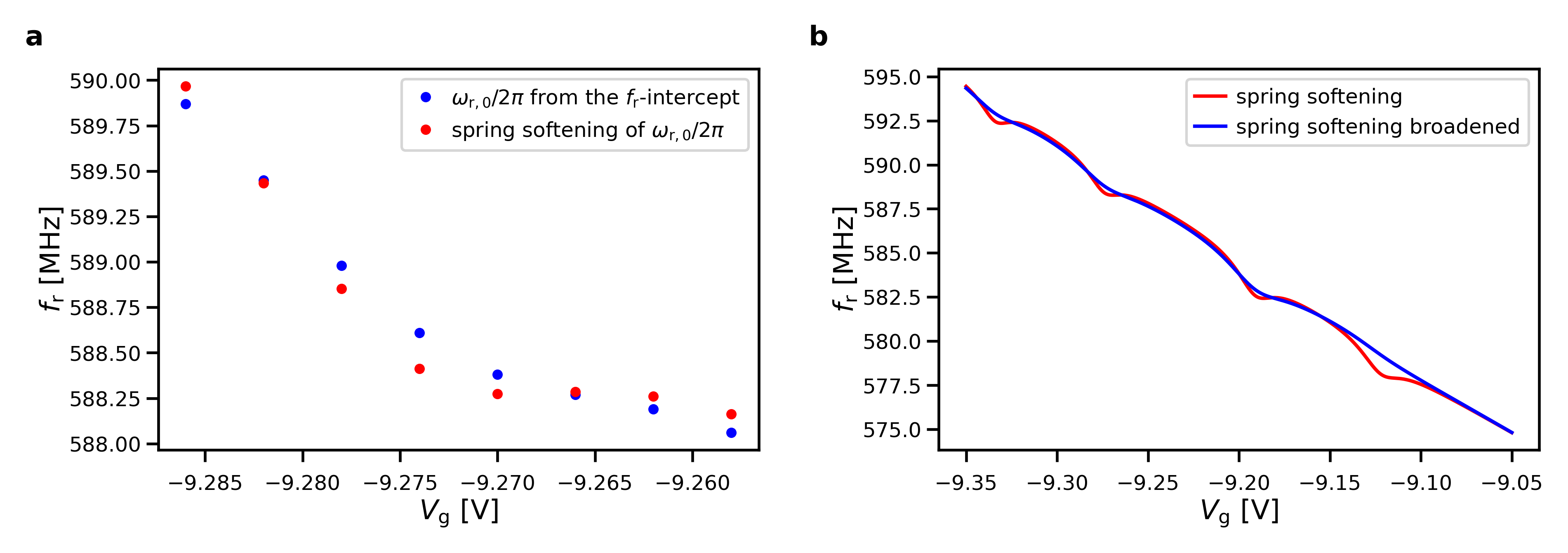}
  \caption{\textbf{Resonance frequency of the  radial  mode around the Coulomb peak at $V_\text{g}^0\,=\,-9.275\,\text{V}$.} \textbf{a} Comparison between the $\omega_{\text{r},0}/2\pi$-intercept and the resonance frequency of the  radial  mode taken at low power. \textbf{b} Comparison between the resonance frequency chosen for the removal of the electromechanical coupling of the  radial  mode and the resonance frequency of the  radial  mode taken at low power.}
  \label{fig:f0t_choice}
\end{figure}

$\omega_{\text{r},0}(V_\text{g}^\text{DC})$ is the $2\pi f_\text{r}$-intercept of $\omega_{\text{r}}$ and is extracted from a frequency-frequency map for a fixed $V_\text{g}^\text{DC}$, as shown in figure \ref{fig:ft_us_fit}. It is a fixed parameter in equation \ref{eq:freq_freq_fit}. \\
The extracted values of $\omega_{\text{r},0}(V_\text{g}^\text{DC})$ differ from the resonance frequency of the  radial  mode taken with one-tone spectroscopy at low RF-driving power, as displayed in figure \ref{fig:f0t_choice}a. The deviation can be explained by the additional second tone $f_\mathrm{t}$ applied at high power, leading to an effective broadening of the spring softening of the  radial  mode at low power.

\subsubsection{Determination of $\alpha_\mathrm{3,r}$}
\label{SI-alpha3t-extractionalpha3t-determinationalpha3t}
\begin{figure}[H]
  \centering
  \includegraphics[width=1\linewidth]{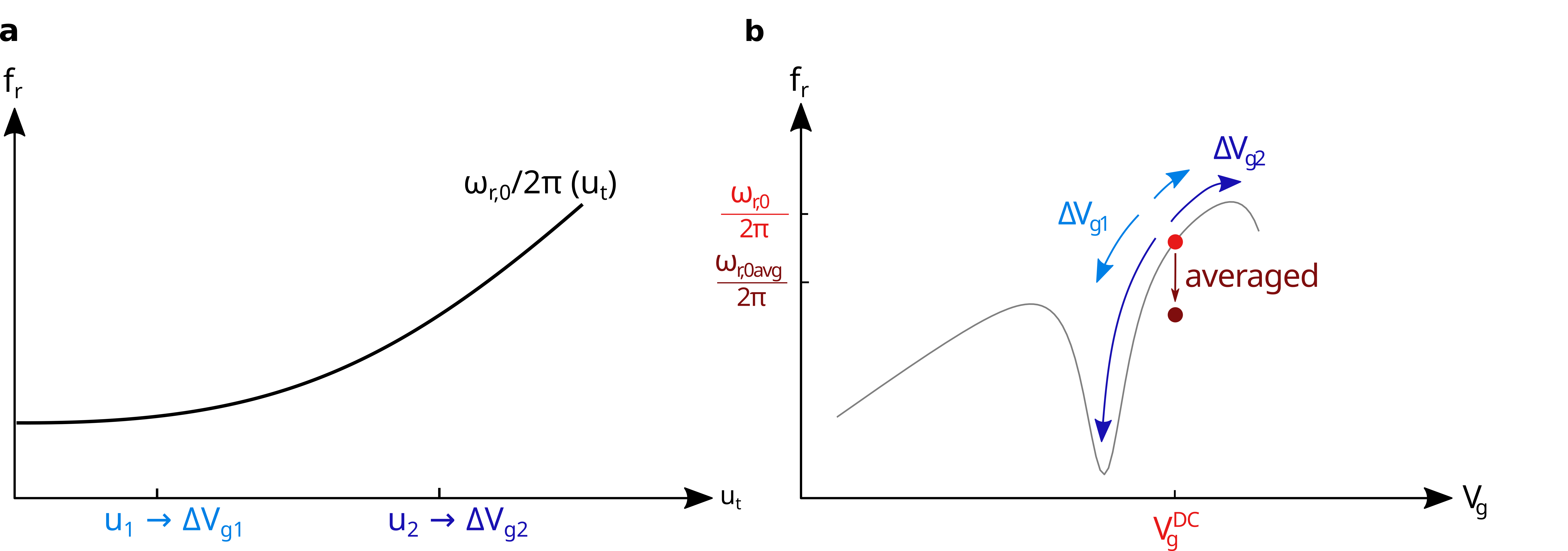}
  \caption{\textbf{Schematic representation of the averaging of the electromechanical spring softening for the  radial  mode.} \textbf{a} Typical amplitude-frequency curve. The oscillation amplitude of the  tangential  mode $u_\mathrm{1/2}$ (weaker amplitude in light and stronger in dark blue respectively) can be converted into the corresponding variation of the gate voltage $\Delta V_\mathrm{g1/2}$. \textbf{b} The assumed gate-dependent mechanical resonance at high power of the  radial  mode is averaged around $\omega_\text{r,0}(V_\text{g}^\text{DC})/2\pi$ with an amplitude $\Delta V_\text{g}$ corresponding to the oscillation amplitude of the  tangential  mode.}
  \label{fig:ss_radial}
\end{figure}

Since $\omega_\mathrm{r,0}$ is fixed as described in the section above, $\alpha_\mathrm{3,r}$ remains the only free parameter in equation \ref{eq:freq_freq_fit}. $\omega_\mathrm{r}(u_\mathrm{t})$ is fitted to determine $\alpha_\mathrm{3,r}$. \\ \\
\textbf{Removal of the electromechanical nonlinearity} \\ 
As detailed in \cite{Sev25}, two main sources of nonlinearity are found in suspended carbon nanotube resonators: electromechanical and intrinsic nonlinearities. Since we only want to fit the intrinsic nonlinearity $\alpha_\mathrm{3,r}$, the effect of the electromechanical coupling of the  radial  mode needs to be removed. In addition to the $\omega_\mathrm{r}$-$u_\mathrm{r}$ electromechanical coupling, which can be neglected due to the small oscillation amplitude $u_\mathrm{r}$, the resonance frequency of the  radial  mode is influenced by the effective gate voltage change arising from the motion amplitude of the  tangential  mode $u_\text{t}$ at a strong RF-drive ($\omega_\mathrm{r}$-$u_\mathrm{t}$ electromechanical coupling). The new value of $\omega_\text{r,0}$ is calculated as the average around $\omega_\text{r,0}(V_\text{g}^\text{DC})$ with an oscillation amplitude $\Delta V_\text{g}$, as shown in figure \ref{fig:ss_radial}. \\ \\
Therefore, the frequency axis of the  radial  mode must be shifted by $\Delta f_\text{r} = 1/2\pi \cdot (\omega_\text{r,0}(V_\text{g}^\text{DC}) - \omega_\text{r,0avg}(V_\text{g}^\text{DC}, u_\text{t}))$ to remove the effect of the electromechanical coupling $\omega_\mathrm{r}$-$u_\mathrm{t}$. As shown in figure \ref{fig:ss_radial}, the frequency $\omega_\text{r,0avg}(V_\text{g}^\text{DC}, u_\text{t})$ is obtained as the sine-weighted average of a gate-dependent mechanical resonance. As the oscillation amplitude of the  tangential  mode continuously increases along the $u_\mathrm{t}$-axis, the resonance dip of the  radial  mode becomes progressively broader. To simplify the computation, we assume a broadened version of the gate-dependent resonance of the  radial  mode at low power (see figure \ref{fig:f0t_choice}b.).

\subsection{Underestimation of the real value of $\alpha_{\mathrm{3,r}}$ due to the estimation of $\omega_{\mathrm{t,0}}(V_\mathrm{g}^\mathrm{DC})$}
\label{SI-alpha3t-bias}
\begin{figure}[H]
  \centering
  \includegraphics[width=0.66\linewidth]{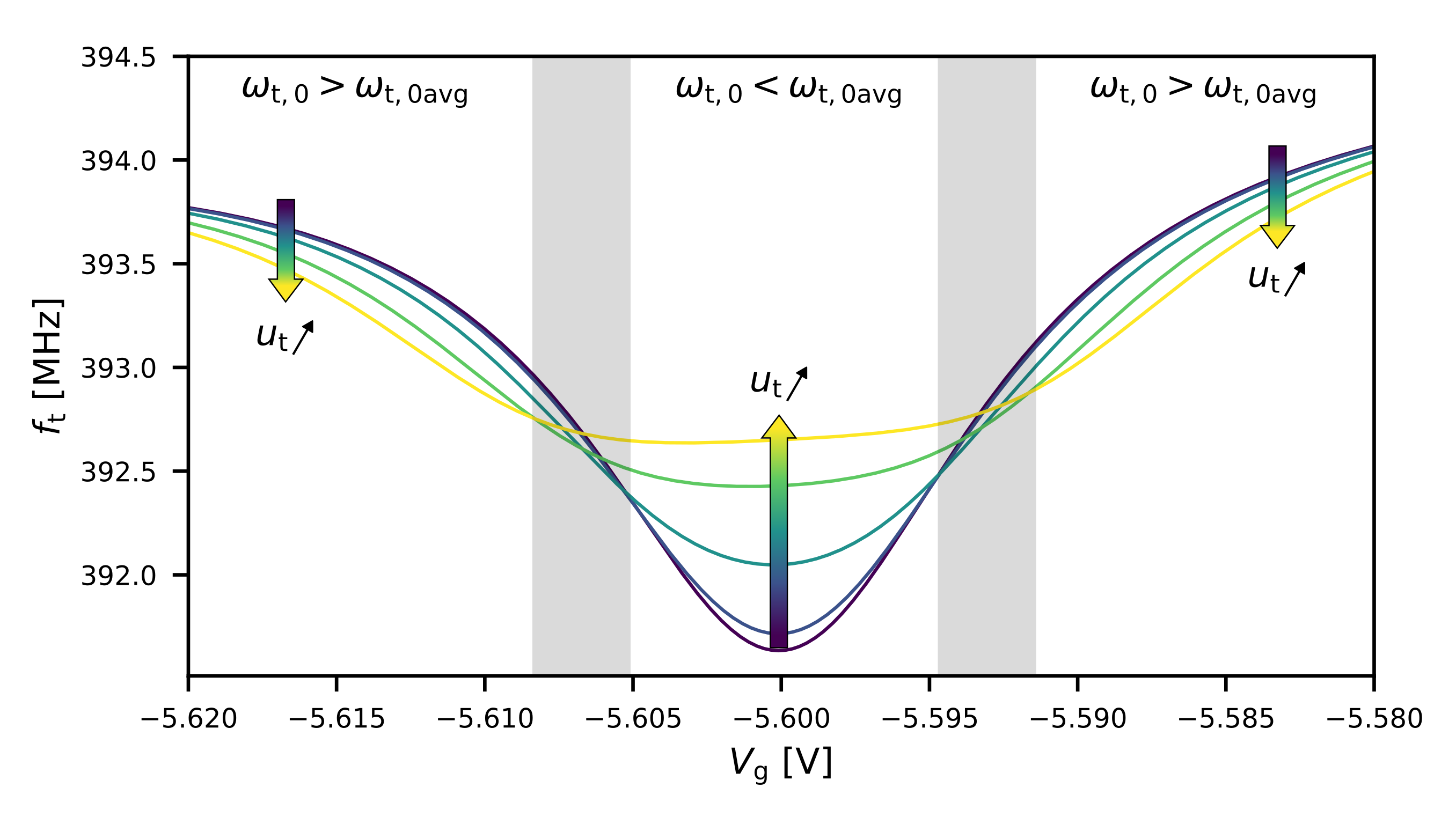}
  \caption{\textbf{Schematic representation of the broadening of the electromechanical spring softening for the  tangential  mode with increasing amplitude of motion $u_\mathrm{t}$.} The oscillation amplitude $u_\mathrm{t}$ increases from purple to yellow.}
  \label{fig:ss_broadening_with_increasing_us}
\end{figure}

This section describes the impact of the choice of $\omega_{\mathrm{t,0}}(V_\mathrm{g}^\mathrm{DC})$ made in Supplementary Section \ref{SI-alpha3t-fsDeltaVg} for the $f_\text{t}\rightarrow u_\text{t}$ conversion  and its consequences for the calculated value of $\alpha_{\mathrm{3,r}}$. The value $\omega_{\mathrm{t,0}}(V_\mathrm{g}^\mathrm{DC})$ needed in equation \ref{eq:amp-f-curve-for-fit}  cannot be accurately determined for high driving powers. For this reason, we use instead $\omega_{\mathrm{t,0}}$ extracted from a low-power $f_\mathrm{t}$-$V_\mathrm{g}$ measurement. As further detailed in this section, this results in an underestimation of the real value of $\alpha_{\mathrm{3,r}}$. 
\\ \\
As explained in Supplementary Section \ref{SI-alpha3t-fsDeltaVg}, the value of $\omega_{\mathrm{t,0}}(V_\mathrm{g}^\mathrm{DC})$ required for the $f_\text{t}\rightarrow u_\text{t}$ conversion depends on the operating point $V_\text{g}^\text{DC}$ of each $f_\text{t}$-$f_\text{r}$ measurement. At high driving power, $\omega_{\mathrm{t,0}}(V_\mathrm{g}^\mathrm{DC})$ cannot be directly measured because the resonance response bifurcates strongly, giving rise to additional effects such as Coulomb peak splitting, which can overlap the measured resonance signal. Nevertheless, it differs from the value of $\omega_{\mathrm{t,0}}(V_\mathrm{g}^\mathrm{DC})$ measured at low driving power. Figure \ref{fig:ss_broadening_with_increasing_us} shows how increasing the oscillation amplitude alters and broadens the frequency response around the spring-softening dip. The depth of the spring-softening dip decreases rapidly with increasing $u_{\mathrm{t}}$, causing the actual averaged resonance frequency $\omega_{\mathrm{t,0avg}}$ to shift upward relative to the value of $\omega_{\mathrm{t,0}}$ extracted from the low-power measurement. On the sides of the spring softening dip, the broadening has the opposite effect: in this case, the average resonance frequency $\omega_{\mathrm{t,0avg}}$ becomes lower than $\omega_{\mathrm{t,0}}$. The absolute difference between $\omega_{\mathrm{t,0avg}}$ and $\omega_{\mathrm{t,0}}$ is smaller on the sides than at the center of the dip as $u_{\mathrm{t}}$ increases.

\begin{figure}[H]
  \centering
  \includegraphics[width=1\linewidth]{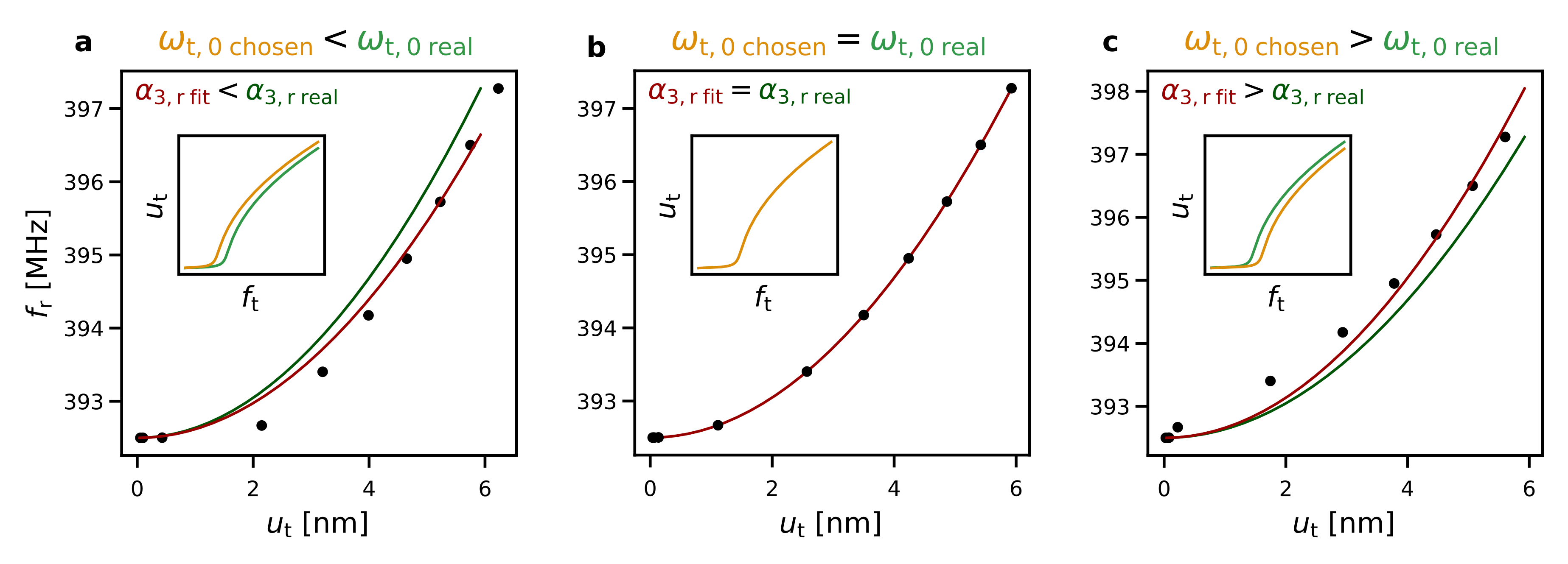}
  \caption{\textbf{Comparison of the extracted $\alpha_\mathrm{3,r}$-values as a function of the $\omega_\mathrm{t,0}$ used in the $f_\mathrm{t}$-$u_\mathrm{t}$ conversion.} The $u_\mathrm{t}$-$f_\mathrm{t}$ dependence for different $\omega_\mathrm{t,0}$ is shown in the inset. $\omega_\mathrm{r}$ (black points) is extracted from the $f_\mathrm{r}$-$f_\mathrm{t}$ measurements before the $f_\mathrm{t}$-$u_\mathrm{t}$ conversion. The fit of $\omega_\mathrm{r}$ (dark red line) is shown for different $\omega_\mathrm{t,0\,chosen}$ values used in the $f_\mathrm{t}$-$u_\mathrm{t}$ conversion (orange). For comparison, $\omega_\mathrm{r}$ (dark green line) using the real $\omega_\mathrm{t,0\,real}$ for the $f_\mathrm{t}$-$u_\mathrm{t}$ conversion (light green) is shown. \textbf{a} $\omega_\mathrm{t,0\,chosen}<\omega_\mathrm{t,0\,real}$, \textbf{b} $\omega_\mathrm{t,0\,chosen}=\omega_\mathrm{t,0\,real}$, \textbf{c} $\omega_\mathrm{t,0\,chosen}>\omega_\mathrm{t,0\,real}$.}
  \label{fig:f0s_choice}
\end{figure}

Not using the true value $\omega_{\mathrm{t,0real}}$ affected by the increased oscillation amplitude $u_{\mathrm{t}}$ alters the estimated value of $\alpha_{\mathrm{3,r}}$.
\\
As shown in Figure \ref{fig:f0s_choice}, the amplitude–frequency response is shifted left or right when $\omega_{\mathrm{t,0chosen}}$ is smaller or larger than $\omega_{\mathrm{t,0real}}$ (see insets). As a result, after the amplitude-frequency conversion, the data points describing $\omega_{\mathrm{r}}$ are shifted in a non-uniform manner. The points at lower amplitude are particularly affected. The whole $f_\mathrm{r}-u_\mathrm{t}$ map is extended or compressed at lower amplitude. \\
Consequently, the fit in equation \ref{eq:freq_freq_fit} provides only an approximated description of $\omega_\mathrm{r}$ (see figure~\ref{fig:f0s_choice}a and \ref{fig:f0s_choice}c). The extracted value of $\alpha_{\mathrm{3,r}}$ is therefore slightly smaller or larger than the true value (see figure~\ref{fig:f0s_choice}b), depending on the case. By considering multiple $V_{\mathrm{g}}^{\mathrm{DC}}$ working points around a single dip of the spring softening and averaging all extracted $\alpha_{\mathrm{3,r}}$ values, this effect is partially compensated for.

\begin{figure}[H]
  \centering
  \includegraphics[width=0.66\linewidth]{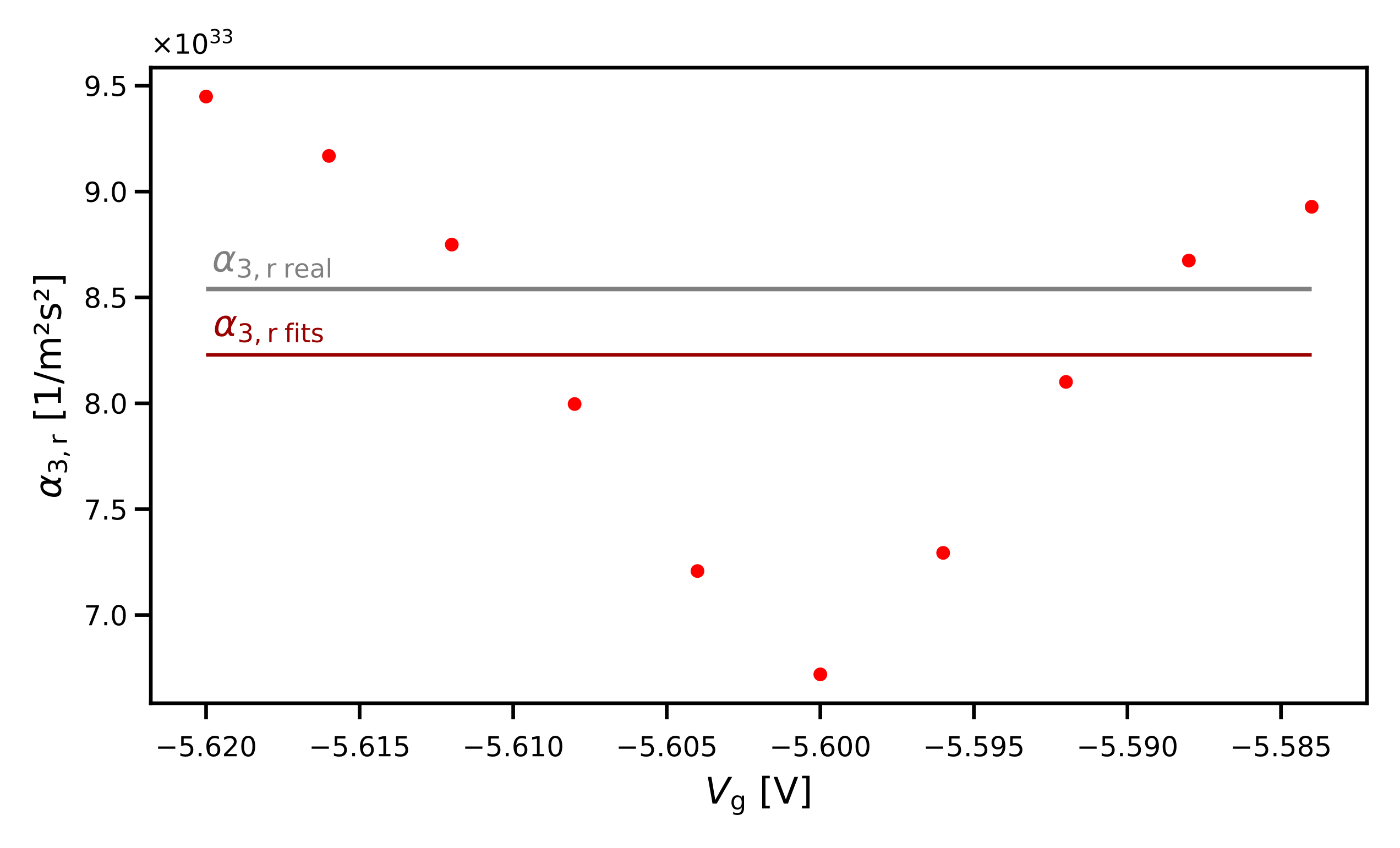}
  \caption{\textbf{Extracted values of the intrinsic cubic nonlinearity $\alpha_\mathrm{3,r}$ for different gate voltages in a dataset.} The mean value of the nonlinearity $\alpha_\mathrm{3,r\,fits}$ for this dataset extracted with our analysis method is represented by a red line. It underestimates the actual value of the nonlinearity $\alpha_\mathrm{3,r\,real}$ (schematized by a gray line).}
  \label{fig:alpha3t_56}
\end{figure}

However, the extracted value of the intrinsic cubic nonlinearity $\alpha_{\mathrm{3,r}}$ is still underestimated despite averaging over multiple gate voltage values. As shown in Figure \ref{fig:ss_broadening_with_increasing_us}, the 
misestimation of $\omega_{\mathrm{t,0}}$ is larger in the center of the spring-softening dip than on its sides (see in particular the comparison between the yellow and purple curves). Consequently, the $\alpha_{\mathrm{3,r}}$ values extracted from the center of the dip shift to much lower values than the flank values shift to higher ones, which ultimately shifts the overall mean of $\alpha_{\mathrm{3,r}}$ to lower values. Figure~\ref{fig:alpha3t_56} shows the values of $\alpha_{\mathrm{3,r}}$ extracted for the working points $V_{\mathrm{g}}^{\mathrm{DC}}$ in the dataset around $-5.6\,\mathrm{V}$. Following the above explanation, the true intrinsic Duffing nonlinearity, schematized by the gray line, is expected to lie above the values shown in Figure~\ref{fig4} of the main text.

\section{Effect of the electromechanical coupling at the maximum of the intrinsic nonlinearity ($V_\mathrm{g}=-6.2\,\mathrm{V}$)}
\label{SI-effectelectromechanicalcoupling}

\begin{figure}[H]
\centering
\includegraphics[width=1\linewidth]{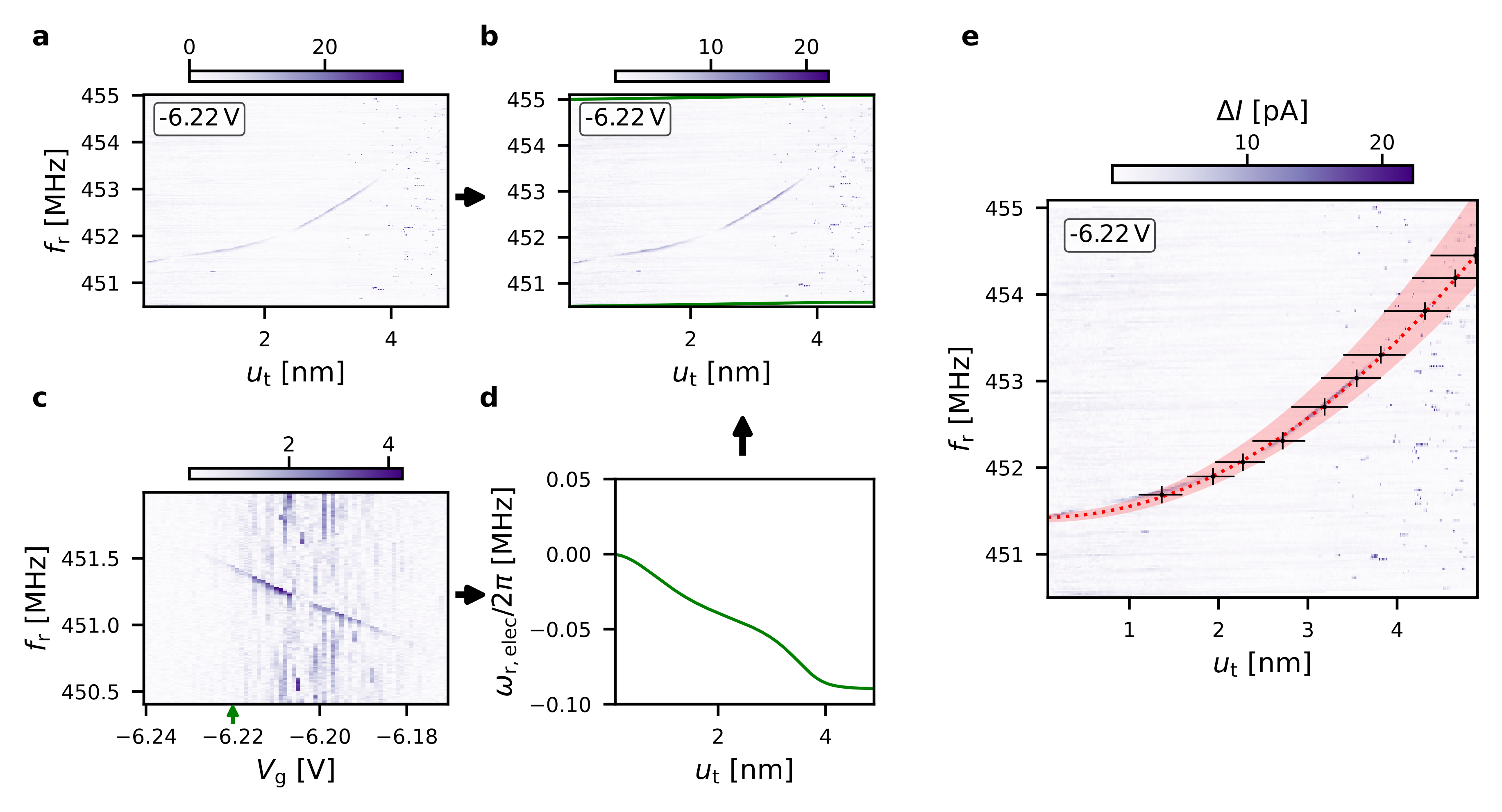}
\caption{
\textbf{Extraction of the intrinsic Duffing nonlinearity at $V_\text{g}\,=\,-6.22\,\text{V}$.}
\textbf{a} $f_\mathrm{r} - u_\mathrm{t}$ two-tone spectroscopy taken at $V_\mathrm{g}=-6.22\,$V ($P_\mathrm{r}=-55\,$dBm and $P_\mathrm{t}=-37\,$dBm). At this gate voltage value, the intrinsic nonlinearity dominates the electromechanical nonlinearity.
\textbf{b} Same two-tone spectroscopy as in \textbf{a} with subtracted electromechanical contribution, leading to a minimal distortion.
\textbf{c} Variation of the current ($|\Delta I| = | I - \langle I \rangle_\text{sweep}|$) as a function of the driving frequency $f_\mathrm{r}$ and the gate voltage at low driving power ($P_\mathrm{r}=-60\,$dBm). The green arrow indicates the gate voltage position $V_\mathrm{g} = -6.22\,\mathrm{V}$.
\textbf{d} Electromechanical contribution to the variations of the  radial  mode resonance frequency  $\omega_{\mathrm{r,elec}}(u_\mathrm{t})$ as a function of the  tangential  mode displacement $u_\mathrm{t}$.
\textbf{e} Fit of $\omega_{\mathrm{r,int}}(u_\mathrm{t})$ using main text equation \ref{eq:omegat}. The selected data points with their associated error bars are shown in black. The dotted red line is the best fit and the light red area corresponds to the uncertainty on the obtained intrinsic nonlinearity $\alpha_{3,\mathrm{r}}$.
}
\label{fig:Dataset-6220mV_removal_electromechanical_coupling}
\end{figure}

Fig.\ref{fig:Dataset-6220mV_removal_electromechanical_coupling} illustrates our method to remove the electromechanical contribution to the frequency shift of the  radial  mode. It is similar to the Fig.\ref{fig3} of the main-text with the only exception that the data shown in Fig.\ref{fig:Dataset-6220mV_removal_electromechanical_coupling} are taken from the dataset around $V_\mathrm{g}=-6.2\,\mathrm{V}$ for which the maximum of the intrinsic nonlinearity is reached. In the Fig.\ref{fig3} of the main-text, we chose the dataset at $V_\mathrm{g}=-9.2\,\mathrm{V}$ because the contributions of the electromechanical coupling and the intrinsic coupling are comparable, which allows for a better illustration of the method. As shown in panel d, the electromechanical coupling induces a modulation $\omega_\text{r,elec}/2\pi$ of less than $0.1\,\mathrm{MHz}$. In contrast, the intrinsic nonlinearity shifts the bare frequency $\omega_{\mathrm{r},0}$ more than $3\,\mathrm{MHz}$, as shown in panel b.


\section{Estimation of the uncertainty in $\alpha_\text{3,r}$}
\label{SI-estimationuncertainty}

The goal of this section is to explain the derivation of the uncertainty in $\alpha_\text{3,r}(V_\text{g}^\text{DC})$ shown in Fig.\ref{fig4}. \\
Multiple sources of error must be considered in order to accurately quantify the uncertainty in $\alpha_\text{3,r}$:
\begin{enumerate}
    \item \textbf{Systematics from the geometry of the system:} The dimensions of the geometry of the electrodes can be extracted from the AFM and SEM measurements presented in Fig.\ref{fig:AFM-SEM}. However, there is an uncertainty due to the measurement accuracy of the AFM and SEM. The size of the quantum dot inside the carbon nanotube is also unknown.
    \item \textbf{Systematics on $z_\text{st}$ and $\theta$:} The uncertainties on the static displacement of the carbon nanotube $z_\text{st}$ and the polarization angle $\theta$ are estimated from the sensitivity analysis of the Euler–Bernoulli model, as presented in Supplementary Section \ref{SI-th-sensitivity-analysis}.
    \item \textbf{Uncertainty in $V_\text{g}$:} 
    This uncertainty includes the measurement uncertainty in $V_\text{g}^\text{DC}$ and $V_\text{g}^0$ resulting from the finite $V_\text{g}$ step size in the $f_\text{t}$-$V_\text{g}$ measurements. An additional uncertainty in $V_\text{g}^\text{DC}$, directly associated with the $f_\text{t}$-$\Delta V_\text{g}$ conversion, arises from the shift of the Coulomb peak induced by the increased oscillation amplitude of the carbon nanotube. It is worth noting that $V_\text{g}^0$ is not affected by this latter contribution, as it corresponds to the original position of the Coulomb peak without carbon nanotube actuation.
    
    \item \textbf{Uncertainty in $f_\text{r}$:} This uncertainty arises from the finite step size used for the sweeping of the  radial  frequency in the $f_\text{t}$-$f_\text{r}$ measurements and the finite linewidth of the resonance.
\end{enumerate}
\subsection{Uncertainty contributions to $\alpha_\text{3,r}$}
\label{SI-estimationuncertainty-contributions}
The sources contributing to the final uncertainty in $\alpha_\text{3,r}$ can be identified from equation \ref{eq:freq_freq_fit}, which can be recast as
\begin{equation}
    f_{\text{r,int}}(V_\text{g}^\text{DC})\,=\,\frac{1}{2\pi} \sqrt{4\pi^2f_{\text{r},0}^2(V_\text{g}^\text{DC})+\frac{\alpha_{3\text{, r}}}{2} u_{\text{t}}^2}\,.
\end{equation}
Since $\alpha_\text{3,r}$ is the free fit parameter in the method, its final uncertainty is determined by the uncertainties in $f_\text{r}$ and $f_\text{r,0}$ (error source 4), together with the uncertainty in $u_\text{t}$ (error sources 1-3). \\
Following the derivation provided in Supplementary Section \ref{SI-alpha3t-DeltaVgus-DeltaVgus}, $u_\text{t}$ may be expressed as
\begin{equation}
    u_\text{t}\,=\,\left|\frac{V_\text{dot, 10}(z_\mathrm{st}(V_\text{g}^\text{DC}))}{E_{10}(z_\mathrm{st}(V_\text{g}^\text{DC}))}\cdot \left(1-\frac{V_\text{g}^0 }{V_\text{g}^\text{DC}}\right) \cdot \frac{1}{\cos(\theta)}\right|\,.
    \label{eq:us_error}
\end{equation}
Three distinct terms associated with different error sources can be identified in equation \ref{eq:us_error}: 
\begin{itemize}
    \item $V_{\text{dot},10}(z_\mathrm{st}(V_\text{g}^\text{DC}))/E_{10}(z_\mathrm{st}(V_\text{g}^\text{DC}))$, which combines the geometric systematics (error source 1) with the systematic uncertainty in $z_\text{st}$ (the first contribution to error source 2)
    \item $\left(1 - V_\text{g}^0 / V_\text{g}^\text{DC}\right)$, which includes the uncertainty in $V_\text{g}$ (error source 3) through $V_\text{g}^0$ and $V_\text{g}^\text{DC}$
    \item $1/\cos(\theta)$, which contains the systematic uncertainty in $\theta$ (the second contribution to error source 2)
\end{itemize}

\subsection{Uncertainty in $u_\text{t}$}
\label{SI-estimationuncertainty-us}

\begin{figure}[H]
  \centering
  \includegraphics[width=0.8\linewidth]{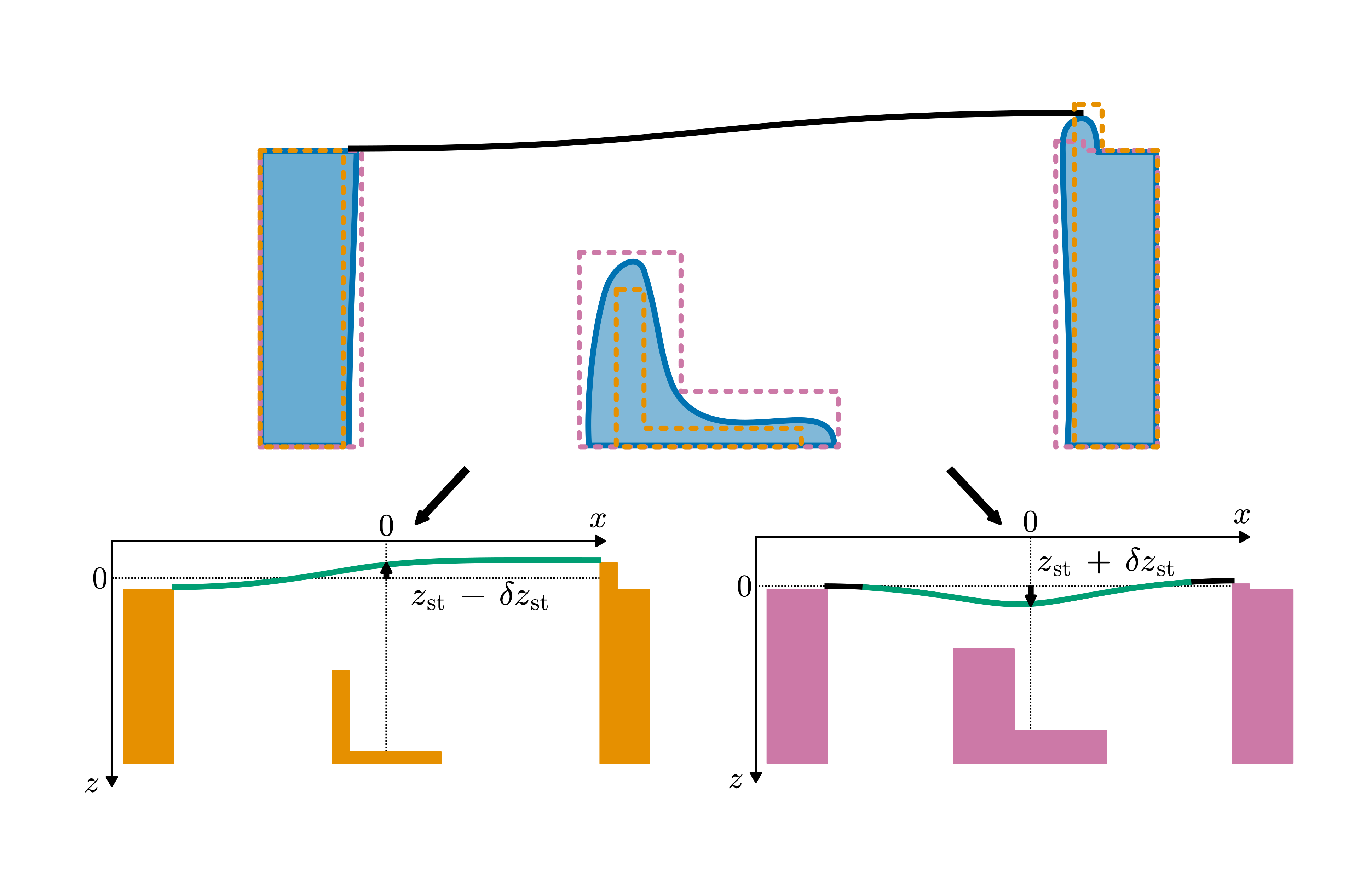}
  \caption{\textbf{Schematic of the estimated geometrical uncertainty of the system.} The true geometry of the system is unknown (illustrated by the blue shapes) and is assumed to lie between the dashed orange and dashed pink boundary geometries (top panel), estimated from the AFM and SEM measurements. In the bottom left panel, the orange boundary geometry A is shown, including the uncertainty in $z_\text{st}$ and the size of the quantum dot (green), chosen to minimize $V_\text{dot}$. In the bottom right panel, the pink boundary geometry B is shown, also including the uncertainty in $z_\text{st}$ and the size of the quantum dot (green), chosen to maximize $V_\text{dot}$.}
  \label{fig:error_geometry}
\end{figure}

Since the uncertainties in $V_{\text{dot},10}(z_\mathrm{st}(V_\text{g}^\text{DC}))$ and $E_{10}(z_\mathrm{st}(V_\text{g}^\text{DC}))$ are correlated through their common dependence on the geometry of the system, it is best to directly evaluate the uncertainty of the ratio $V_{\text{dot},10}(z_\mathrm{st}(V_\text{g}^\text{DC}))/E_{10}(z_\mathrm{st}(V_\text{g}^\text{DC}))$, which is proportional to $u_\text{t}$. This uncertainty can be determined by evaluating the two extreme cases that produce the minimum and maximum values of the ratio, while accounting for both the uncertainty in the geometry of the system and that in $z_\mathrm{st}$ ($\delta z_\text{st} = \pm 1.1635\,\text{nm}$ from Supplementary Section \ref{SI-th-sensitivity-analysis}). This error-estimation approach corresponds to an envelope procedure, also referred to as a Type B standard uncertainty evaluation \cite{JCGMGUM}, in which upper and lower bounds are derived for the quantity of interest. \\\\
The two boundary geometries considered in this analysis are constrained by the values extracted from the AFM and SEM measurements and are illustrated in Figure \ref{fig:error_geometry}. The uncertainty in the relevant geometric dimensions of the device (the spacing between the contact electrodes, the size of the gate electrode, and the height difference between the two contact electrodes) is estimated to be $\pm 10\,\text{nm}$. This uncertainty is attributed to the finite precision of the AFM and SEM measurements, which only allows an approximate reconstruction of the exact shape of the device components. The boundary geometries are assumed to consist of piecewise straight-edged shapes, whereas the actual device geometry may deviate from this idealization (see blue shapes in Figure \ref{fig:error_geometry}). However, it is assumed to lie somewhere within the two following boundary cases:
\begin{itemize}
    \item \textbf{Boundary case A:} small and narrow gate electrode, quantum dot extending over the full suspended length of the carbon nanotube, nanotube shifted away from the gate electrode by $-\delta z_\text{st}$ (orange in Figure \ref{fig:error_geometry}).
    \item \textbf{Boundary case B:} large gate electrode, reduced suspended length, quantum dot extends over only 80\% of the suspended carbon-nanotube length, nanotube shifted toward the gate electrode by $\delta z_\text{st}$ (pink in Figure \ref{fig:error_geometry}).
\end{itemize}
These two boundary cases were chosen to provide an envelope for the ratio between the quantum dot potential and the electric field experienced by the quantum dot, since both quantities vary substantially between the two cases.
For each of the two cases, the electrostatic potential of the system is simulated using a finite-element model in COMSOL \cite{comsol}. The mean electrostatic potential of the quantum dot and the electric field experienced by the quantum dot, for an applied gate voltage of $V_\text{g} = 10\,\text{V}$, are then calculated from
    \begin{equation}
        V_\text{dot, 10}(z_\mathrm{st})\,=\, \int_{x\,=\,-l/2}^{x\,=\,l/2}V_{10}(x,\,z_\text{st}(x))\,\mathrm{d}x,
    \end{equation}
    and
    \begin{equation}       
    E_{10}(z_\mathrm{st},\,V_\text{g}^\text{DC})\,=\,\frac{\partial V_\text{dot, 10}(z_\mathrm{st})}{\partial z_\mathrm{st}}.
    \end{equation}

Importantly, even if the uncertainties on $V_{\text{dot},10}(z_\mathrm{st}(V_\text{g}^\text{DC}))$ and $E_{10}(z_\mathrm{st}(V_\text{g}^\text{DC}))$ can be large, the strong correlation between these two quantities leads to a comparatively smaller uncertainty in their ratio.
\\\\
Two boundary cases for the amplitude of motion of the  tangential  mode in the $z$-direction $u_{\text{t,}z}$ are derived from the upper and lower bounds of $V_{\text{dot},10}(z_\mathrm{st}(V_\text{g}^\text{DC}))/E_{10}(z_\mathrm{st}(V_\text{g}^\text{DC}))$ with

\begin{equation}
    u_{\text{t,}z\text{ up/low}}\,=\,\left|\frac{V_\text{dot, 10}(z_\mathrm{st}(V_\text{g}^\text{DC}))}{E_{10}(z_\mathrm{st}(V_\text{g}^\text{DC}))}\right|_\text{up/low}\cdot\,\left|1-\frac{V_\text{g}^0}{V_\text{g}^\text{DC}}\right|.
\end{equation}
Since the uncertainty in the gate voltage $\delta V_\text{g}$ is uncorrelated with both the uncertainty in the geometry and that in $z_\text{st}$, it can be added independently in quadrature to the uncertainty obtained from the difference of the lower and upper bounds to the $u_{\text{t,}z}$ obtained from the mean geometry. The uncertainties in $V_\text{g}^\text{DC}$ and $V_\text{g}^0$ are assumed to be $\pm 5\,\text{mV}$ and $\pm 1\,\text{mV}$, respectively. They can be treated as uncorrelated and can therefore be propagated independently, leading to
\begin{equation}
    \delta u_{\text{t,}z\text{ up/low}}\,=\,\sqrt{(u_{\text{t,}z\text{ up/low}}-u_{\text{t,}z})^2\,+\,\left(\frac{\partial u_{\text{t,}z}}{\partial V_\text{g}^\text{DC}}\, \cdot \, \delta{ V_\text{g}^\text{DC}}\right)^2\,+\,\left(\frac{\partial u_{\text{t,}z}}{\partial V_\text{g}^0}\, \cdot \, \delta{ V_\text{g}^0}\right)^2}
\end{equation}
\\
To propagate the uncertainties $\delta u_{\text{t},z\text{ up/low}}$ to the total  tangential  mode amplitude, the uncertainty in the conversion factor from $u_{\text{t},z}$ to $u_\text{t}$, given by $1/\cos(\theta)$, must be determined. For this purpose, the uncertainty in $\theta$, $\delta \theta = \pm 3.4^\circ$, is propagated through $\cos(\theta)$:
\begin{equation}    \delta{\cos(\theta)}\,=\,\sin(\theta)\,\cdot \, \delta\theta.
\end{equation}
The resulting uncertainty is subsequently added in quadrature to the one in $u_{\text{t},z\text{ up/low}}$ to obtain upper and lower bounds on $u_\text{t}$:
\begin{equation}
    \delta u_\text{t up/low} \,=\, \sqrt{\left(\frac{\partial u_{\text{t}}}{\partial u_{\text{t,}z}}\, \cdot \, \delta {u_{\text{t,}z\text{ up/low}}}\right)^2\,+\,\left(\frac{\partial u_{\text{t}}}{\partial \cos(\theta)}\, \cdot \, \delta{\cos(\theta)}\right)^2}.
\end{equation}

\subsection{Uncertainty in $\alpha_\text{3,r}$}
\label{SI-estimationuncertainty-alpha3t}
\begin{figure}[H]
  \centering
  \includegraphics[width=0.7\linewidth]{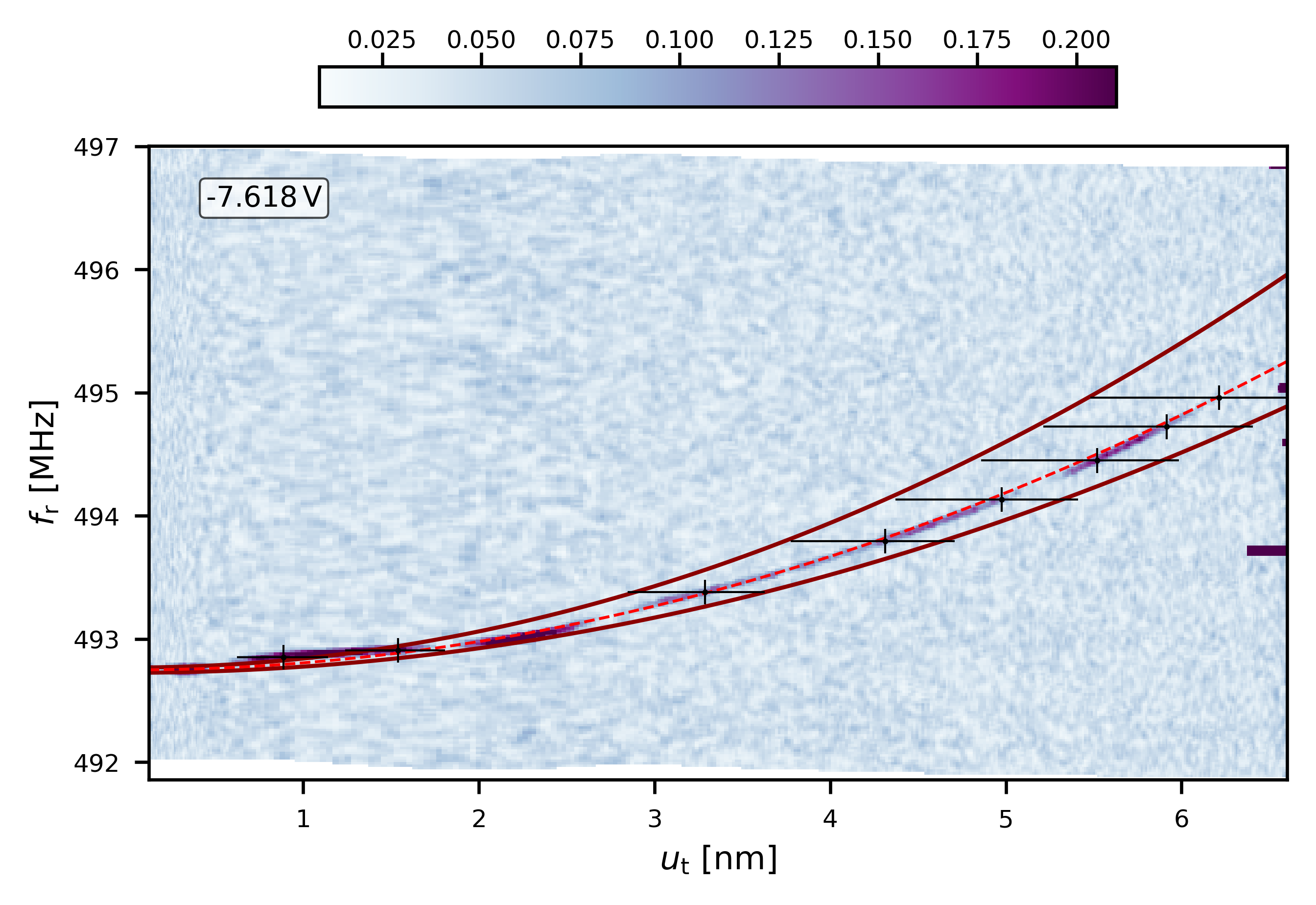}
    \caption{\textbf{Fit of $\omega_\text{r,int}/2\pi$ used to determine the upper and lower bounds of $\alpha_\text{3,r}$.} $\Delta G$ is measured in \unit{\micro\siemens} (colorbar). The central value of $\alpha_\text{3,r}$ is obtained by fitting the manually extracted black data points, resulting in the dashed red line. The upper and lower bounds of $\alpha_\text{3,r}$ are determined by accounting for the black error bars in both the $u_\text{t}$ and $f_\text{r}$ directions. In particular, the upper and lower bounds of $u_\text{t}$ are fitted separately, while the uncertainty in $f_\text{r}$ and $f_{\text{r},0}$ are included as errors in the fit procedure. The values of $\alpha_\text{3,r}$ obtained from the dark red lines define the upper and lower bounds of $\alpha_\text{3,r}$.}
  \label{fig:fit_error_geom}
\end{figure}

The uncertainty in $\alpha_\text{3,r}$ is determined by fitting $\omega_\text{r,int}/2\pi$ with respect to $u_\text{t}$ (data points manually extracted from the $f_\text{t}$-$f_\text{r}$ measurements), while replacing $u_\text{t}$ by $u_\text{t} + \delta u_\text{t up}$ and $u_\text{t} - \delta u_\text{t low}$ (see Figure \ref{fig:fit_error_geom}). This procedure results in conservative upper and lower bounds in $\alpha_\text{3,r}$, corresponding to the two boundary cases of the $f_\text{t}$-$u_\text{t}$ conversion. The finite linewidth of the resonance and the finite step size in $f_\text{r}$ of the measurement limit the uncertainties in $f_\text{r}$ and $f_{\text{r},0}$ to $\pm 0.1\,\text{MHz}$. Since these uncertainties are not systematic contributions, they are included as errors in the fit procedure.
Applying this procedure to all measurements within a given dataset yields $N$ upper and lower bounds for $\alpha_\text{3,r}$, where $N$ denotes the number of measurements in the dataset. Since the same uncertainty-propagation procedure is applied consistently to all measurements within a dataset, the uncertainties in $\alpha_\text{3,r}$ are combined linearly. The resulting upper and lower bounds of $\alpha_\text{3,r}$ for a single dataset are therefore given by
\begin{equation}
    \alpha_\text{3,r up/low}
    =
    \frac{1}{N}
    \sum_{i=1}^{N}
    \left(
    \alpha_\text{3,r up/low}
    \right)_i
\end{equation}
and are shown in Figure \ref{fig4}.
\subsection{Influence of the uncertainties from the Euler-Bernoulli model}
\label{SI-estimationuncertainty-influenceEBmodel}
\begin{figure}[H]
  \centering
  \includegraphics[width=1\linewidth]{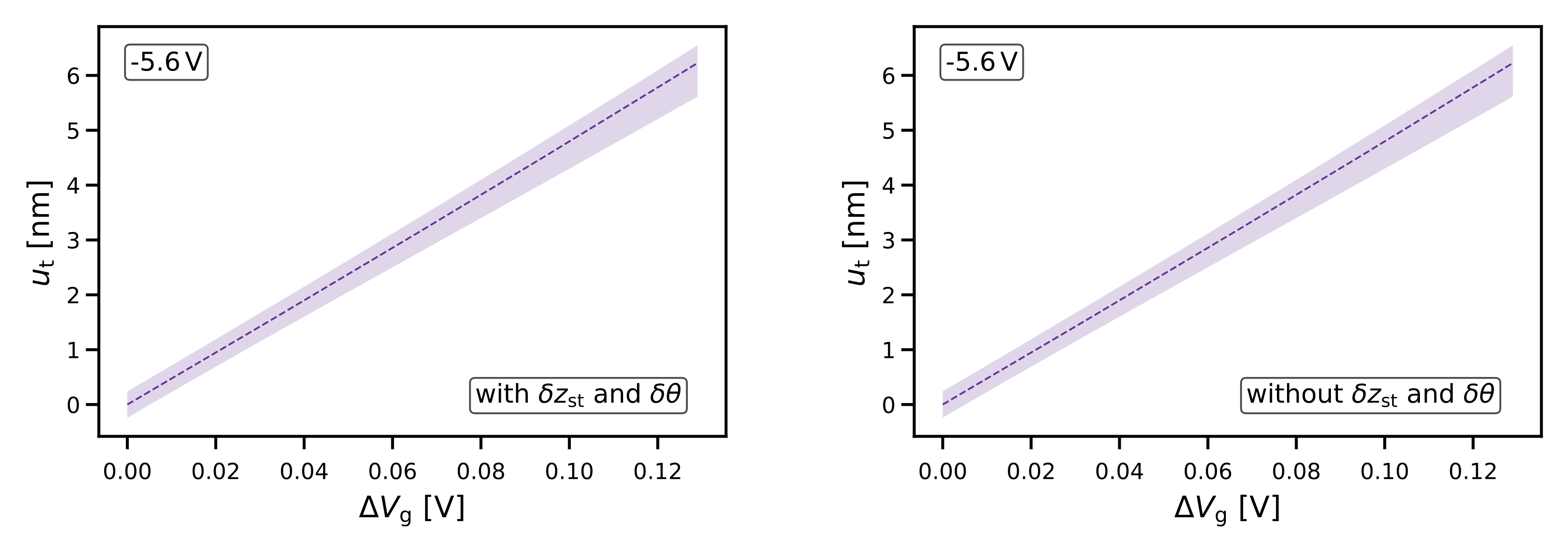}
  \caption{\textbf{Effect of $\delta z_\text{st}$ and $\delta \theta$ on the uncertainty of the $\Delta V_\text{g}$--$u_\text{t}$ conversion for the dataset at $V_\text{g}\,=\,-5.6\,\text{V}$.} On the left, the uncertainty analysis is performed with the uncertainties in $z_\text{st}$ and $\theta$ included, whereas on the right these contributions are omitted. The uncertainty region is represented by the purple band, while the conversion used in the analysis is shown as the dashed dark-purple line. The uncertainty region does not change substantially between the two plots, indicating that the contributions from $\delta z_\text{st}$ and $\delta \theta$ to the uncertainty band are much smaller than those from the other error sources.}
  \label{fig:effect_zst_theta}
\end{figure}

Finally, the impact of the uncertainties arising from the Euler-Bernoulli model (error source 2), as introduced in Supplementary Section \ref{SI-th-sensitivity-analysis}, is examined. To evaluate the contribution of the uncertainties in the static displacement at the center of the carbon nanotube, $\delta z_\text{st}$, and in the polarization angle, $\delta \theta$, these contributions are either included or omitted in Figure \ref{fig:effect_zst_theta}. The comparison shows that the contribution of this error source to the final conversion is negligible, demonstrating that the presented model is sufficiently consistent and stable.

\section{Nonlinearity near the quantum ground state}
\label{SI-alpha3t-nonlinearity}
\subsection{Ratio $u_\text{nl}/u_\text{zp}$}
\label{SI-alpha3t-nonlinearity-unluzp}
\begin{figure}[H]
\centering
\includegraphics[width=1\linewidth]{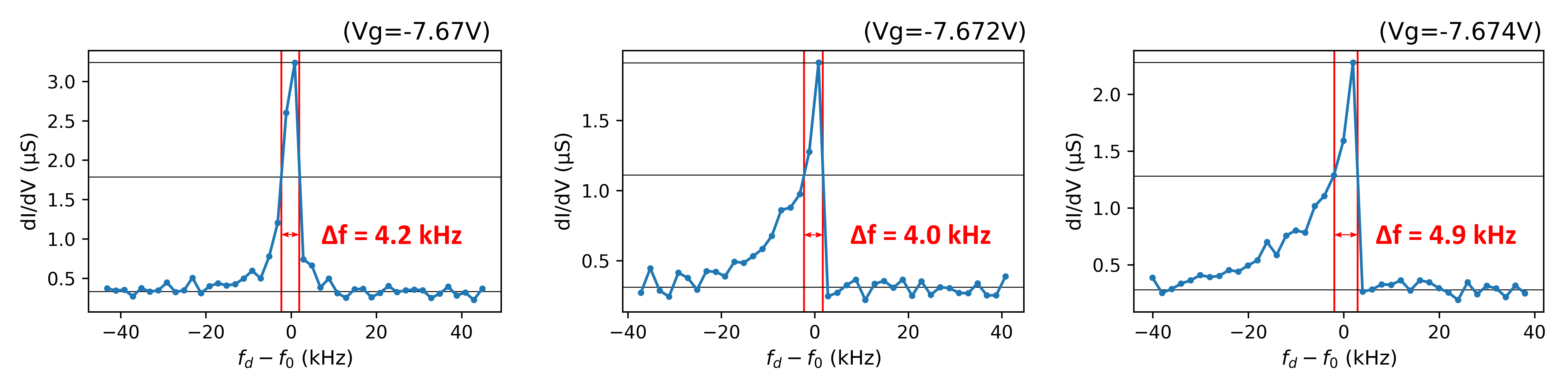}
\caption{\textbf{Lowest measured frequency linewidth in the weakly-driven regime ($P=-65 \,$dBm)}
}
\label{fig:estimation_linewidth_radialmode}
\end{figure}

One way to quantify the nonlinearity of our system is to define $u_\text{nl}$ which is the amplitude where the amplitude-frequency response begins to exhibit two stable steady-state solutions. According to \cite{Lif08}, it is defined as
\begin{equation}
    u_\text{nl}\,=\,\sqrt{\frac{8}{3\sqrt{3}}\frac{\Delta \omega_{\text{r},0} \,\omega_{\text{r},0}(V_\text{g}^\text{DC})}{|\alpha_{3,\text{r}}|}}
    \label{eq:xnl}
\end{equation}
in the absence of nonlinear damping. $\Delta \omega_\text{r,0}$ is defined as the full width at half maximum of the resonance of the  radial  mode. In the calculation of the critical phonon number, we considered the frequency linewidth to be $\Delta \omega_{\text{r},0} = 2 \pi \times 4 \,$kHz, which corresponds to the lowest measured frequency linewidth of the  radial  mode (Figure \ref{fig:estimation_linewidth_radialmode}). \\
Importantly, the measurement setup used in the present work does not allow to measure reliably the mechanical quality factor of bending mode of the carbon nanotube and only leads to an underestimation of the quality factor. 
Indeed, the measurement technique used here is based on a DC readout relying on the rectification behavior of the carbon nanotube when driven on-resonance \cite{Hut09}\cite{Steele09}. The low sensitivity of this detection method requires to drive the nanotube into large displacement. In ref.\cite{Sam23}, the authors show that such large displacement strongly reduced the measured quality factor, and they observe an increase of the quality factor by a factor 200 ($Q=1700$ to $Q=4 \times 10^5$) by changing the measurement technique from a driven method (FM mixing technique) to a non-driven method (thermal vibration measurement). They propose that this strong quality factor enhancement could be explained by nonlinear damping forces.
\\
In Figure \ref{fig:estimation_linewidth_radialmode} we show three resonance curves of the  radial  mode measured on the device studied in this work. These measurements are taken at low RF power applied on the antenna ($P=-65\,$dBm). The extracted FWHM for the three resonances are in the 4-5 kHz range.

As discussed above, the observed FWHM are a strong over-estimation of the intrinsic mechanical linewidth. Moreover, all three resonances are obtained in the nonlinear regime (the low displacements required to reach the linear regime were not detectable), which further artificially increases the linewidth. For these two reasons, the measured value $\Delta \omega_{\text{r},0} = 2 \pi \times 4 \,$kHz can be taken as an upper bound of the intrinsic frequency linewidth.
\\ \\
Strong nonlinearities near the quantum ground state are linked to the ratio $u_\text{nl}/u_\text{zp}$ (also known as critical phonon number) being smaller than one, with $u_\text{zp}$ the vibration displacement fluctuations or zero-point amplitude, defined as
\begin{equation}
    u_\text{zp}\,=\,\sqrt{\frac{\hbar}{2\omega_\text{r,0}(V_\text{g}^\text{DC})m_\text{r}}}
    \label{eq:xzp}
\end{equation}    
with $m_\text{r}\,=\,0.5 \cdot m$ the modal mass of the  radial  mode and $m$ the mass of the carbon nanotube, which is defined as \cite{Laurent10}
\begin{equation}
    m\,[\,\text {kg}\,]\,=\,\frac{10^{-3}}{1315}\pi \,L\,d.
    \label{eq:mass}
\end{equation}
The length of the carbon nanotube $L$ and the diameter $d$ are listed in table \ref{table:fixparam}. \\ \\
The minimum amplitude for the onset of bistability is found for the maximal intrinsic nonlinearity of the  radial  mode, which in our case is found in the dataset around $V_\text{g}\,=\,-6.2\,\text{V}$. The value of the cubic nonlinearity for this dataset is $9.4\times10^{33}\, \mathrm{m^{-2}s^{-2}}$. \\
In this case, the ratio $u_\mathrm{nl}/u_\mathrm{zp}$, which provides an upper bound for the critical phonon number, is equal to 
\begin{equation}
    \frac{u_\text{nl}}{u_\text{zp}}\,=\,34.0.
\end{equation}
This value is approximately twice that reported in Ref.~\cite{Sam23} and is two orders of magnitude below previous experiments for pure intrinsic nonlinearities \cite{Eichler12}. \\ 
The uncertainty on the ratio is calculated using Gaussian propagation of uncertainty on $\alpha_\text{3,r}$ from the previous section.

\subsubsection{Estimation of the uncertainty in $u_\text{nl}/u_\text{zp}$}
\label{SI-alpha3t-nonlinearity-unluzp-uncertainty}
To estimate the uncertainty in the ratio $u_\text{nl}/u_\text{zp}$, standard error propagation is applied to
\begin{equation}
    \frac{u_\text{nl}}{u_\text{zp}}
    \,=\,
    \sqrt{
    \frac{16}{3\sqrt{3}}
    \frac{\Delta \omega_{\text{r},0}\,\omega_{\text{r},0}^2\,m_\text{r}}
    {\hbar\,|\alpha_\text{3,r}|}
    }.
\end{equation}

The uncertainty in the mass $m_\text{r}$ depends on the uncertainties in the nanotube diameter $d$ and suspension length $L$. From the sensitivity analysis presented in Supplementary Section \ref{SI-th-sensitivity-analysis}, the diameter uncertainty is determined to be $\delta d = \pm 0.4\,\text{nm}$. The uncertainty in the suspension length is estimated as $\delta L = \pm10\,\text{nm}$, arising from the uncertainty in the spacing of the contact electrodes, as discussed in Supplementary Section \ref{SI-estimationuncertainty-us}. This results in
\begin{equation}
    \delta m_\text{r}
    \,=\,
    \sqrt{
    \left(
    \frac{10^{-3}}{1315}\pi d\,\delta L
    \right)^2
    +
    \left(
    \frac{10^{-3}}{1315}\pi L\,\delta d
    \right)^2
    }.
\end{equation}

The uncertainty in the resonance frequency of the  radial  mode is constrained by the finite $f_\text{r}$-step size in the $f_\text{t}$-$f_\text{r}$ measurements and the linewidth of $\omega_{\text{r},0}$ to $\delta \omega_{\text{r},0} = \pm 0.1\,\text{MHz}$. \\ \\
The uncertainty in $\alpha_\text{3,r}$ is determined following the procedure presented in Supplementary Section \ref{SI-estimationuncertainty} for the dataset at $V_\text{g} = -6.2\,\text{V}$, yielding $\delta \alpha_\text{3,r up} = +2.3\times10^{33}\,\mathrm{m^{-2}s^{-2}}$ and $\delta \alpha_\text{3,r low} = -1.1\times10^{33}\,\mathrm{m^{-2}s^{-2}}$. \\ \\
We assume the uncertainty in the frequency linewidth $\Delta \omega_{\text{r},0}$ to be dominated by the measurement step size employed in Figure \ref{fig:estimation_linewidth_radialmode}. Accordingly, $\delta \Delta \omega_{\text{r},0} =  \pm 2 \pi \times2 \,$kHz is used in the calculation of the error in the critical phonon number. \\ \\
The final uncertainty in the critical phonon number is obtained by standard error propagation:
\begin{align}
    \delta \left( \frac{u_\text{nl}}{u_\text{zp}}\right)_\text{up/down}\,=\, & \left(\left(\frac{1}{2 \,|\alpha_\text{3,r}|}\cdot \frac{u_\text{nl}}{u_\text{zp}}\cdot \delta\alpha_\text{3,r down/up}\right)^2\,+\,\left(\frac{1}{2 \,m_\text{r}}\cdot \frac{u_\text{nl}}{u_\text{zp}}\cdot \delta m_\text{r}\right)^2\right. \\
    & \left. \,+\,\left(\frac{1}{2 \,\Delta \omega_{\text{r},0}}\cdot \frac{u_\text{nl}}{u_\text{zp}}\cdot \delta \Delta \omega_{\text{r},0}\right)^2 
    \,+\,\left(\frac{1}{\omega_{\text{r},0}}\cdot \frac{u_\text{nl}}{u_\text{zp}}\cdot \delta\omega_{\text{r},0}\right)^2 \right)^{1/2}. \nonumber
\end{align}
The values $\delta (u_\text{nl}/u_\text{zp})_\text{up} = 8.9$ and $\delta (u_\text{nl}/u_\text{zp})_\text{down} = 9.7$ are obtained. Since they are nearly identical, the value of the uncertainty is taken as $9.7$, resulting in
\begin{equation}
    \frac{u_\text{nl}}{u_\text{zp}} = 34.0 \pm 9.7\,.
\end{equation}
\\\\

\subsubsection{Extrapolation of $u_\text{nl}/u_\text{zp}$}
\begin{figure}[H]
  \centering
  \includegraphics[width=0.67\linewidth]{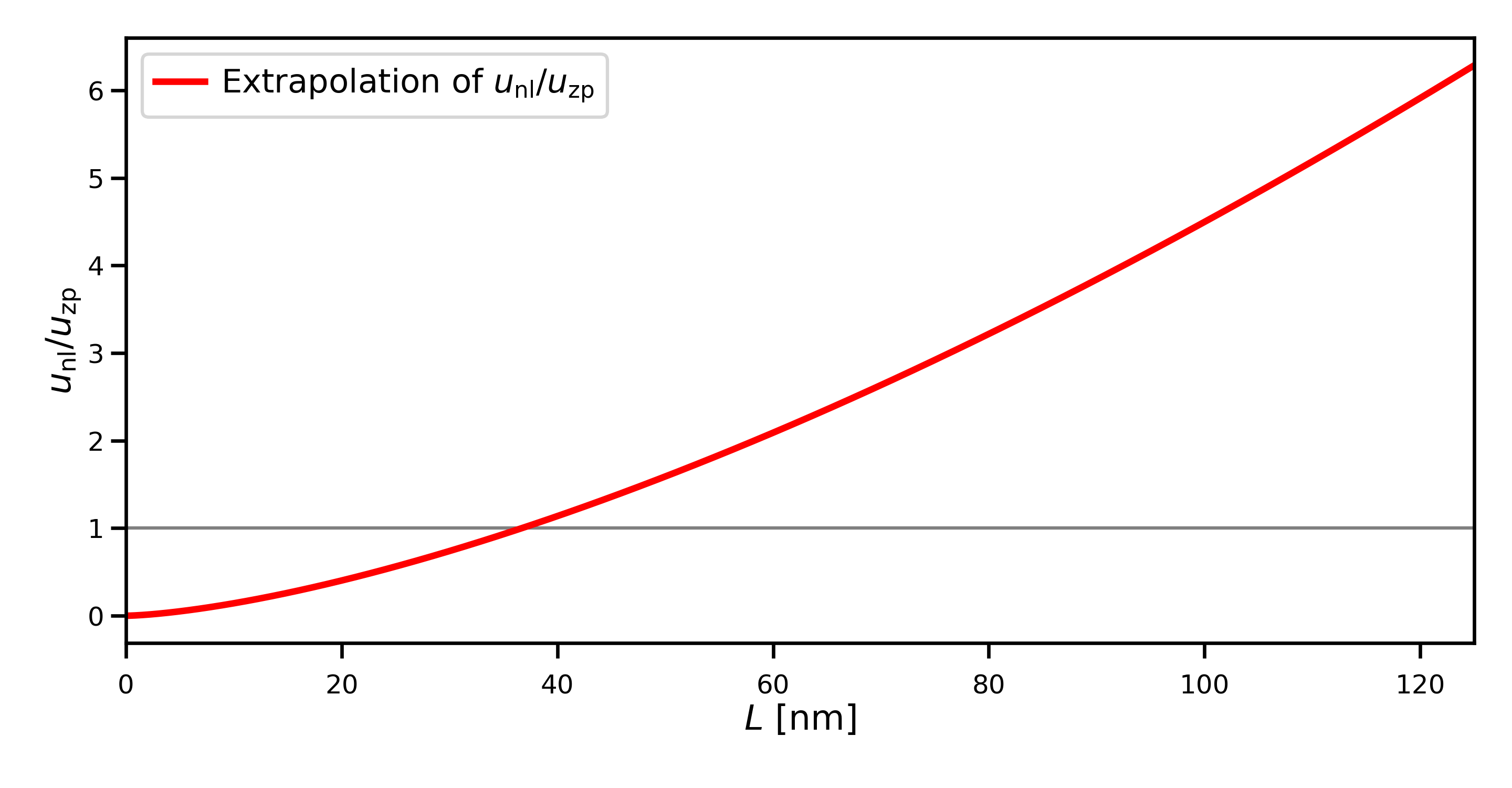}
  \caption{\textbf{Extrapolation of the ratio $u_\text{nl}/u_\text{zp}$ for shorter resonators}. $u_\text{nl}/u_\text{zp}\,=\,1.0$ is indicated by an horizontal gray line.}
  \label{fig:xnl-xzp}
\end{figure}

To extrapolate this ratio for realistically shorter resonators, the length dependence of the involved parameters is determined:
\begin{align}
   & m_\text{r} \;\propto\; L                && \text{\cite{Laurent10}}\\
   & \alpha_{3,\text{r}}\;\propto\; 1/L^4   && \text{(from Euler–Bernoulli theory)}\\
   & \omega_{\text{r,0}}\;\propto\; 1/L      && \text{(tension regime) \cite{Saz06}} \label{eq:length_dependencies_omega}\\
   & \Delta\omega_\text{r,0} \;\propto\; \text{const}. &&
\end{align}
By implementing these dependencies in the equations \ref{eq:xnl} and \ref{eq:xzp}, we find
\begin{align*}
    & u_\text{nl}\;\propto\;L^{3/2} \\
    & u_\text{zp}\;\propto \;\text{const}. \\
\end{align*}
Therefore, the ratio $u_\text{nl}/u_\text{zp}$ is proportional to $L^{3/2}$. The ratio $u_\text{nl}/u_\text{zp}\,=\,34.0$ found for our resonator with $L\,=\,385\,\text{nm}$ is extrapolated for shorter resonators in figure \ref{fig:xnl-xzp}.

\noindent Following this extrapolation, the resonator length needed to obtain a ratio $u_\text{nl}/u_\text{zp}\,=\,1$ is slightly smaller than $40\,\text{nm}$. It should be noted that this value represents only a lower bound on the true value, since the extracted ratio $u_\mathrm{nl}/u_\mathrm{zp}$ constitutes an upper bound on the critical phonon number.

\subsection{Kerr nonlinearity}
\label{SI-alpha3t-nonlinearity-Kerr}
Another way to characterize the nonlinearity of a nanomechanical system is through its single-phonon anharmonicity \cite{yang2024mechanical}, which corresponds to the Kerr frequency shift $K$ produced by a single phonon.
The Kerr nonlinearity is found in Hamiltonians with corresponding equations of motion showing characteristics of a Duffing oscillator. The link between the intrinsic Duffing nonlinearity $\alpha_\text{3,r}$ computed in our work and the Kerr constant $K$ is made in this section. \\ \\
As a first step, we isolate a single term in equation \ref{eq:EOM1} of the main text that involves $\alpha_\text{3,r}$. Because each term contributing to the intrinsic cubic nonlinearity in equation \ref{eq:EOM1} carries the same prefactor $\alpha_\text{3,r}$, considering e.g. only $\alpha_\text{3,r}u_\text{r}^3$ is enough to derive the Kerr constant. This term will be mapped into the potential part of the Hamiltonian and is derived from $\alpha_\text{3,r}u\subr{r}^3$.\\
By expressing this term in its corresponding energy form, we have
\begin{equation}
 \hat{\mathcal{H}}_\text{NL}\,=\,\frac{m_\text{r}\alpha_\mathrm{3,r}}{4}\hat{u_\mathrm{r}}^4.
\end{equation}
We introduce the standard phonon creation and annihilation operators $\hat{a}$ and $\hat{a}^\dagger$ used for the quantization of the harmonic oscillator, with
\begin{equation}
    \hat{u_\mathrm{r}}\,=\,u_\text{zp}\,(\hat{a}^\dagger+\hat{a}). 
\end{equation}
The nonlinear Duffing term in the Hamiltonian $\hat{\mathcal{H}}_\text{NL}$ can be expressed in terms of $\hat{a}$ and $\hat{a}^\dagger$. Applying the rotating wave approximation and ignoring constant energy shifts it holds
\begin{equation}
    \hat{\mathcal{H}}_\text{NL}\,=\, \frac{3m_\text{r}\alpha_\mathrm{3,r}u_\text{zp}^4}{2} (\hat{a}^\dagger)^2\hat{a}^2.
\end{equation}
The Kerr constant is defined as
\begin{equation}
    \frac{K}{2\pi}\,=\,\frac{3\alpha_\text{3,r}\hbar}{8\pi m_\mathrm{r} \omega_\text{r,0}^2},
    \label{eq:kerr}
\end{equation}
leading to a term of the form of a Kerr nonlinearity describing the phonon-phonon interaction strength
\begin{equation}
 \hat{\mathcal{H}}_\text{Kerr}\equiv\hat{\mathcal{H}}_\text{NL} \,=\,\frac{\hbar K}{2} (\hat{a}^\dagger)^2\hat{a}^2.
\end{equation}

\subsubsection{Extrapolation of the Kerr constant}
By using equation \ref{eq:kerr}, the Kerr constant can be estimated for multiple extrapolations of the geometry of the resonator (length and diameter of the carbon nanotube). \\
The Kerr constant depends on the length and diameter of the resonator via equations \ref{eq:mass} and \ref{eq:length_dependencies_omega}
\begin{equation}
    K\;\propto\,\frac{1}{L^3d}.
\end{equation}
Hence, smaller system geometries lead to an enhanced Kerr constant.
\\
For a resonator of diameter $d\,=\,2\,\text{nm}$ and length $L\,=\,100\,\text{nm}$, the Kerr constant equals $K/2\pi\,=\,913 \,\text{Hz}$. 
For even smaller dimensions of the resonator with $L\,=\,60\,\text{nm}$ and $d\,=\,1.5\,\text{nm}$, we find $K/2\pi\,=\,5636\,\text{Hz}$. \\ \\
Comparing this value with a decoherence rate of a carbon nanotube resonator of $\kappa\,=\,9\,\text{Hz}$, corresponding to a state-of-the-art carbon nanotube resonator as presented in \cite{moser2014nanotube}, we get a ratio of $K/\kappa\,=\,101$ for the first geometry presented, and $K/\kappa\,=\,626$ for the second one. Both ratios are much greater than one, indicating that a purely mechanical system could operate in the single-phonon Kerr regime. Furthermore, the ratio for the smaller geometry even approaches the anharmonicity levels observed in superconducting transmon qubits \cite{krantz2019quantum}.
\newpage

\end{document}